\pdfoutput=1 

\documentclass[twoside,11pt]{PhDthesisPSnPDF}

\def\Tr{{\rm Tr}}

\def\R{\boldsymbol{R}}

\def\R{R_{\rm AdS}}

\newcommand{\Dslash}{D\mkern-11.5mu/\,} 
\newcommand{\delslash}{{\partial\mkern-9mu/}}

\newcommand{\dd}[2]{\frac{d #1}{d #2}}

\def\Dbarslash{\,\,{\raise.15ex\hbox{/}\mkern-12mu {\bar D}}}
\def\Dslash{\,\,{\raise.15ex\hbox{/}\mkern-12mu D}}
\def\delslash{\,\,{\raise.15ex\hbox{/}\mkern-9mu \partial}}
\def\delbarslash{\,\,{\raise.15ex\hbox{/}\mkern-9mu {\bar\partial}}}

\def\a{\alpha}
\def\b{\beta}

\def\d{\delta}
\def\e{\epsilon}

\def\m{\mu}
\def\n{\nu}
\def\o{\theta}
\def\p{\pi}
\def\r{\rho}
\def\s{\sigma}
\def\t{\tau}

\def\w{\omega}
\def\x{\chi}

\def\G{\Gamma}
\def\D{\Delta}
\def\L{\Lambda}

\def\W{\Omega}

\newcommand{\EQ}[1]{\begin{equation} #1 \end{equation}}
\newcommand{\AL}[1]{\begin{subequations}\begin{align} #1
\end{align}\end{subequations}}
\newcommand{\SP}[1]{\begin{equation}\begin{split} #1
\end{split}\end{equation}}

\ifpdf  
    \pdfcatalog { /PageMode (/UseOutlines)
                  /OpenAction (fitbh)  }
\fi

\title{Time-dependent Backgrounds in String Theory and Dualities}

\ifpdf
  \author{\href{mailto:jhutasoi@andrew.cmu.edu}{Jimmy A. Hutasoit}}
  \collegeordept{\href{http://www.cmu.edu/physics/index.html}{Department of Physics}}
  \university{\href{http://www.cmu.edu}{Carnegie Mellon University}}

  \crest{\includegraphics[width=1.5in]{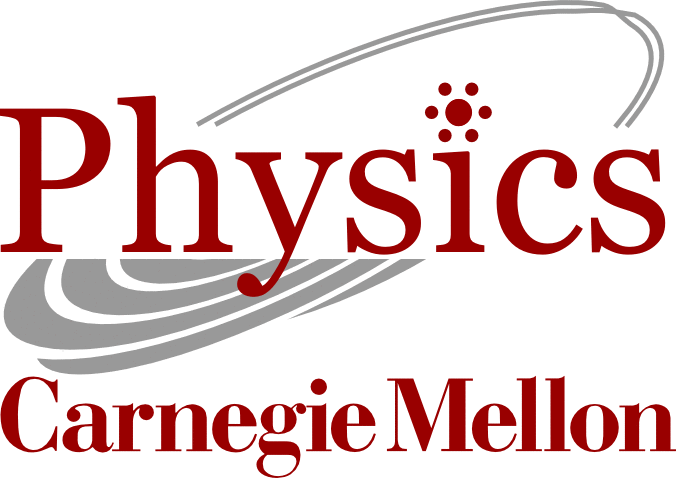}}
  
\else
  \author{YourName}
  \collegeordept{CollegeOrDept}
  \university{University}
  \crest{\includegraphics[width=4cm]{logo.png}}
\fi

\degree{Doctor of Philosophy}
\degreedate{July 2010}

\begin{document}
\frontmatter

\renewcommand\baselinestretch{1.2}
\baselineskip=18pt plus1pt

\maketitle  


\newpage
\thispagestyle{empty}
\vspace{10mm}
Advisor: Prof. Richard Holman

\vspace{5mm}
Committee Members: \begin{itemize} \item[] \hspace{12.5mm}Prof. Daniel Boyanovsky
\item[] \hspace{12.5mm}Prof. Adam Leibovich
\item[] \hspace{12.5mm}Prof. Ira Rothstein
\end{itemize}

\vspace{20mm}
Date of the defense: July $27^{\rm th}$, 2010

\vspace{20mm}
\hfill{Signature from thesis advisor:}



\begin{abstracts}        
This thesis consists of two parts. The first part deals with gauge/gravity duality in the context of anti de Sitter (AdS) spacetimes with de Sitter (dS) boundary, which can be used to study issues concerning strongly coupled field theory on de Sitter space, such as the issue of vacuum ambiguity. By calculating the symmetric two point function of the strongly coupled ${\cal N} = 4$ supersymmetric Yang-Mills theory on de Sitter space, we show that the vacuum ambiguity persists at strong coupling. Furthermore, the extra ambiguity in the strong coupling correlator seems to suggest that transition between two different vacua is allowed. The second part of this thesis deals with the duality between the rolling tachyon backgrounds in superstring theory and the Dyson gas systems. This duality can be interpreted as a reformulation of non-BPS D-branes in superstring theory in terms of statistical systems in thermal equilibrium, whose description does not include time. We argue that even though the concept of time is absent in the statistical dual sitting at equilibrium, the notion of time can emerge at the large number of particles $N \rightarrow \infty$ limit.
\end{abstracts}

\begin{dedication}
\Large{$\tau\tilde{\omega}\iota$ $\sigma o \varphi \tilde{\omega}\iota$}
\end{dedication}

\setcounter{secnumdepth}{4} 
\setcounter{tocdepth}{4}    
\tableofcontents            









\begin{savequote}[15pc]
\sffamily
If I have seen further, it is only by standing on the shoulders of giants.
\qauthor{Isaac Newton (1643 - 1727)}
\end{savequote}
\chapter{\textit{Foreword}}

\vspace{1in}
Several years ago, some graduate students in my department started a new ``tradition" where students who succeeded defending their theses earn the right to write the titles on the wall of our graduate student lounge. Some adopted this tradition, some did not. Among those who left their marks on the white concrete walls of the graduate student lounge, some did so because they felt that they had worked hard during their time here at Carnegie Mellon and they felt that it would be such a waste for the final results of their hard work and tears (!) to be left on a shelf in the library, forgotten and accruing dust. By writing the title on the wall, their masterpiece might still be forgotten, but at least there is a bigger chance that a future graduate student during her procrastinating time might stumble upon that title (and perhaps wonder, ``Why on earth would someone spend 5-6 years of their life doing research on \textit{that}?"). 

\begin{figure}[h]
\begin{center}
\includegraphics[width=5in]{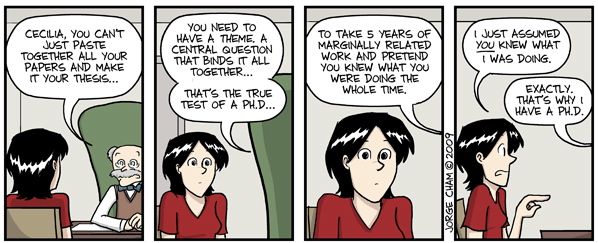} 
\includegraphics[width=5in]{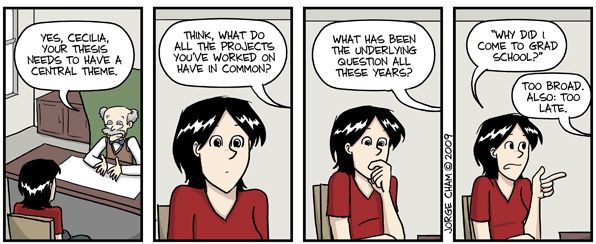} 
\end{center}
\caption{\footnotesize Taken from PhD Comics, originally published on April $24^{\rm th}$, 2009 and May $1^{\rm st}$, 2009.}
\label{phdcomics}
\end{figure}

Yet for some other students, they wrote the titles of their theses on that wall because as they were preparing their theses, they realized that the hardest part of writing a thesis is actually finding its title. I think this is true for many fields of research (see Fig. \ref{phdcomics}), but this is especially true for graduate students working on theoretical physics. To be able to compete in the job market, not only do we need to have a deep knowledge of our field, but we also need to have a broad set of skills. In order to achieve that, our advisors advise us to work on many different projects, most of which might not relate to the others. An extra advantage of working on many different projects is that once you receive a job offer and are ready to leave the student's world, you will have enough materials to write a thesis. The disadvantage is that you have to work hard to create a connection between parts of your thesis to make it \emph{look} like a congruous work. Some times, this connection is real, but other times, it is superficial. This is why for a lot of the theses of graduate students in theoretical physics, the titles can always be replaced by ``Two or Three Things I Did as a Graduate Student." 

In my case, this process of finding a title was not too bad. After some procrastination, I decided that I wanted to have my first single author paper \cite{Hutasoit:2009sc} to be a part of this thesis. As this work was motivated by a result of a work in collaboration with S. Prem Kumar and Jim Rafferty \cite{Hutasoit:2009xy}, I have also included the latter in this thesis and those two will make the bulk of this thesis. As for the rest of the thesis, at first, I wanted to include a series of work in unparticle physics that I did with Dan Boyanovsky and Rich Holman. Both these works and the works I mentioned earlier involve a conformal sector, but I feel that the connection is a bit too weak. After much consideration, I decided to include a work I did with Niko Jokela \cite{Hutasoit:2007wj}. This project with Niko and those that make the bulk of this thesis involve time-dependent backgrounds in string theory and some kinds of dualities\footnote{These works are supported in part by the DOE through Grant No. DE-FG03-91-ER40682. I was also supported by the De Benedetti Family Fellowship in Physics.}. Thus, the issue of finding a title is settled.

This thesis opens with a brief survey of the study of time-dependent backgrounds in string theory. I will start by overviewing the progress of string theory, starting from its conception to the present time. This overview is by no mean a complete one, as I will focus on developments that are relevant to how researchers approach the problem of studying string theory on time-dependent spacetimes. I will refer to some of the representative works on this broad subfield of string theory, but it is inevitable that I will miss some works, even though I am sure their results are significant. For this, I sincerely apologize.

As I mentioned above, the body of this thesis consists of two parts. In the first part, I will talk about backgrounds with de Sitter boundary in the context of gauge/gravity duality \cite{Maldacena:1997re, Aharony:1999ti}. In Chapter \ref{chapter:ads/cftrev}, I will start by reviewing some aspects of gauge/gravity duality that will be relevant to the work presented in this thesis. Chapter \ref{chapter:geom} will set up the duality, namely the duality between strongly coupled ${\mathcal N}=4$ Super Yang-Mills (SYM) theory conformally coupled to a de Sitter background and classical type IIB supergravity living on asymptotically locally anti de Sitter (AdS) spacetimes whose boundary is a de Sitter space. These spacetimes are the so-called ``topological AdS black hole" of \cite{Cai:2002mr,Ross:2004cb,Balasubramanian:2005bg} and the AdS ``bubbles of nothing" found in \cite{Birmingham:2002st,Balasubramanian:2002am}. In Chapter \ref{chapter:phases}, by calculating the retarded correlators of some scalar operators of the strongly coupled boundary theory, in particular the correlators of the scalar glueball operators, I will identify which phases of the boundary theory the aforementioned geometries are dual to. It turns out that the (small) AdS bubble of nothing corresponds to the confining phase, while the topological AdS black hole corresponds to the deconfined plasma phase. In Chapter \ref{chapter:plasma}, I will then study further the transport properties of this plasma. After that, I will close the first part of this thesis by trying to use the gauge/gravity duality to address the issue of vacuum ambiguity in de Sitter space.

In the second part, I will talk about the rolling tachyon backgrounds in superstring theory. These backgrounds are given by the D-branes with the ``wrong" dimensionality, which are unstable non-BPS objects. In Chapter \ref{chapter:gas}, I will introduce a duality between the non-BPS D-branes and a statistical system in thermal equilibrium, namely the paired Dyson gas system. In Chapter \ref{chapter:time}, using this duality, I will argue that the notion of time is not fundamental but emerging in the regime where the number of particles in the Dyson gas system is large. 

This thesis is the curtain that closes a (long) chapter of my life. It is the symbol of the beginning of a new era -- an era I am looking forward to step into -- and also a symbol of an end of an era, an era that ends well with me still in one piece. The latter should not be taken lightly nor for granted, and for this I have a lot of people to thank.

I would like to thank my thesis committee, not only for their questions and suggestions concerning research materials, but also for their advice on how I can better present my results to other people in the scientific community. I am indebted to my advisor, Rich Holman, not only for giving me interesting ideas for research projects, but also for giving me the freedom to chase after other topics that I am interested in. I am also thankful to him for encouraging me to go to conferences, workshops and summer schools, where I could learn more stuff from others and find interesting collaborators. I would like to thank Dan Boyanovsky who started me on the whole neutrino business. Without his introduction to neutrino physics, I would have never thought that a high energy physicist should care about the collapse of wave functions. I am grateful to Cliff Burgess, who helped me start on string cosmology research and got me thinking about the relation between string theory and condensed matter physics. Many thanks to Vijay Balasubramanian, who gave long extra lectures on AdS/CFT at the summer school at Perimeter Institute and who got me interested in the concept of emergent spacetime.

I am thankful for the wonderful research collaborations I have had with Niko Jokela, S. Prem Kumar, James Rafferty and Jun Wu. Special thanks to Prem for his patience in teaching me tricks on how to do holographic calculations. I am grateful to all the friends and colleagues I met through conferences, workshops and summer schools I attended. In particular, I would like to thank Carlos Hoyos and Andy O'Bannon for discussions on topics in AdS/CFT, Borun Chowdhury for discussions on fuzzball and also Luca Grisa for letting me stay at his apartment every time I visited New York and for his crazy idea on how to get more citations\footnote{Luca's idea is to make a pact with a number of young physicists, in which every member of the pact must cite one of everyone else's papers every time she publishes despite the relevance of the works.}. 

Many thanks to other graduate students and postdocs in the HEP Theory group at Carnegie Mellon for discussions on physics (and other topics), especially Chris Anersen, Duff Neil and Ambar Jain, who have also served as Mathematica and $\LaTeX$ help desk for me. I would also like to thank my office mates for making my office life so much better, especially Chang-Yu Hou, Feng Wu and Jason Galyardt, whose friendships I have enjoyed beyond the concrete walls of our offices. Many thanks also go to graduate students from the Department of Physics who have played in an intramural sports team with me: \textit{mens sana in corpore sano}.

\begin{figure}[h]
\begin{center}
\includegraphics[width=5in]{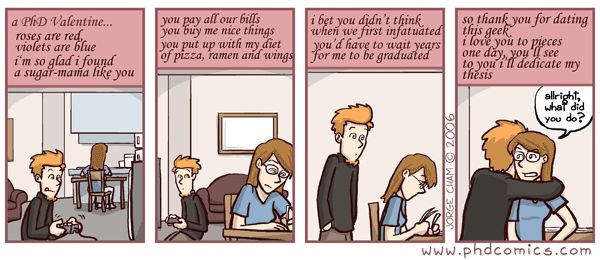} 
\end{center}
\caption{\footnotesize Taken from PhD Comics, originally published on Valentine's Day, 2006.}
\label{phdpoem}
\end{figure}

When a bunch of graduate students in the aforementioned summer school asked him what the secret to success in physics is, Vijay answered that the secret is to have a stable private life outside physics. I agree with him and I would like to thank the following people to help me create such life. I would like to thank my wife, Leeann, for sticking by my side through the high and the low, especially the low of not getting any job offers for a while. I do not think that all aspects of this poem is applicable for every married Ph.D. students, but I think every married Ph.D. students should include Fig. \ref{phdpoem} in their theses.
I would like to thank my family, who support me even though they barely know what I am doing: the Hutasoits and Sindapatis in Indonesia and the Watkins and Kites in northwestern Pennsylvania. Many thanks to my friends who become my ``family" in Pittsburgh: Andrew and Ella Rishikof -- especially Andrew for his prank calls, where for example he pretends to be a dictator from a warring country who wants to hire me to build him a weapon using black hole (?!) -- John and Cheryl Riley, the Weebers, Mike and Robin Namisnak, Clint and Su-Lin Harshman, Matt and Bethany Julian, Sam Kwak, Ruth Helmus, the Hoffmans, Joel Brewton, Steph Kelly, Bethany Santos, Chris and Christian Olivieri, Jason Smith, the Verms, the Fries, Jake Haberman and the Merrys. I would also like to thank the Pittsburgh Steelers for winning two Super Bowls during my stay in Pittsburgh. Believe it or not, my productivity level went up during those two years. Last but not least, I would like to thank my dog, Tigger, for his willingness to supply me with the ultimate Ph.D. student's excuse (just in case I need it) ``The dog ate my thesis!"


\mainmatter

\renewcommand{\chaptername}{} 


\begin{savequote}[15pc]
\sffamily
Those who do not learn from history are doomed to repeat it.
\qauthor{George Santayana (1863 - 1952)}
\end{savequote}
\chapter{Overview} \label{overview}

\section{A Brief History of String Theory}

Nowadays, most people view string theory as a contender for a ``theory of everything," a quantum theory that unifies gravity with other fundamental forces (for an introductory textbook, see for example \cite{Polchinski:1998rq, Polchinski:1998rr}). However, at its birth, the motivation for string theory was a lot more modest. It was invented as an effort to explain strong interaction. 

As early as 1940, it was clear that unlike electrons, strongly interacting protons and neutrons were not point-like particles. One observation that lead to this conclusion is the measurement of their magnetic moment. Their magnetic moment is very different from that of a point-like spin-1/2 electromagnetically charged particles and this difference is so large that it cannot be explained by a small perturbation.

In 1958, Tullio Regge discovered that bound states in quantum mechanics can be organized into families of different angular momentum \cite{Regge:1959mz,Regge:1960zc}. These families are called Regge trajectories. Two years later, Geoffrey Chew and Steven Frautschi recognized that the mesons, which are also strongly interacting particles, made Regge trajectories in straight lines \cite{Chew:1961ev}. In 1968, Gabriele Veneziano noted that the scattering amplitude of four particles on this Regge trajectory can be described using the Euler beta function \cite{Veneziano:1968yb}. The interpretation of this Euler beta function came forth two years later, when Yoichiro Nambu \cite{Nambu:1970si}, Holger Bech Nielsen \cite{Koba:1969rw,Nielsen:1970bc}, and Leonard Susskind \cite{Susskind:1970xm} proposed a description of nuclear forces as vibrating, one-dimensional strings. This is the birth of string theory.

Quantizing a vibrating string moving on a space-time, many predictions were made for the strong interaction. However, most of these predictions directly contradicted experimental findings. A famous example of these predictions is the existence of a massless spin-2 boson. As more and more experimental results contradicted string theory predictions for strong interaction, the physics community lost interest in string theory as a theory of strong interaction and quantum chromodynamics became the main focus of theoretical research on strong interaction.

In 1974, John Schwarz, Joel Scherk \cite{Scherk:1974ca} and independently Tamiaki Yoneya \cite{Yoneya:1974jg} found that the massless spin-2 boson mentioned above had properties that exactly matched those of the graviton. They argued that string theory had failed to catch on because physicists had underestimated its scope: string theory is not a theory of strong interaction, it is a theory of gravity. This led to the development of bosonic string theory, which only has bosons in its spectrum of particles. However, by introducing supersymmetry into the worldsheet description of string theory, it was found that string theory may include fermions in its spectrum. String theories that include fermions in their spectra are now known as heterotic string theories and superstring theories. Since these theories include not only graviton, but also fermions and gauge bosons, it was argued that these theories may lead to the unification of all fundamental forces in nature.

Until early 1990s, the research in string theory focused on the perturbative aspects of string theory and the effective theory descriptions that can be obtained from the worldsheet theory. However, as string theorists see more evidences that the different string theories are actually dual to each other, the focus of the research shifted toward non-perturbative aspects of string theory. In mid 1990s, Joseph Polchinski discovered solitonic objects in superstring theories, called D-branes. D-branes can be thought of as hypersurfaces where open strings can end \cite{Polchinski:1995mt} (for an introductory textbook, see for example \cite{Johnson:2003gi}). In the worldsheet description, D-branes are described by Dirichlet boundary conditions for the fields living on the worldsheet, while in the effective theory description, D-branes are described as extended objects charged under the higher-form gauge fields. The latter led to classification of D-branes using cohomology. Later, it was realized that the more complete way of describing and classifying D-branes is not by cohomology, but by K-theory \cite{Witten:1998cd,Horava:1998jy,GarciaCompean:1998rg,Gukov:1999yn,Hori:1999me,Sharpe:1999qz,Olsen:1999dw}. Using the K-theory classifications allows us to put D-branes, which are supersymmetric objects, on the same footing with other non-perturbative objects that are not supersymmetric. Some of these non-supersymmetric objects are not stable and will decay to the stable D-branes (for a review, see \cite{Sen:1999mg}).

In 1997, inspired by the study of string scattering off D-branes, Juan Maldacena conjectured the gauge/gravity duality \cite{Maldacena:1997re}. The conjecture comes from the realization that in some cases, the gauge theory on the D-branes is decoupled from the gravity living in the bulk. In other words, the open strings that are attached to the D-branes are not interacting with closed strings on the bulk. Such a situation is termed a \textit{decoupling limit}. In these cases, the D-branes have two independent alternative descriptions. From the point of view of closed strings, the D-branes are gravitational sources, and thus we have a gravitational theory on some spacetimes with some background fields. From the point of view of open strings, the physics of the D-branes is described by the appropriate gauge theory. Therefore in such cases, it is often conjectured that the gravitational theory on a class of spacetimes with the appropriate background fields is dual to the gauge theory on the boundary of these spacetimes.

The gauge/gravity duality should be viewed as a full non-perturbative prescription for a quantum theory of gravity, in which the gravity path integral is described with all the fields, including the metric, having fixed boundary values, but fluctuating in the bulk. This path integral therefore includes processes in which the topology of spacetime changes and it should also include regimes in which the notion of spacetime itself is no longer valid. Interestingly, this complicated path integral is equivalent to something simpler and more familiar: a generating functional of a gauge theory living on a manifold, which is the boundary value of the metric.

In a certain regime, the (non-perturbative) gravity path integral is reduced to a sum of path integrals of quantum field theories living on classical spacetimes with the given boundary. This regime corresponds to the boundary field theory having a large rank and a strong coupling, and the spacetimes correspond to different phases of the strongly coupled gauge theory. This means that in this regime, the gauge/gravity duality can be used as a tool to study strongly coupled gauge theories in which the difficult calculation of strongly coupled gauge theories is reduced to a relatively simple classical gravity calculation. In this regime, the gauge/gravity duality has become a very powerful tool and its application, which includes understanding topics such as quark gluon plasma (for reviews, see for example \cite{Gubser:2009md,Janik:2010we}) and strongly correlated condensed matter systems (for reviews, see for example \cite{Hartnoll:2009sz,McGreevy:2009xe}), has enjoyed a large interest from the theoretical physics community.

The discovery of D-branes in mid 1990s not only lead to the conjecture of gauge/gravity duality, but it also helped the progress of finding (semi-)realistic cosmological and particle physics models from string theory. In the framework of flux compactifications, D-branes have been the essential components for stabilizing the size and shapes of the extra dimensions \cite{Giddings:2001yu}. D-branes have also been the essential ingredients to build models of particle physics that include standard model (for a review, see \cite{Uranga:2007zz}).

\section{Time-dependent Backgrounds in String Theory}

Ever since the proposal of John Schwarz, Joel Scherk \cite{Scherk:1974ca} and independently Tamiaki Yoneya \cite{Yoneya:1974jg}, string theory has been viewed as a candidate for quantum theory of gravity. The naive dimensional analysis tells us that the scale of quantum gravity is at the so-called Planck scale of the order of $10^{19}$ GeV, which is about $10^{16}$ order of magnitudes higher than the energy accessible to our current accelerator\footnote{There are however, scenarios where the string scale is at an intermediate scale of the order of $10^{11}$ GeV. See for example \cite{Burgess:1998px,Gioutsos:2006fv,Conlon:2006tj}.}. This means that our best chance to confront string theory with observational/experimental data is through cosmological observations of the early universe, which can be sensitive to the Planck scale \cite{Easther:2001fi,Easther:2001fz,Easther:2002xe,Burgess:2002ub}. 

Furthermore, an important topic which any fundamental theory of gravity eventually has to address 
is that of the evolution of the universe. This entails not only the challenge of finding signatures of string inspired early universe models, but also the challenges of finding answers to questions related to the fate of cosmological singularities and understanding the quantum nature of de Sitter space, which includes answering the question concerning the cosmological constant (``Why is it small but not zero?") and the issue of instability of de Sitter space.

Another big challenge for a quantum theory of gravity is understanding the dynamics of quantum gravity processes, such as black hole evaporation process and tunneling of spacetimes with semi-classical instabilities.

All of the above require us to understand string theory in non-trivial time-dependent (and possibly singular) spacetime backgrounds. Following the historical progress of string theory, there are three main approaches to this issue of string theory on time-dependent spacetimes. Broadly speaking, these approaches are based on the world-sheet description, gauge/gravity duality and the low-energy effective action.

The first approach relies on the world-sheet description of string theory with time-dependent background geometries. Several main avenues of research in this directions are the investigation of time-dependent orbifolds (for a review, see for example \cite{Cornalba:2003kd}), the open string rolling tachyons (for a review, see for example \cite{Sen:2004nf}) and the closed string rolling tachyons (for a review, see for example \cite{Headrick:2004hz}). It is worth mentioning that open string rolling tachyons can be described as statistical systems in thermal equilibrium \cite{Balasubramanian:2006sg,Jokela:2007wi,Hutasoit:2007wj}, the so-called Dyson gases. This duality might lead to further understanding on how time can emerge from a system that does not contain it. We will discuss the case of superstring rolling tachyon in further details in the second part of this thesis. It is also worth mentioning that closed string rolling tachyon condensations connect bosonic, Type 0 and Type II string theories of different dimensions \cite{Hellerman:2006ff,Hellerman:2006hf,Hellerman:2007fc}.

The second approach relies on the insight from gauge/gravity duality to address the issue of cosmological singularity (see for example \cite{Craps:2005wd,Craps:2006yb,Berkooz:2007nm,Craps:2007ch,Awad:2009bh}), the dynamics of black hole evaporation process (see for example \cite{Iizuka:2008hg,Iizuka:2008eb,Chowdhury:2010ct}) and to understand the quantum gravity of de Sitter space (see for example \cite{Strominger:2001pn,Sekino:2009kv}). Gauge/gravity duality also enables us to address the issues we encounter when putting quantum field theories on a fixed de Sitter background, such as the issue of vacuum ambiguity \cite{Ross:2004cb,Hutasoit:2009sc}. We will discuss this in more details in Chapter \ref{chapter:alpha} of the first part of this thesis.

The last approach is based on the low-energy four-dimensional effective action of string theory, \textit{i.e.}, supergravity augmented by higher derivative corrections. Its main goals are to build string inspired cosmological models and to find their signatures, which then can be compared to the data coming from cosmological observations. Inflationary models from this approach can be categorized according to the sector in which the inflatons originated. Closed string moduli inflation models include \cite{Conlon:2005jm,BlancoPillado:2006he,Holman:2006ek,Bond:2006nc,Holman:2006tm,Grimm:2007hs,Cicoli:2008gp,McAllister:2008hb,Badziak:2010qy} while open string moduli inflationary models include \cite{Alishahiha:2004eh,Bean:2008na,Burgess:2008ir,Baumann:2010sx}.

In this thesis, we will focus on the first two approaches. In the first part, we will consider gauge/gravity duality on time-dependent backgrounds with de Sitter boundary. Even though there are interesting open questions concerning the bulk dynamics, we will focus on using gauge/gravity duality to understand the boundary field theory which lives on de Sitter space. In the second part, we will be considering the superstring rolling tachyon backgrounds and their connection to understanding the nature of time.

\part{Spacetimes with de Sitter Boundary and Gauge/Gravity Duality}

\begin{savequote}[16pc]
\sffamily
Physics is like sex. Sure, it may give some practical results, but that's not why we do it.
\qauthor{Richard Feynman (1918 - 1988)}
\end{savequote}
\chapter{Introduction}

Time dependent backgrounds in gravity and in string theory are of
great interest from  the standpoint of the AdS/CFT correspondence
\cite{Maldacena:1997re, Aharony:1999ti} and related holographic dualities
between gauge theories and gravity. Time dependent classical gravity
backgrounds, in locally asymptotically Anti-de-Sitter spacetimes, can
potentially provide a fully nonperturbative description of
non-equilibrium phenomena in the strongly coupled dual gauge
theories. Such non-equilibrium physics in field theories arises, most
notably, in cosmology and in heavy ion
collisions at RHIC. To understand how gauge/gravity dualities work for
such processes, it is important to investigate how holography applies
in various examples with explicit time dependence. In this part of the thesis, we
attempt the holographic computation of real time 
correlators of the boundary gauge theory dual to the time  
dependent, asymptotically {\em locally} AdS backgrounds found in
\cite{Birmingham:2002st, Balasubramanian:2002am, Cai:2002mr, Ross:2004cb, Balasubramanian:2005bg}. The boundary of these geometries has a de Sitter factor and thus, they are useful in studying the behaviors of strongly coupled field theory living in a de Sitter space. Understanding de Sitter space is important because de Sitter space plays a central role in cosmology, not only at the early universe, during the inflationary epoch when the seeds of cosmic structure were generated, but also in the late universe as the cosmological constant dominates over other matter contents in the universe. The ultimate goal is to understand the full quantum gravity of de Sitter space, but even at the semi-classical level, where one considers quantum field theory living in a fixed de Sitter background, many interesting features appear. 

Concerning the bulk geometry, the authors of \cite{Aharony:2002cx} studied the double
analytic continuations of vacuum solutions such as Schwarzschild and
Kerr spacetimes providing examples of smooth, time dependent solutions
called ``bubbles of nothing'' 
\cite{Witten:1981gj, Myers:1986un, Dowker:1995gb}. These
asymptotically flat solutions were generalized to asymptotically
locally AdS spacetimes in
\cite{Birmingham:2002st,Balasubramanian:2002am}, by considering the
double analytic continuations of AdS black holes \footnote{For the classifications of solutions obtained by analytically continuing black hole solutions, see \cite{Astefanesei:2005eq}.}. The bubbles are
obtained by analytically continuing the time coordinate to Euclidean
signature $t\to i\chi$ where $\x$ is periodically identified, 
and  a polar angle $\theta \to i\t$. In addition, the $\x$ circle has
supersymmetry breaking boundary conditions for fermions. The 
resulting ``bubbles''  undergo exponential de Sitter expansion (and
contraction). For the asymptotically locally $AdS_5 \times S^5$ case
\cite{Balasubramanian:2002am}, the conformal boundary of the geometry is $dS_3
\times S^1$. The corresponding dual field theory, ${\cal N}=4$ SYM,
is thus formulated on $dS_3 \times S^1$ with antiperiodic
boundary conditions for the fermions around $S^1$. Each of the two
AdS-Schwarzschild black holes (the small and big black holes) yield an
AdS bubble of nothing solution, only one of which is stable. The
bubble of nothing geometries are vacuum solutions with cosmological
horizons \cite{Aharony:2002cx} and particle creation effects.

It was realized in \cite{Cai:2002mr, Ross:2004cb, Balasubramanian:2005bg} that
there is another spacetime with the same AdS asymptotics as the
bubble geometries, with $dS_3\times S^1$ conformal boundary. This is
the so-called ``topological black hole'' 
\footnote{The term ``topological AdS black hole'' 
has also been used to refer to black holes
with a hyperbolic horizon having a non-trivial topology. In the
AdS/CFT context these have been studied in 
\cite{Emparan:1999gf,Alsup:2008fr,Koutsoumbas:2008yq,Koutsoumbas:2008wy}
and references therein.}
-- 
a quotient of AdS space obtained by an
identification of global $AdS_5$ along a boost 
\cite{Banados:1997df, Banados:1998dc}. It is the five dimensional
analog of the BTZ black hole \cite{Banados:1992gq,
  Banados:1992wn}. The topological AdS black hole can also be obtained by a
Wick rotation of thermal AdS space. As Euclidean thermal AdS space can
be unstable to decay to the big AdS black hole via the first order Hawking-Page
transition \cite{Hawking:1982dh, Witten:1998qj}, 
a similar instability is associated to the topological AdS
black hole with antiperiodic boundary conditions for the fermions around $S^1$. In this case, the topological AdS black hole is unstable to
semiclassical decay via the nucleation of an AdS bubble of nothing. The
associated bounce solution is the Euclidean small AdS-Schwarzschild
black hole which has a non-conformal negative mode. The topological
black hole becomes 
unstable only when the radius of the spatial circle becomes smaller
than a critical value. In the Euclidean thermal setup, this corresponds to the case when the
temperature exceeds a critical value. Precisely such an instability
to decay to ``nothing'' was, of course, first noted for flat space times a
circle having antiperiodic boundary conditions for fermions
\cite{Witten:1981gj}. We note that when the fermions in the topological AdS black hole have periodic boundary conditions around the circle, this geometry is stable. We will review in more details the properties of these geometries in Chapter \ref{chapter:geom}.

The two different geometries described above are dual to two
different phases of strongly coupled, large $N$ gauge theory
formulated on $dS_3\times S^1$.  As in the more widely known thermal interpretation, where the field theory lives on $S^3 \times S^1$, the two phases are
distinguished by the expectation value of the Wilson loop around the
$S^1$. In the bubble of nothing phase, the circle shrinks to zero size
in the interior of the geometry and the Wilson loop is non-zero,
indicating the spontaneous breaking of the ${\mathbb Z}_N$ symmetry of
the gauge theory. The topological black hole phase is
${\mathbb Z}_N$ invariant. Unlike the thermal situation however, the
spontaneous breaking of ${\mathbb Z}_N$ invariance is not a
deconfinement transition since the circle is a spatial direction and
not the thermal circle.

The primary motivation of Chapter \ref{chapter:phases} is to understand how the behaviour
of real time correlators in the two geometries reflects the properties
and distinguishes the two phases of the ${\cal N}=4$ theory on
$dS_3\times S^1$. Since the de Sitter boundary has its own
cosmological horizon accompanied by a Gibbons-Hawking radiation
\cite{Gibbons:1977mu}, this should also be reflected in the properties of
the boundary correlation functions. An interesting feature of both the
geometries in question is that infinity is connected, {\it i.e.} the
asymptotics is unlike the AdS-Schwarzschild black hole whose
asymptotics consists of two disconnected boundaries. This means that 
the Schwinger-Keldysh approach in 
\cite{Herzog:2002pc} is not
directly applicable. It would be interesting to understand how to
apply that idea and also the recently proposed prescription of \cite{Skenderis:2008dh, Skenderis:2008dg} in the present context. 
Instead, we will simply use the Son-Starinets
prescription \cite{Son:2002sd, Policastro:2002se, Son:2007vk} to
compute real time correlators in the topological AdS black 
hole geometry, by requiring infalling boundary conditions at the
horizon of the black hole.

We will find that in the topological black hole phase, retarded
 scalar glueball
correlators have a simple description in frequency space. Here, we only consider the case that is homogeneous on spatial $S^2$ slices of $dS_3$. However, as can be seen from Chapter \ref{chapter:plasma}, generalizing it to the inhomogeneous case should not be too hard. The scalar glueball correlators we found have an
infinite number of poles in the lower half of the complex frequency
plane. As in the case of the BTZ black hole and other well known
examples, these poles represent the black hole quasinormal
frequencies \cite{Son:2002sd}. 
The Green's functions have imaginary parts and display
features closely resembling thermal physics. These features are naturally
associated to the Gibbons-Hawking temperature due to the cosmological
horizon of de Sitter space. This suggests that the ${\cal N}=4$ theory
on $dS_3\times S^1$ is in a plasma-like or deconfined state in the
exponentially expanding universe. 

When the spatial circle is small compared to the radius of curvature
of $dS_3$, below a certain critical value, the unstable topological black hole decays
into the  AdS bubble of nothing. In this geometry, correlation
functions are not analytically calculable. However, scalar glueball
propagators can be calculated in a WKB approximation. We show that in
this approximation, the correlation functions have an infinite set of
isolated poles on the real axis in the frequency plane. We interpret
this naturally as high mass glueball-like bound states of the field
theory. The transition from the topological black hole to the bubble
of nothing by tunelling is interpreted as a hadronization process. A
related picture of hadronization was discussed in
\cite{Horowitz:2006mr}. 

In Chapter \ref{chapter:plasma}, we further investigate the properties of the plasma or deconfined phase of the ${\cal N}=4$ Super Yang-Mills on $dS_3 \times S^1$. We study the real time correlators involving spatial spherical
harmonics of conserved R-currents to find whether they 
exhibit transport properties, {\it i.e.,} if they relax via
diffusion on the expanding spatial $S^2$ slices of $dS_3$. Applying
the Son-Starinets recipe, we find that the retarded propagator of the
R-current does not appear to relax hydrodynamically. This is likely
 due to the ``rapid'' expansion of de Sitter space,
the expansion rate of $dS_3$ being of the same order as the
Gibbons-Hawking temperature. The real time correlators are
represented in the form of a de Sitter mode expansion, which allows to
identify a natural frequency space correlator. This latter object has
isolated poles in the lower half plane and at the origin,
and its imaginary part exhibits the features characteristic of
a thermal state.

We would like to note that in applying the Son-Starinets recipe to the case that is inhomogeneous on the spatial $S^2$ slices of $dS_3$,  we have to acount for a certain subtlety involving discrete normalizable mode functions in de Sitter space. Later in Chapter \ref{chapter:alpha}, we will learn that this subtlety originates from having to include contributions from both the temporally separated points and the spatially separated point in de Sitter space.


In Chapter \ref{chapter:alpha}, we will focus on the vacuum structure of the boundary quantum field theory, namely ${\cal N}=4$ supersymmetric Yang-Mills theory, on a three-dimensional de Sitter space at the strong coupling regime. In particular, we will be interested in the fate of vacuum ambiguity in de Sitter space at strong coupling.

At the level of free theory, by considering the symmetric two-point function of a free massive scalar theory, it can be shown that there is an infinite family of vacua that are invariant under the isometries of de Sitter space. This is different from Minkowski space, where the symmetries of the theory determine a unique Poincar\'{e} invariant vacuum. The existence of this vacuum ambiguity in de Sitter space was first emphasized by Mottola \cite{Mottola:1984ar} and Allen \cite{Allen:1985ux}.  This class of de Sitter invariant vacua is often parametrized by a complex parameter $\a$, and thus are usually called the $\a$-vacua.

One of the $\a$-vacua, the Bunch-Davies \cite{Bunch:1978yq} vacuum, stands prominently among others as it is the only one that behaves thermally when viewed by an Unruh detector \cite{Birrell:1982ix} and reduces to the standard Minkowski vacuum when the de Sitter radius is taken to infinity. The correlators in the Bunch-Davies vacuum can be obtained by analytical continuation from the Euclidean theory, thus it is also known as the ``Euclidean" vacuum. The difference between a correlator in an $\a$-vacuum and the one in the Bunch-Davies vacuum is often thought of as arising from an image source at the antipodal point, behind the event horizon.

The early studies of a weakly interacting scalar theory in an $\a$-vacuum found that new divergences appear which, unlike in the Bunch-Davies vacuum, cannot be renormalized \cite{Danielsson:2002mb,Collins:2003zv,Collins:2003mk}. Self-energy graphs in a theory with a cubic interaction produce pinched singularities \cite{Einhorn:2002nu} or require peculiar non-local counterterms \cite{Banks:2002nv}\footnote{Beside the Bunch-Davies vacuum, Ref. \cite{Banks:2002nv} points out another special vacuum, in which the Green's function can be interpreted as living on elliptic de Sitter space \cite{Parikh:2002py}.}. However, when one modifies the generating functional of the theory, these divergences disappear \cite{Collins:2003mj} (see also \cite{Goldstein:2003ut,Goldstein:2003qf,Einhorn:2003xb}). 

With the advent of the gauge/gravity duality \cite{Maldacena:1997re,Aharony:1999ti}, it is only natural to ask what happens with this vacuum ambiguity as one goes to the strong coupling\footnote{For earlier efforts in trying to connect this issue to holography, see for example \cite{Danielsson:2002qh}.}. A context in which we can ask such a question is the ${\cal N}=4$ supersymmetric Yang-Mills theory living in a three-dimensional de Sitter space (times a circle). 

\begin{figure}[h]
\begin{center}
\includegraphics[width=3.5in]{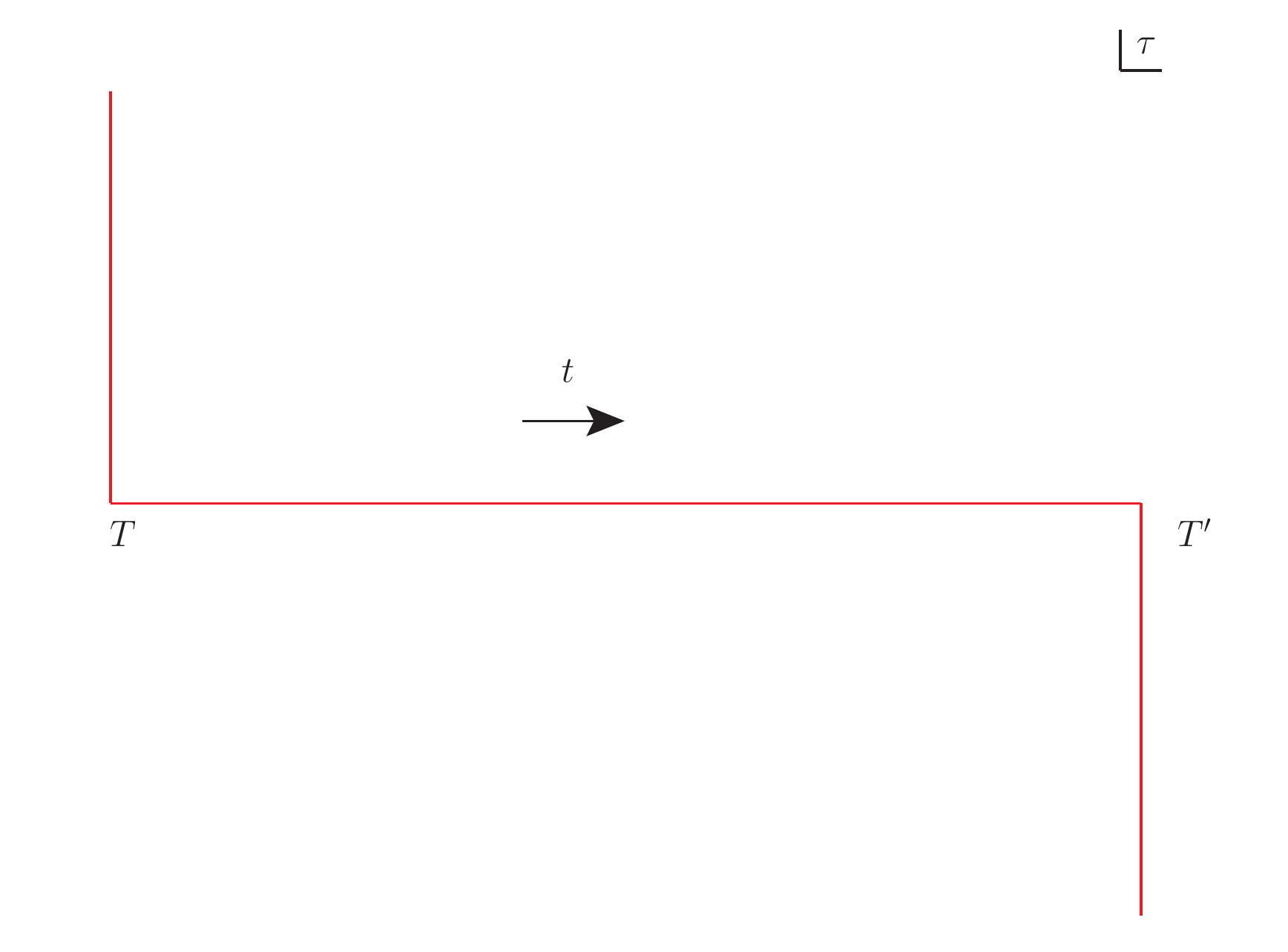} 
\end{center}
\caption{\footnotesize Son-Starinets prescription is equivalent to performing path integral on vacuum-to-vacuum contour in the complex time plane, as depicted above. The vacuum state is obtained by projection in the Euclidean path integral.}
\label{Euclid}
\end{figure}

Concerning the issue of vacuum ambiguity in de Sitter space, in Chapter \ref{chapter:phases}, we find that Son-Starinets prescription \cite{Son:2002sd}, which is used to calculate the retarded propagators, automatically implies that the boundary theory is in the Bunch-Davies vacuum. This is due to the fact that having the relevant boundary conditions in Son-Starinets prescription is equivalent to preparing the states in the boundary field theory by the mean of Euclidean projection \cite{vanRees:2009rw} (see Fig. \ref{Euclid}). Therefore, one can only obtain the propagators for the Euclidean vacuum using this prescription.

In Chapter \ref{chapter:alpha}, using gauge/gravity duality, we calculate the symmetric two-point function of the strongly coupled ${\cal N}=4$ Super Yang-Mills as a function of the geodesic distance. Here, we will focus only on the ${\mathbb Z}_N$-invariant phase, which corresponds to the topological black hole geometry. We find that there is an ambiguity in the two-point function that arises from the vacuum ambiguity of the boundary field theory. From the bulk point of view, this ambiguity comes from the fact that there are an infinite family of bulk radial wave functions that are normalizable deep into the bulk, at the horizon of the topological AdS black hole \cite{Balasubramanian:1998de}. 

Thus, the strongly coupled ${\cal N}=4$ Super Yang-Mills on $dS_3$ has an infinite class of symmetric two-point functions, parametrized by a set of complex parameters $\{\a_{\n}\}$. This is different from the case of the weakly coupled theory, where the infinite family of symmetric two-point functions is parametrized by a single complex parameter $\a$. One possible explanation is that as one increases the coupling constant of the theory, going from the free theory toward the strongly coupled one, the $\a$-vacua mix with one another. 

In order to be self-contained, before going to the detailed calculations on the asymptotically locally anti de Sitter spacetimes with de Sitter boundary, we will start this part of the thesis by reviewing some aspects of gauge/gravity duality that will be important in our calculation. We will then end this part of the thesis with some discussions and open questions in Chapter \ref{chapter:disc1}.

\begin{savequote}[15pc]
\sffamily
Just as we have two eyes and two feet, duality is part of life.
\qauthor{Carlos Santana (b. 1947)}
\end{savequote}
\chapter{Aspects of Gauge/gravity Duality} \label{chapter:ads/cftrev} 

\section{The Correspondence}\label{section:corr}

The gauge/gravity duality is defined by the relation
\begin{equation}
\int_{\Phi\rightarrow\Phi_0}\,  {\cal D}\Phi \, e^{i \, S_{\rm grav}[\Phi]} = Z_{\rm gauge}[\partial M, \Phi_0], \label{eq:defadscft}
\end{equation}
where the right hand side is the generating functional of the gauge theory. The boundary values $\Phi_0$ of the fields $\Phi$ of the gravity theory are kept fixed and they become the source for their dual operators ${\cal O}$ in the gauge theory side. In particular, the boundary value of the expectation value of the metric becomes the spacetime $\partial M$ the gauge theory lives in.

\begin{figure}[h]
\begin{center}
\includegraphics[width=3in]{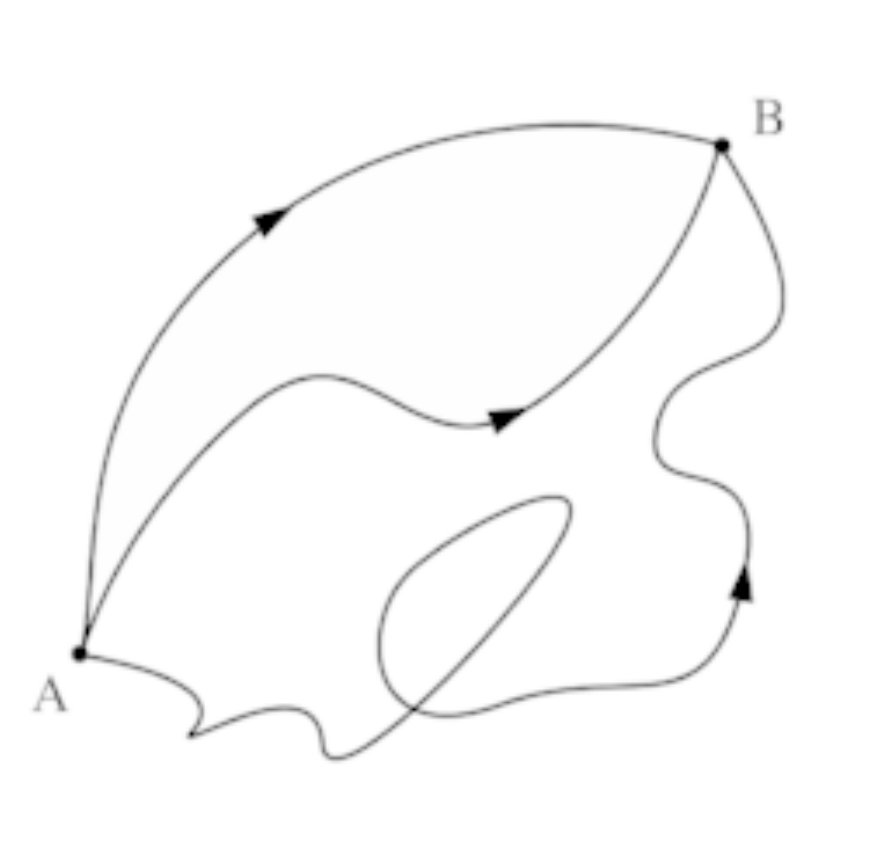} 
\end{center}
\caption{\footnotesize Path integral of a point particle moving in space starting at point A and ending at point B.}
\label{fig:path}
\end{figure}

\begin{figure}[h]
\begin{center}
\includegraphics[width=3.5in]{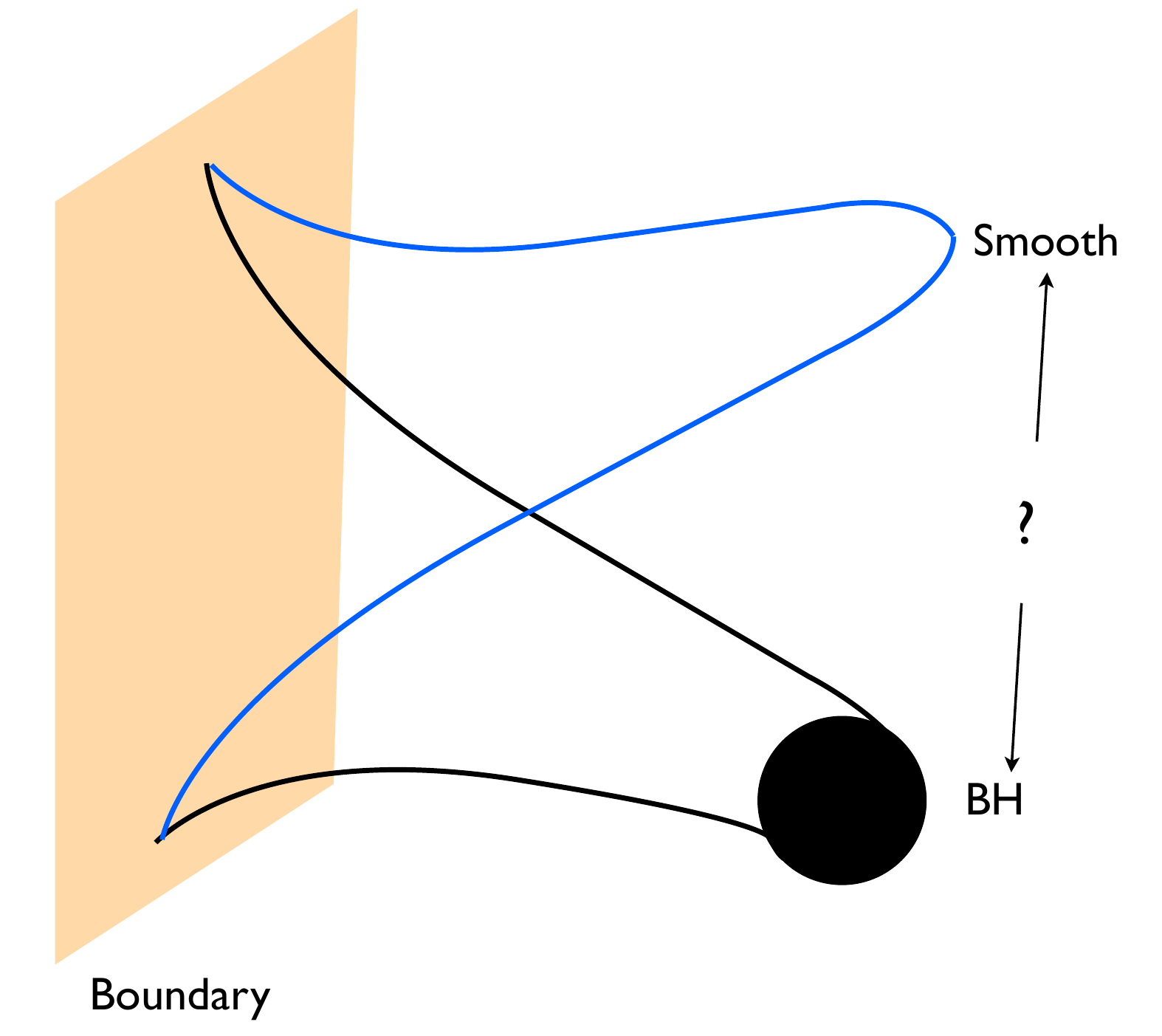} 
\end{center}
\caption{\footnotesize Path integral of spacetimes with boundary. Topology changing processes are included and such processes might go through a regime where the notion of spacetime itself is not valid.}
\label{fig:adscftpath}
\end{figure}

The path integral on the left hand side of Eq. \ref{eq:defadscft} is the full non-perturbative description of a quantum theory of gravity. In order to appreciate this more fully, let us recall the definition of path integral for a point particle moving in space. The boundary values for the position field at initial and final time are given (\textit{i.e.}, kept fixed) while in between (\textit{i.e.}, deep in the ``bulk"), the position field is allowed to fluctuate and to take any possible values. This is depicted in Fig. \ref{fig:path}. Similarly, in gauge/gravity duality, the boundary spacetime and boundary values of other fields are kept fixed, but they are allowed to fluctuate in the bulk (see Fig.  \ref{fig:adscftpath}). In particular, the bulk spacetime can be smooth or it can have a horizon. The process in which the bulk spacetime changes its topology from one to another is encoded in the path integral. It is also possible that this transition will go through a regime in which the notion of spacetime itself is no longer valid.

Remarkably, the complicated gravity path integral can be expressed as something that is rather simpler and somewhat more familiar: a generating functional of a gauge theory. The latter is a quantum field theory on a fixed classical spacetime, \textit{i.e.}, it is a theory without gravity. Therefore, it can be said that gravity is emergent and that parts of the bulk geometry are emergent too. For example, let us consider the gauge/gravity duality that involves Type IIB superstring theory on $AdS_5 \times S^5$. The gravity theory is dual to ${\cal N}=4$ Super Yang-Mills living on a four-dimensional Minkowski space. The radial direction of the $AdS_5$ emerges from the renormalization group flow of the boundary theory, while the five-dimensional sphere emerges from the R-symmetry of the ${\cal N}=4$ Super Yang-Mills. For further study on how spatial dimensions and gravity might emerge from gauge theories, see for example \cite{Berenstein:2005aa,Gursoy:2007np,Berenstein:2007wi,Berenstein:2007kq}. On the issue of how the locality of the bulk physics might emerge, see for example \cite{Gary:2009ae,Heemskerk:2009pn}.

Gauge/gravity duality has also been used to understand the quantum gravity of black holes, in particular the issue of information paradox. In this approach, it has been realized that the black hole horizon geometry itself arises as an effective description. This is the key to solving the information paradox. For reviews, see for example \cite{Mathur:2008nj,Balasubramanian:2008da}.

In the semi-classical regime, we can evaluate the gravity path integral using saddle point approximation
\begin{equation}
\int_{\Phi\rightarrow\Phi_0}\,  {\cal D}\Phi \, e^{i \, S_{\rm grav}[\Phi]}  \approx \sum_i \int_{\Phi[\partial M]=\Phi_0}\,  {\cal D}\Phi \, \, \exp\left(i \, \int_{M_i} {\cal L}[\Phi]\right),
\end{equation}
where the sum in the right hand side is over bulk classical geometries $M_i$, whose boundary is $\partial M$. In other words, in this regime, the full non-perturbative gravity path integral has been reduced to the sum of perturbative gravity path integrals around fixed spacetimes. These spacetimes corresponds to different phases of the boundary field theory.

This regime corresponds to the gauge theory having a large rank gauge group and strong coupling constant. In particular, using this correspondence, we are able to obtain the leading order results of the boundary gauge theory, by merely perform classical gravitational calculations. Here, leading order means leading order in the $1/N$ expansion, where $N$ is the rank of the gauge group. Therefore, in this regime, gauge/gravity duality has become a powerful tool to study strongly coupled gauge theories. Its application encompasses the study of quark gluon plasma and strongly correlated condensed matter systems. Later in Chapter \ref{chapter:alpha}, we will see how to use gauge/gravity duality to try to address an issue in understanding inflationary cosmology, namely the issue of vacuum ambiguity in de Sitter space.
 
\section{Building Dualities}

In the previous section, we have given a general definition of gauge/gravity duality. However, in order to be able to perform calculations, we need a specific example of a specific gravity theory that is dual to a specific gauge theory. One way to build such a duality is by brane construction. We can construct a given gauge theory from the open string sector of a brane configuration by stacking and/or intersecting branes on a certain spacetime. If we increase the number of branes, the branes will backreact and the near horizon geometry of the backreacted spacetime will be the geometry that the (strongly coupled) field theory is dual to. If the field theory has more than one phase, the other phases will correspond to other geometries that have the same boundary as the near horizon geometry we just found.

\subsection{Type IIB Superstring in $AdS_5 \times S^5$ and ${\cal N}=4$ Super Yang-Mills}

A canonical example involves Type IIB superstring theory. This theory has ${\cal N}=2$ supersymmetry and lives in 10 dimension. The low energy spectrum consists of:
\begin{itemize}
\item metric/graviton $g_{\m\n}$, 
\item dilaton $\phi$, which is a scalar field,
\item axion $C_0$, which is a pseudo-scalar field,
\item 2-form field $B_{\m\n}+i A_{2\mu\n}$,
\item 4-form field $A_{4\m\n\r\s}$,
\item gravitino $\psi_\m$, which is a Majorana-Weyl fermion,
\item dilatino $\lambda$, which is also a Majorana-Weyl fermion.
\end{itemize} 
For more details, see for example \cite{Polchinski:1998rr}.

It is natural to couple a $(p+1)$-form to an object $\Sigma_{p+1}$ of spacetime dimension $p+1$. The action can be constructed as follows
\begin{equation}
S_{p+1} = T_{p+1} \int_{\Sigma_{p+1}} A_{p+1},
\end{equation}
where $T_{p+1}$ is the charge of this object under the $U(1)$ gauge group of the $(p+1)$-form. When this $(p+1)$-form is the Ramond-Ramond form field, the object $\Sigma_{p+1}$ is the D$p$-brane, a solitonic object at which open string can end (for more details, see for example \cite{Johnson:2003gi}). Related to the rank of the Ramond-Ramond form fields, D-branes in Type IIA superstring theory have even number of spatial dimensions, while those in Type IIB superstring theory have odd number of spatial dimensions.

Let us now consider a stack of $N$ D3-branes on a ten dimensional Minkowski space. Analyzing the open strings that end on this stack of D3-branes, we can see that the physics of these D3-branes is described by an ${\cal N}=4$ Super Yang-Mills  living in a four-dimensional Minkowski space. The gauge group is $U(N)$, with the $U(1)$ factor corresponds to the overall position of the stack of the D3-branes. The gauge coupling constant is related to the string coupling constant by
\begin{equation}
g_{YM}^2 = g_s.
\end{equation}

For large $N$, the ten dimensional spacetime is no longer flat and the backreacted geometry is given by
\begin{equation}
ds^2 = \left(1+\frac{L^4}{y^4}\right)^{-1/2} \eta_{\mu\n}\, dx^\m dx^\n + \left(1+\frac{L^4}{y^4}\right)^{1/2} \left(dy^2 + y^2 d\Omega_5^2 \right),
\end{equation}
where
\begin{equation}
L^4 = 4\pi\, g_s \, N \, (\alpha')^2.
\end{equation}
Here, $1/2\pi\a'$ is the string tension.

The near horizon geometry is obtained by taking a large $u$ limit, where
\begin{equation}
z \equiv \frac{L^2}{y},
\end{equation}
and it is given by
\begin{equation}
ds^2 = L^2 \left[ \frac{dz^2 + \eta_{\m\n} \, dx^\m dx^\n}{z^2}+ d\Omega^2_5 \right].
\end{equation}
This is a product geometry with one component is given by the hyperbolic space $AdS_5$, while the other component is a five dimensional sphere $S^5$, with L becoming the radius for both the AdS space and the sphere. Therefore, the physics close to the surface of the stack of D3-branes is given by Type IIB superstring theory living in $AdS_5 \times S^5$. This should be equivalent to the physics described by the open string, modulo the spatial translation perpendicular to the D3-branes. Therefore, Type IIB superstring theory living in $AdS_5 \times S^5$ is dual to ${\cal N}=4$ Super Yang-Mills  living in a four-dimensional Minkowski space with $U(N)/U(1) \sim SU(N)$ gauge group.

The ${\cal N}=4$ Super Yang-Mills is a conformal theory. Combining the conformal symmetry and supersymmetry, it can be shown that the continuous global symmetry of this theory is the so-called superconformal group $SU (2,2|4)$. The maximal bosonic subgroup of the superconformal group is $SU(2,2) \times SU(4)_R \sim SO(2,4) \times SO(6)_R$. The conformal symmetry $SO(2,4)$ matches to the isometry of $AdS_5$, while the R-symmetry $SO(6)_R$ matches to the isometry of $S^5$. Combining this with the supersymmetry of the bulk theory, it can be shown that the gravity side also enjoys the full $SU (2,2|4)$ symmetry. Since the symmetry matches on both sides of the correspondence, it is then straightforward to establish a dictionary by matching a representation of the symmetry group $SU (2,2|4)$ of one side of the correspondence to the representation of the other side.

We note that when 
\begin{equation}
\frac{L^4}{(\alpha')^2} = 4\pi\, g^2_{YM} \, N \, \gg 1
\end{equation}
the gravity theory is well described by the Type IIB supergravity, which is the low energy description of the perturbative string theory. Therefore, at large $N$ and large t'Hooft coupling $\lambda = g^2_{YM} \, N$, the boundary field theory is dual to the Type IIB supergravity living in $AdS_5 \times S^5$. In general, when the scale of the bulk geometry is larger than the string scale, or correspondingly at  large $N$ and large t'Hooft coupling $\lambda$, we do not need the full non-perturbative string theory description and the gravity path integral is reduced to the sum of perturbative string theory/supergravity on fixed backgrounds.

\subsection{AdS Black Hole, Thermal AdS Space and Thermal ${\cal N}=4$ Super Yang-Mills}

A more complicated example can be obtained by putting the ${\cal N}=4$ Super Yang-Mills on a three-dimensional sphere and by switching on the temperature. The boundary geometry now becomes $S^3 \times S^1$ with antiperiodic boundary conditions along the $S^1$ for the fermions. 

There are two known geometries that have an $S^3 \times S^1$ boundary, identified by Hawking and Page \cite{Hawking:1982dh}. They are the so-called thermal AdS space and the Euclidean AdS Schwarzschild black hole. The thermal AdS space is a quotient of the Euclidean AdS space and the metric is given by
\begin{equation}
ds^2 = f(r) \, dt^2 + f(r)^{-1} \,dr^2+ r^2 d\Omega_3^2, \label{eq:thermalads}
\end{equation}
where
\begin{equation}
f(r) = 1+ \frac{r^2}{R^2_{\rm AdS}},
\end{equation}
and $t$ is a periodic coordinate with an arbitrary period $\beta$. As $\b$ is the period of the circle at the boundary, it is identified as the inverse temperature of the boundary field theory.

The metric of the Euclidean AdS Schwarzschild black hole has the same form as the metric of thermal AdS \ref{eq:thermalads}, but with
\begin{equation}
f(r) = 1+ \frac{r^2}{R^2_{\rm AdS}} - \frac{M}{r^2}.
\end{equation}
$M$ is related to the mass of the black hole $m_{\rm BH}$ as
\begin{equation}
M = \frac{8 \, G_N}{3\pi} \, m_{\rm BH}.
\end{equation}
There are two solutions to the equation $f(r) = 0$: the larger solution $r_+$ is the radial position of the horizon of the so-called large AdS black hole and the smaller solution $r_-$ becomes the radial coordinate of the horizon of the small AdS black hole. It turns out that the small AdS black hole is unstable to perturbations. Furthermore, when viewed as a ten dimensional, asymptotically $AdS_5 \times S^5$ solution smeared on the $S^5$, Ref. \cite{Hollowood:2006xb} shows that  the small AdS black hole is also unstable to localization on $S^5$, \textit{i.e.}, it suffers the Gregory-Laflamme instability.

The metric of the large AdS black hole is smooth and complete if and only if the period of $t$ is given by
\begin{equation}
\b_0 = \frac{2\pi R^2_{\rm AdS} r_+}{2r_+^2 + R^2_{\rm AdS}}.
\end{equation}
This is the inverse Hawking temperature of the black hole and is also identified with the temperature of the boundary field theory.

It turns out that the large AdS black hole corresponds to the plasma or deconfined phase of the thermal ${\cal N} = 4$ Super Yang-Mills, while the thermal AdS space corresponds to the confined phase. Furthermore, the confinement-deconfinement phase transition of the boundary field theory corresponds to the Hawking-Page transition of the gravity theory. For more details, see for example \cite{Witten:1998zw,Brandhuber:1998er,Aharony:1999ti,Aharony:2003sx}.

\section{Examples of Holographic Calculations}

Above, we have given some examples of gauge/gravity duality. Now, let us consider some examples of holographic calculations in which we perform classical gravity calculations to obtain results for strongly coupled field theories.

\subsection{Euclidean Calculation}

Let us start with the simpler case of gauge/gravity duality with Euclidean signature. In Euclidean signature, the gauge/gravity correspondence defined in Eq. \ref{eq:defadscft} becomes
\begin{equation}
Z_{\rm gauge}[\partial M, \Phi_0] = \int_{\Phi\rightarrow\Phi_0}\,  {\cal D}\Phi \, e^{- \, S_{\rm grav}[\Phi]} \approx \sum_i Z_{\rm grav}[M_i, \Phi],
\end{equation}
where the left hand side is the partition function of the gauge field living on $\partial M$ with source $\Phi_0$, while the right hand side is the gravitational partition function summed over geometries $M_i$ whose boundary is $\partial M$ and the boundary values of the fields in the gravity theory are given by $\Phi(\partial M) = \Phi_0$.

Our goal is to study the strongly coupled field theory by calculating the correlation functions of its operators. The latter can be obtained by taking the appropriate derivatives of the gauge theory partition function with respect to the source $\Phi_0$ and then setting the source to zero. Even though it is very difficult to calculate the strongly coupled gauge theory partition function directly, the gauge/gravity correspondence tells us that this partition function is nothing but the gravity partition function with the fields in the gravitational theory having appropriate boundary conditions, namely $\Phi(\partial M) = \Phi_0$. We can obtain this gravity partition function by first solving the classical equations of motion of the fields with the boundary condition $\Phi(\partial M) = \Phi_0$.

To summarize, in order to study the strongly coupled field theory using holographic techniques, our strategy is as follows:
\begin{enumerate}
\item Solve the classical equations of motion of the bulk fields $\Phi$ subject to the boundary conditions $\Phi(\partial M) = \Phi_0$.
\item Substitute the solutions of step 1 into the Euclidean path integral.
\item The result of step 2 is the partition function for the gauge theory, where $\Phi_0$ being the source for the dual operator. By taking derivatives with respect to $\Phi_0$ and then setting $\Phi_0=0$, we obtain the correlation functions of the field theory operators ${\cal O}$.
\end{enumerate}

Let us now consider a specific example of the duality between strongly coupled ${\cal N} = 4$ Super Yang-Mills on a four-dimensional Euclidean flat space with classical Type IIB supergravity theory on a five-dimensional Euclidean AdS space (times the five-dimensional sphere).

For a reminder, the $AdS_5$ metric in  is given by
\EQ
{
ds^2= {R_{\rm AdS}^2\over z^2}\left(dz^2 +\delta_{\m\n}\,dx^\m dx^\n\right).
}
This metric does not describe the full $AdS_5$ geometry, but only the so-called Poincar$\acute{\rm e}$ patch. In this parametrization, the conformal boundary is at $z=0$, while the
``origin'' of the space is at $z\rightarrow \infty$.

Let us consider a scalar field $\phi$ with mass $m$ living in the bulk. The action is given by
\begin{equation}
S_E= \frac{N^2}{16 \pi^2 R_{\rm AdS}} \int d^4x \int dz \; \sqrt{g}
\left(g^{MN}\, \partial_M \phi\,\partial_N\phi+m^2\phi^2 \right).
\end{equation}
Its equation of motion is given by
\begin{equation}
\left(\Box-m^2\right)\phi = 0,
\end{equation}
which translates to
\begin{equation}
z^5\, \partial_z \left({1\over z^3}\,\partial_z\phi\right) +z^2\,\partial_\m \partial^\m \phi - m^2 R^2_{\rm AdS} \, \phi=0. 
\end{equation}

There are two independent solutions
\begin{equation}
\phi= z^2 \;e^{-i k\cdot x} \;I_{\n}(kz) \qquad{\rm and}\qquad \phi= 
z^2 \;e^{-i k\cdot x}\; K_{\n}(kz),
\end{equation}
where the index of the modified Bessel functions $I_\n$ and $K_\n$ is given by
\begin{equation}
\nu=\sqrt{4+(m\, R_{\rm AdS})^2},
\end{equation}
and $k$ in their argument is understood to be the magnitude $k = \sqrt{\delta_{\m\n}\, k^\m k^\n}$ of the boundary momentum. 

Neither of the above solutions is square integrable, but only the second solution is smooth in the interior. Since there is no singularity in the interior, the physical solution will be
\begin{equation}
\phi= z^2 \; \int {d^4k\over (2\pi)^4} \, C(k)\, e^{-i k\cdot x}\; K_{\n}(kz),
\end{equation}
where $C(k)$ is chosen such that the solution satisfies the boundary condition $\phi(z=0,x) = \phi_0(x)$. 
Substituting this solution into the Euclidean action, we get
\begin{equation}
S_E = \frac{N^2}{16 \pi^2 R_{\rm AdS}}\int d^4x \;g^{z z}\sqrt g\, \phi(x,z)\,\partial_z\phi(x,z)
\big|_{z=0},
\end{equation}
which diverges due to the divergent volume element at the boundary. We can regulate this by explicitly cutting off the AdS bulk geometry at $z=\e$ and then taking $\e \rightarrow 0$.

Denoting
the boundary value of the field $\phi$ as
\EQ
{
\phi_0(x) = \phi(\epsilon,x) = \int {d^4k\over (2\pi)^4} \, \bar\phi_0(k)
\, e^{-ik\cdot x},
}
it is easy to see that the solution is now given by
\EQ
{
\phi(z,x)=\int {d^4k\over (2\pi)^4} \, 
{z^2 K_\nu(kz)\over \epsilon^2 K_\nu(k\epsilon)}\,e^{-ik\cdot x}\,\bar\phi_0(k).
}
Substituting this solution into the Euclidean action, we get
\begin{eqnarray}
S_E &=& \frac{N^2}{16 \pi^2 R_{\rm AdS}}\int d^4x \;g^{z z}\sqrt g\, \phi(x,z)\,\partial_z\phi(x,z) \big|_{z=\e} \nonumber \\
&=& \frac{N^2}{16 \pi^2 R_{\rm AdS}}\,{1\over \epsilon^3} \int {d^4p \over (2\pi)^4}\,{d^4q\over (2\pi)^4} \, \bar \phi_0(p)\Bigg[(2\pi)^4\,\delta^{(4)}(p+q)\, {d\over d\epsilon}\left(\ln \left[\epsilon^2  K_\nu(p\epsilon)\right]\right) \Bigg] \bar\phi_0(q). \nonumber \\
\end{eqnarray}
It is now straightforward to read off the Fourier transform of the regulated two-point function of the boundary field theory from this on-shell action and it is given by
\begin{eqnarray}
\langle{\cal O}(p){\cal O}(q)\rangle &=& {\delta^2 S_E\over \delta\bar\phi_0(p)\delta\bar\phi_0(q)} \nonumber \\
&= & (2\pi)^4\, \delta^{(4)}(p+q)\, \frac{N^2}{8 \pi^2 R_{\rm AdS}}\,{1\over \epsilon^3} \, {d\over d\epsilon}\left(\ln \left[\epsilon^2 K_\nu(p\epsilon)\right]\right).
\end{eqnarray}

For example, the lowest KK modes of the dilaton and axion of the Type IIB superstring theory, which are massless, minimally coupled scalar fields in $AdS_5$, lead to 
\begin{equation}
\langle{\cal O}(p){\cal O}(-p)\rangle  = \frac{N^2}{32 \pi^2 R_{\rm AdS}} \left(-{p^2\over \epsilon^2}-{p^4\over 2}\ln\left[p \epsilon\right] \right) + {\rm constants}, \label{eq:2ptcft}
\end{equation}
which is the two-point function the scalar glueball operators $\Tr\, F_{\mu\nu}F^{\mu\nu}$ and $\Tr \, F_{\mu\nu}\tilde F^{\mu\nu}$ of the ${\cal N} = 4$ Super Yang-Mills theory. Fourier transforming back to position space, we get
\begin{equation}
\langle{\cal O}(x){\cal O}(x')\rangle = \frac{N^2}{64 \pi^2 R_{\rm AdS}}\, \frac{1}{|x-x'|^8},
\end{equation}
which is a two-point function for dimension four operators in the conformal field theory. This confirms further that the dilaton and axion of the Type IIB superstring theory are dual to the scalar glueball operators  of the ${\cal N} = 4$ Super Yang-Mills theory $\Tr\, F_{\mu\nu}F^{\mu\nu}$ and $\Tr \, F_{\mu\nu}\tilde F^{\mu\nu}$, respectively.

In general, using the asymptotic behavior of the non-normalizable mode near the boundary, it can be shown that the relation between the mass $m$ of the bulk scalar field and the dimension $\D$ of the boundary field theory operator it is dual to is given by
\begin{equation}
m^2 = \D (\D-4).
\end{equation}

The divergent terms in Eq. \ref{eq:2ptcft}, which depend on the UV cutoff of the boundary field theory, are the so-called contact terms. Contact terms are Fourier transforms of derivatives acting on Dirac delta function. For example, $p^2$ is the Fourier transform of $\Box \delta^{4}(x-x')$. The divergences from these contact terms manifest themselves when the field theory operators we are considering lay on top of each other, \textit{i.e.}, when $x = x'$.

\subsection{Lorentzian Calculation} \label{reviewson}

In the Euclidean case, with the help of the physical argument that the solution of the equation of motion should be smooth in the interior of the bulk, the solution to the equation of motion for the given boundary condition is unique. However, for the Lorentzian anti de Sitter spacetimes, fixing the boundary value of the field is not enough to warranty a unique solution. This is because there are two linearly independent solutions that are smooth in the interior \cite{Avis:1977yn}. The variety of the radial parts of the solution to the equation of motion of the gravity theory is related to the variety of real time correlators of the boundary field theory. The latter is the result of two things:
\begin{itemize}
\item the multitude of real time Green's functions, \textit{i.e.}, retarded, advanced, \textit{etc};
\item the possibility of vacuum ambiguity, especially if the boundary manifold is time-dependent.
\end{itemize}

In this section, we will review Son and Starinets' prescription on how to obtain the retarded correlators for the boundary field theory \cite{Son:2002sd}. If the boundary field theory suffers from vacuum ambiguity, this prescription will also choose the vacuum obtained by Euclidean projection \cite{vanRees:2009rw}, as depicted in Fig. \ref{Euclid}. This prescription is especially suitable for calculations involving black holes. In order to obtain the retarded propagator, the prescription requires infalling boundary condition at the horizon, while the advanced propagator corresponds to the outgoing boundary condition at the horizon.

\subsubsection{Retarded Correlators for the Scalar Glueball Operators}

To study the Lorentzian AdS/CFT duality, let us consider the near-horizon geometry of a stack of three-dimensional near extremal branes. It is equivalent to the Lorentzian AdS Schwarzschild black hole in the large mass limit and the metric is given by
\begin{equation}
ds^2 = \frac{R_{\rm AdS}^2}{z^2} \left(- f(z) dt^2 + d\vec{x}^2 + f(z)^{-1} dz^2 \right), \label{eq:nearext}
\end{equation}
where
\begin{equation}
f(z) = 1 - \frac{z^4}{z^4_H}.
\end{equation}
The boundary is at $z=0$ and the horizon is at $z=z_H$, with $z_H = 1/\pi T$, $T$ being the Hawking temperature.

The Lorentzian $AdS_5$ can then be considered as the $T \rightarrow 0$ limit of the geometry \ref{eq:nearext} and we can apply the Son-Starinets prescription to obtain the retarded propagator of the strongly coupled ${\cal N}=4$ Super Yang-Mills.

For spacelike boundary momenta $k^2 > 0$, the calculation is identical to the Euclidean case we considered before, but with an extra minus sign in front of the Lorentzian action. The answer is given by 
\begin{equation}
G_R(k) = \frac{N^2\,k^4}{64 \pi^2 R_{\rm AdS}}\, \ln k^2.
\end{equation}

For timelike boundary momenta $k^2 < 0$, we can obtain the retarded correlator by requiring the solution to take the incoming wave form at $z = 0$, which can be viewed as the ``point" or vanishing horizon. Introducing $q = \sqrt{- k^2}$, the solution for the equation of motion is
\begin{equation} 
   \phi(z,x)  \;=\; \int {d^4k\over (2\pi)^4} \,e^{-ik\cdot x}\,\bar\phi_0(k)\,
\begin{cases} 
 \displaystyle{  { z^2 H_{\n}^{(1)}(qz)
\over \epsilon^2  H_{\nu}^{(1)}(q\epsilon ) }\,,}
                          &  \omega > 0,  \cr
           \noalign{\vskip 4pt}
          \displaystyle{  { z^2 H_{\n}^{(2)}(qz)
\over \epsilon^2  H_{\n}^{(2)}(q\epsilon ) }\,, }
 &  \omega < 0.   \cr
\end{cases}
\end{equation}
As before,
\begin{equation}
\nu=\sqrt{4+(m\, R_{\rm AdS})^2},
\end{equation}
and $\w$ is the time component of $k$. 

Substituting the solution into the action, we can read the retarded propagator
\begin{equation}
 G_R (k) =  {N^2 k^4\over 64 \pi^2} (\ln{k^2} - 
 i\pi\, \mbox{sgn} \, \omega)\,.
\label{timelike_0}
\end{equation}
From this result, we can then obtain the Feynman propagator
\begin{equation}
G_F (k) =  {N^2 k^4\over 64 \pi^2} (\ln{|k^2|} -   i\pi\,  \theta (-k^2))\,,
\end{equation}
which can be obtained from the Euclidean correlator
\begin{equation}
 G_{E} (k_E) = - {N^2 k_E^4\over 64 \pi^2} \ln{k_E^2}
\end{equation}
by analytical continuation.

\subsubsection{Retarded R-charge Current Correlators} \label{revplasma}

The massive AdS Schwarzschild black hole geometry \ref{eq:nearext} is interesting in its own right as it can be used to learn the transport property of the strongly coupled plasma of the ${\cal N} = 4$ Super Yang-Mills. To do so, let us consider a $U(1)$ gauge field living in the bulk whose action is given by
\begin{equation}
S = -\frac{N^2}{64 \p^2 R_{\rm AdS}} \int d^5 x \sqrt{-g} g^{\m \a} g^{\n
  \b} F_{\m \n} F_{\a \b}\,. \label{action} 
\end{equation}
This field is obtained by compactifying around the $S^5$ the gauge field whose gauge field strength is the Hodge dual of the field strength of the 2-form field. This $U(1)$ gauge field $A^\m$ corresponds to the R-current $j^\m$ of the boundary field theory, which is a conserved current. From the R-charge correlator, we can then read the diffusion constant of the strongly coupled plasma.

Without losing any generality, we can align the R-charge perturbation to carry the spatial momentum along the $x^3$-axis, \textit{i.e.}, $p = (\w,0,0,k)$. Furthermore, we can use gauge freedom to set the radial component of the gauge potential $A_z = 0$. Hence, from the symmetries of the configuration, there are two remaining bulk gauge fields that are non zero and they can be written as
\begin{equation}
A_0 = e^{- i p \cdot x} \, \tilde{A}_0(z,p) \qquad {\rm and} \qquad A_3 = e^{- i p \cdot x} \, \tilde{A}_3(z,p).
\end{equation}

It is convenient to introduce an alternative radial coordinate $u = z_H^2/z^2$. In this parametrization, the Maxwell equations can be written as
\begin{eqnarray}
  w \tilde{A}_0' + q f \tilde{A}_3' &=& 0\,,\\
  \tilde{A}_0'' - \frac1{uf} (q^2 \tilde{A}_0 + wq \tilde{A}_3) &=& 0\,,\label{eq-A0}\\
  \tilde{A}_3'' + \frac{f'}f \tilde{A}_3' + \frac1{uf^2}(w^2\tilde{A}_3+wq\tilde{A}_0) &=& 0\,,
\label{eq-A3}
\end{eqnarray}
 where $'$ denotes a derivative with respect to $u$ and we have introduced dimensionless quantities
\begin{equation}
w = \frac{\w}{2 \pi T} \qquad {\rm and} \qquad q = \frac{k}{2 \pi T}.
\end{equation}
We can eliminate $\tilde{A}_3$ and obtain a third-order differential equation for $\tilde{A}_0$,
\begin{equation}\label{Atprime}
  \tilde{A}_0''' + \frac{(uf)'}{uf} \tilde{A}_0'' + \frac{w^2-q^2f}{uf^2}\tilde{A}_0' = 0\,,
\end{equation} 
which we only need to solve for $\tilde{A}_0'$ due to the fact that the action only depends on $\tilde{A}_0'$ and not $\tilde{A}_0$. The full solution is not known, but fortunately, we do not need to know the full solution to learn about the transport property of the strongly coupled plasma. To do so, we only need to know the solution for small energy and momentum, which we can obtain perturbatively in $w$ and $q^2$.

Near the horizon $u = 1$, there are two independent solutions, $\tilde{A}_0'\propto(1-u)^{\pm
iw/2}$, but the incoming wave boundary condition singles out the solution that behaves like $(1-u)^{-iw/2}$ near the horizon. Our ansatz is then 
\begin{equation}
\tilde{A}_0' = (1-u)^{-iw/2} \left( g_{0,0} + g_{1,0}\,\frac{iw}2+ g_{0,1}\,q^2 + \cdots\right),
\end{equation}
whose coefficients can be obtained by requiring that the solution has no extra singularities at the horizon.

At leading order, we have
\begin{equation}
\frac{1-3u^2}{u-u^3} \, g_{0,0}'+ g_{0,0}'' = 0,
\end{equation}
whose solution is
\begin{equation}
 g_{0,0} = C_1 + \frac{C_2}{2}\ln\frac{u^2 -1}{u^2}.
\end{equation}
The regularity condition at the horizon enforces $C_2 = 0$. 

At ${\cal O}(w)$, we have
\begin{equation}
-(1 + 2u)C_1 + (3u^2 - 1)\,g'_{1,0} + u(u^2 - 1)\,g_{1,0}'' = 0,
\end{equation}
whose solution is
\begin{equation}
g_{1,0} = C_1 \ln(u -1) + C_3 + \frac{C_4}{2} \ln\frac{u^2 -1}{u^2}.  
\end{equation}
In order for the solution not to have extra singularities, we must have
\begin{equation}
C_3 = C_1 \ln 2 \qquad {\rm and} \qquad C_4 = -2 C_1.
\end{equation}

Finally, at ${\cal O}(q^2)$, we have
\begin{equation}
C_1 + (3u^2-1)\,g_{0,1}' + u(u^2- 1) \,g_{0,1}''= 0,
\end{equation}
whose solution is
\begin{equation}
g_{0,1}= C_5 - \frac{C_1-C_6}{2} \ln(u -1) + \frac{C_1+C_6}{2} \ln(u + 1) - C_6 \ln u.
\end{equation}
Again, requiring the same condition as above, we get
\begin{equation}
C_5 = - C_1 \ln 2 \qquad {\rm and} \qquad C_6 = C_1.
\end{equation}

Using Eq.~(\ref{eq-A0}), we can express $C_1$ in terms of the boundary values
 of $\tilde{A}_0$ and $\tilde{A}_3$ at $u=0$, namely
\begin{equation}
C_1 = \left. \frac{q^2 \tilde{A}_0 + wq \tilde{A}_3}{iw-q^2}\right|_{u=0}\,.
\end{equation}
Differentiating the on-shell Maxwell action with respect to the boundary values, we can find the R-current correlators. In particular,
\begin{equation}\label{JJAdS}
\langle j_0(p)\, j_0(-p)\rangle = \frac{N^2T}{16\pi} \frac{k^2}{i\omega-Dk^2}\,,
\end{equation}
where
\begin{equation}
  D = \frac1{2\pi T}\,.
\end{equation}
The correlator given by Eq.~(\ref{JJAdS}) has a hydrodynamic diffusive pole and $D$ is the R-charge diffusion constant (for a treatment of relativistic fluid dynamics, see for example \cite{LL6}).

\begin{savequote}[15pc]
\sffamily
Boundaries are actually the main factor in space, just as the present, another boundary, is the main factor in time.
\qauthor{Eduardo Chillida (1924 - 2002)}
\end{savequote}
\chapter{Anti de Sitter Spacetimes with de Sitter Boundary} \label{chapter:geom}

In this chapter, we review the geometries involved in the holographic calculation. Before reviewing the anti de Sitter spacetimes with de Sitter boundary, let us start by reviewing some properties of de Sitter space.

\section{The de Sitter Space} \label{desitterrev}

The de Sitter space is the maximally symmetric solution to Einstein equation with positive cosmological constant. It can be viewed as a ``sphere" described by
\begin{equation}
-x_0^2+x_1^2+\cdots+x_d^2= R_{\rm dS}^2
\label{hyper}
\end{equation}
in a flat $d{+}1$-dimensional Minkowski space, where $R_{\rm dS}$ is a parameter with unit of length, which we will call the de Sitter radius. The relation between the de Sitter radius and the cosmological constant $\L$ is given by
\begin{equation}
\Lambda= {(d-2)(d-1) \over 2 \, R_{\rm dS}^2}.
\label{lambda}
\end{equation}

We note that in the embedding \ref{hyper} the O$(d,1)$ symmetry, which is the isometry group of $dS_d$, is manifest. We will sometime refer to this symmetry as the de Sitter symmetry. 

The metric for the global patch can be obtained by setting
\begin{eqnarray}
x^0 &=& R_{\rm dS} \, \sinh \frac{t}{R_{\rm dS}},\cr
x^i &=& R_{\rm dS} \, \omega^i\, \cosh \frac{t}{R_{\rm dS}}, ~~~~~ i=1,\ldots,d,
\end{eqnarray}
\label{ccv}
where $\w_i$'s form a coordinate for a $d-1$-dimensional sphere $S^{d-1}$
\begin{eqnarray}
\omega^1 &=& \cos \theta_1,\cr
\omega^2 &=& \sin \theta_1 \cos \theta_2,\cr
&\vdots&\cr
\omega^{d-1} &=& \sin \theta_1 \cdots \sin \theta_{d-2} \cos \theta_{d-1},\cr
\omega^{d} &=& \sin \theta_1 \cdots \sin \theta_{d-2} \sin \theta_{d-1},
\label{omegas}
\end{eqnarray}
where $0 \le \theta_i < \pi$ for  $1 \le i < d-1$, but $0 \le \theta_{d-1} < 2
\pi$. The metric is given by
\EQ
{ds^2 = - dt^2 +R_{\rm dS}^2 \cosh^2 \left({t\over R_{\rm dS}}\right)  \;d\Omega_{d-1}^2 \label{dsdmetric}.
}
In these coordinates, the $d$-dimensional de Sitter space looks like a $d{-}1$-dimensional sphere which starts out infinitely large at $t = -\infty$, then shrinks to a minimal finite size at $\tau = 0$, then grows again to infinite size as $t \to +\infty$.

We note that $\partial_t$ is not a Killing vector and there is no globally time-like Killing vector in de Sitter space. This absence of a globally time-like Killing vector in de Sitter space is tightly related to the issue of vacuum ambiguity of field theories on de Sitter space. We will discuss this issue of vacuum ambiguity in Chapter \ref{chapter:alpha}.

\section{The Topological AdS Black Hole \label{toprev}}
The so-called topological black hole in $AdS_5$ (times a five-dimensional sphere) can be obtained as a near horizon limit of $N$ D3-branes filling a boost orbifold ${\mathbb R}^{1,1}/{\mathbb Z}$ \cite{Balasubramanian:2005bg}. It is an orbifold of the $AdS_5$ space, obtained by an identification of points along the orbit of a Killing vector
\EQ
{\xi = \frac{r_\chi}{R_{\rm AdS}} \left(x_4 \partial_5 + x_5 \partial_4 \right),
}
where $r_\x$ is an arbitrary real number and the $AdS_5$ is described as the universal covering of the hypersurface
\EQ
{-x_0^2 + x_1^2 + x_2^2 + x_3^2 + x_4^2 - x_5^2 = - R^2_{\rm AdS},
}
$R_{\rm AdS}$ being the $AdS_5$ radius. In Kruskal-like coordinates, which cover the whole spacetime, the metric has the form
\EQ
{
ds^2= {4R^2_{\rm AdS}\over (1-y^2)^2}\;dy^\mu dy^\nu \eta_{\mu\nu}+ 
{(1+y^2)^2\over (1-y^2)^2}\;r_\x^2d\chi^2,
\label{kruskal}
}
where $\chi$ is a periodic coordinate with period $2\pi$. The four coordinates $y^\mu$, with $\mu=0,\ldots 3$, are non-compact, with the Lorentzian norm $y^2= y^\mu y^\nu\eta_{\mu\nu}$ is between -1 and 1. Locally, the spacetime is anti de Sitter with a periodic identification of the $\chi$ coordinate,
\EQ
{
\chi\sim \chi+2\pi.
}
The conformal boundary of the spacetime is approached as $y^2\rightarrow 1$, and it is $dS_3 \times S^1$. The boundary conformal field theory is therefore formulated on a three dimensional de Sitter space with radius of curvature $R_{\rm AdS}$ times a spatial circle of radius $r_\x$. 

The geometry has a horizon at $y^2=0$, which is the three dimensional hypercone,  
\EQ
{y_0^2= y_1^2+y_2^2+y_3^2,}
and a singularity at $y^2=-1$. The hyperboloid $y^2=-1$ is a singularity since timelike geodesics end there and the Killing vector $\partial_\chi$ generating the orbifold identification has vanishing norm at $y^2=-1$. This singularity appears because the region where the Killing vector has negative norm needs to be excised from the physical spacetime to eliminate closed timelike curves. 

It is interesting to see that 
we can get a better understanding of the geometry in the vicinity of
the singularity at $y^2=-1$ by zooming in on the 
the metric \eqref{kruskal} in this region.
Introducing the coordinates
\EQ
{y_0= (1-\delta) \, \cosh\varepsilon\,,\qquad Y =(1-\delta)\,\sinh\varepsilon
\qquad0< \delta\ll 1,}
we find
\EQ
{
ds^2\approx R^2_{\rm AdS}(-d\delta^2 + d\varepsilon^2 + \sinh^2\varepsilon
\,d\Omega_2^2)+ r_\x^2 \delta^2\,d\chi^2,
}
which is a product of geometries, with the factor in the $\delta, \chi$ directions is describing the Milne spacetime.
 
The topology of the spacetime is ${\mathbb R}^{3,1}\times S^1$, in contrast to that of the  AdS-Schwarzschild black hole, which has the topology ${\mathbb R}^{1,1}\times S^3$. For this reason, in this geometry, infinity is connected, unlike in the usual Schwarzschild black hole which has two disconnected asymptotic regions.

\begin{figure}[h]
\begin{center}
\includegraphics[width=3.0in]{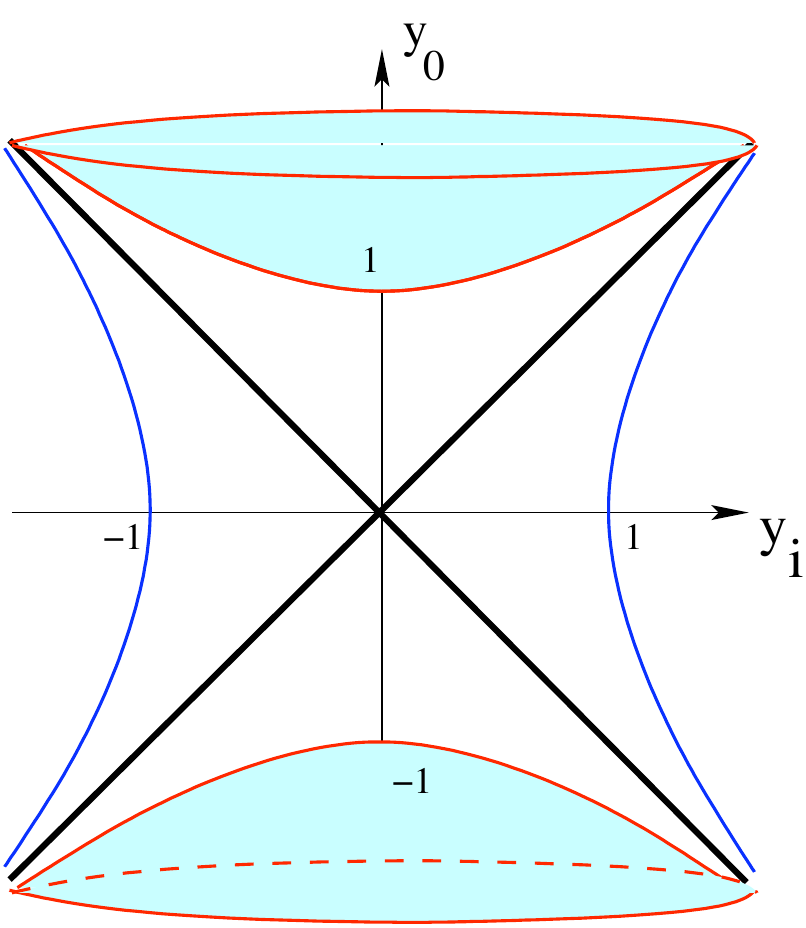} 
\end{center}
\caption{\footnotesize The global structure of the topological AdS black hole spacetime. The singularity is the hyperboloid $-y_0^2+y_i y_i=-1$ and the horizon is at the cone $y_0^2 = y_i y_i$.}
\label{spacetime}
\end{figure}

It is possible to rewrite the metric in Schwarzschild-like coordinates by introducing the following coordinate transformations
\EQ
{
Y^2=\sum_{i=1}^3y_iy_i\,;\qquad {Y\over y_0}=
\coth\left({t\over R_{\rm AdS}}\right)\,;
\qquad {r^2\over R^2_{\rm AdS}} = 4 {(Y^2-y_0^2)\over (1+y_0^2-Y^2)^2}.
}
These coordinates only  cover the exterior of the topological black hole $y^2\geq 0$. Locally, the metric takes the form
\SP
{
ds^2= {R^2_{\rm AdS} \, dr^2\over (r^2+R^2_{\rm AdS})} + 
\left({r_\x\over R_{\rm AdS}}\right)^2
(r^2+R^2_{\rm AdS})d\chi^2+
\frac{r^2}{\R^2}
\left(- dt^2+ \R^2 \, \cosh^2 \left({t\over R_{\rm AdS}}\right)
  \;d\Omega_2^2\right). 
\label{schw}
}
The Euclidean continuation of this metric yields the thermal AdS space. Hence, the metric of  the exterior region of the topological black hole can also be obtained following a double Wick rotation of global AdS spacetime and a periodic identification of the  $\chi$ coordinate. 

In the Schwarzschild-like coordinates, the horizon of the topological black hole is at $r=0$, while the boundary is at $r \rightarrow \infty$. It is clear that each slice of constant $r$ is a $dS_3\times S^1$ geometry, where the three-dimensional de Sitter space is in the global patch. The metric (\ref{schw}),  while locally describing AdS space, differs from it globally due to the identification ${\chi\sim\chi + 2\pi}$. We note also that the spatial $S^1$ remains finite sized at the horizon, with radius  $R_{S^1}= r_\x$.

It will also be convenient to introduce a dimensionless radial coordinate $z=R_{\rm AdS}/r$, in which the metric for the exterior region is now given by
\EQ
{
ds^2= R^2_{\rm AdS} \left[{ dz^2\over z^2 (z^2+1)} +
{1 \over z^2}
\left(-{dt^2\over \R^2}+\cosh^2 \left({t\over R_{\rm AdS}}\right)
  \;d\Omega_2^2\right) + \frac{1+z^2}{z^2}
\left({r_\x\over R_{\rm AdS}}\right)^2 
d\chi^2 \right]. 
\label{in z}
}
Here, the boundary is at $z=0$, while the horizon is at $z \rightarrow \infty$.

If the bulk fermions have anti-periodic boundary conditions in the $\x$-direction, then the topological black hole has a semiclassical instability when
\EQ
{
r_{\x}<{R_{\rm AdS} \over 2\sqrt{2}}\, ,
}
which causes it to decay into the AdS bubble of nothing \cite{Balasubramanian:2005bg}.  The instability only occurs if fermions have antiperiodic
boundary conditions in the $\chi$-direction. With periodic boundary
conditions for both bosons and fermions, the topological AdS black
hole is absolutely stable. This instability can also be seen in the boundary field theory by considering the effective potential of the Polyakov loop of the Euclidean field theory \cite{Hollowood:2006xb}.

\section{The AdS Bubble of Nothing}

As originally pointed out in \cite{Witten:1981gj} (and
\cite{Balasubramanian:2005bg}, in the present context), the 
decay of a false vacuum in semiclassical gravity
is computed by the Euclidean bounce which has the same
asymptotics as the false vacuum in Euclidean signature. The bounce is
a solution to the Euclidean equations of motion with a non-conformal 
negative mode. In the context of the asymptotically locally AdS
spaces in question, the small Euclidean Schwarzschild solution
represents such a bounce solution. In Lorentzian signature, the
semiclassical picture of the decay process at say, time $t=0$ involves
replacing the $t>0$ part of the false vacuum solution, \textit{i.e.}, the topological
black hole solution, with the appropriate analytic continuation 
of the Euclidean bounce to
Lorentzian signature. The analytic continuation of the small Euclidean
AdS black hole bounce which leads to 
$dS_3\times S^1$ boundary asymptotics is the (small) AdS bubble of nothing.

\begin{figure}[h]
\renewcommand{\figurename}{Fig.}
\begin{center}
\includegraphics[width=2.5in]{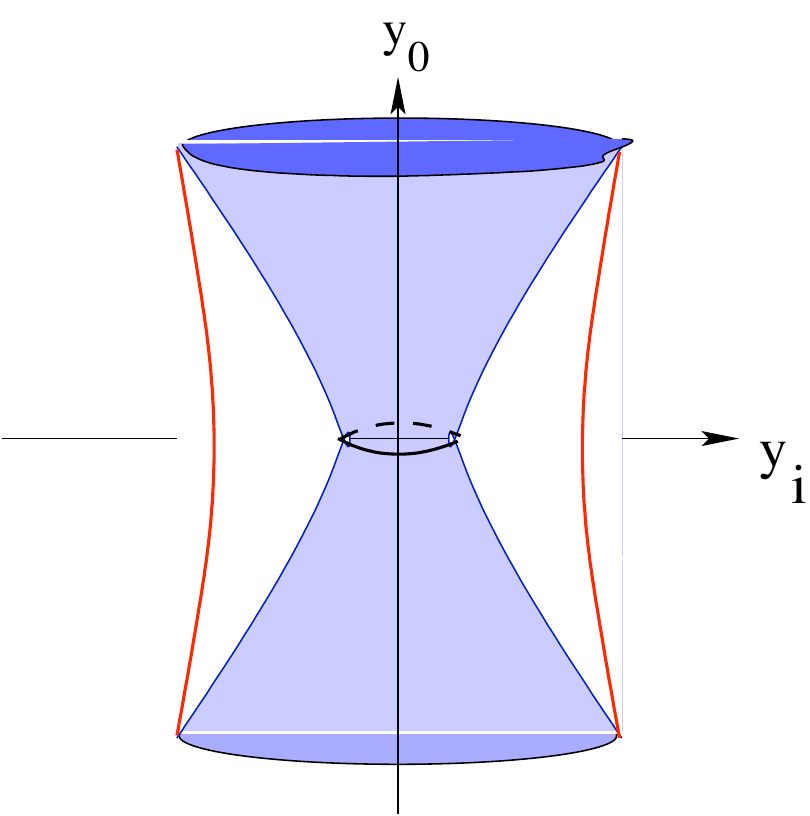}
\end{center}
\caption{\footnotesize The global structure of the bubble
  geometry. The region inside the shaded region is empty, and its
  surface represents the de Sitter expansion of the bubble at $r=r_h$.
}\label{bubble}
\end{figure}

The metric for the AdS bubble of nothing solution is
\EQ
{ds^2=
  f(r)\;r_\x^2\;d\chi^2+f(r)^{-1}\;dr^2
+{r^2\over R_{\rm AdS^2}}\left(-  dt^2+ 
\cosh^2\left({t\over R_{\rm AdS}}\right)d\Omega_2^2\right),
}
where
\EQ
{f(r)=1+{r^2\over R_{\rm AdS}^2}-{r_h^2(R_{\rm AdS}^2+r_h^2)\over
  r^2}.
} 
In order to avoid a conical singularity in the interior, the
periodicity of the compact $\chi$ coordinate is related to $r_h$ as
\EQ
{
2\pi r_\x =2\pi R^2_{\rm AdS} \;{r_h\over r_h^2+ R_{\rm AdS}^2}.
} 
Passing to the dimensionless coordinates
\EQ
{
\rho=\sqrt{(r/R_{\rm AdS})^2+1}, \qquad\tau={t\over R_{\rm
    AdS}},\qquad \tilde r_h = {r_h\over R_{\rm AdS}},
}
the metric becomes
\EQ
{
ds^2= R^2_{\rm AdS}\left[\tilde f(\rho) \;{r_\x^2 \over R^2_{\rm AdS}}
\;d\chi^2 + \tilde f (\rho)^{-1}
\;{\rho^2\over \rho^2-1}\;d\rho^2+ (\rho^2-1)(-d\tau^2+\cosh^2\tau d\Omega_2^2)
\right],
}
where
\EQ
{ \tilde f(\rho)=\rho^2- {1\over \rho^2-1}\;{\tilde r_h^2}\;
\left(\tilde r_h^2+1\right).
}
In the AdS bubble of nothing 
spacetime, a slice of constant $\rho$ is $dS_3\times S^1$. 
The $S^1$, however, shrinks to zero size smoothly at $\rho=
\sqrt{\tilde r_h^2 +1}$. The shrinking circle is the cigar of the Euclidean
Schwarzschild solution. 
The boundary of the spacetime is approached as $\rho\rightarrow \infty$.
The semiclassical decay of the topological black hole at $\tau=0$,
results in the sudden appearance of a bubble of nothing in the
region of spacetime, $\rho^2 \leq 1+ \tilde r_h^2$, see Fig. \ref{bubble}

\begin{savequote}[15pc]
\sffamily
You go through these little phases and fads, \\
and it never turns out the way you think it's going to turn out. 
\qauthor{Will Sergeant (b. 1958)}
\end{savequote}
\chapter{Phases of ${\mathcal N}=4$ Super Yang-Mills Theory on de Sitter Space} \label{chapter:phases}

In this chapter, in order to further establish the duality between the strongly coupled ${\mathcal N}=4$ Super Yang-Mills Theory on de Sitter Space and the Type IIB supergravity on the asymptotically locally AdS spacetimes introduced in the previous chapter, we are going to obtain the real time correlators using holographic calculation and use the results to distinguish the two phases of the boundary field theory.

\section{Real Time Correlators in the Topological AdS Black Hole}

We will compute the real time correlators in the 
Yang-Mills theory on the boundary of the
topological $AdS_5$ black hole following the  
recipe of Son and Starinets \cite{Son:2002sd}, which we have reviewed in Section \ref{reviewson}, in the Schwarzschild-like patch
\eqref{schw} of the black hole. 

Viewing the topological black hole as 
a Wick rotation of the thermal AdS space, one expects that 
such correlators can also
be obtained by an appropriate analytic continuation of 
Euclidean Yang-Mills correlators on $S^3 \times S^1$ 
in the confined phase (the ${\mathbb Z}_N$ symmetric phase) with
anti-periodic boundary conditions for fermions. 
Since the relevant Wick rotation  
turns the polar angle on $S^3$ into the time
coordinate of de Sitter space, a complete knowledge of the angular
dependence of Euclidean correlators on $S^3\times S^1$would be necessary.
However, finite temperature Yang-Mills correlators on $S^3$ at strong coupling 
have not been calculated explicitly, so we will not follow the route of
analytic continuation. Instead, we will directly calculate the real
time correlators  
using the holographic prescription of Son and Starinets 
applied to the topological AdS black hole geometry. 

\subsection{Scalar Wave Equation in the Topological Black Hole}
To extract the field theory correlators, we first need to look for solutions to the
equation of motion in the region exterior to the horizon of the
topological black hole. 

It is instructive to write the metric for the black hole 
in the Schwarzschild form \cite{Balasubramanian:2005bg} of
\EQ
{
ds^2= \R^2 \left[{d\rho^2\over (\rho^2 -1)} + 
\left({r_\x\over \R}\right)^2
\rho^2d\chi^2+
(\rho^2-1)
\left(-d\tau^2+\cosh^2 \tau \;d\Omega_2^2\right)\right],
\label{schw1}
}
where we have introduced the dimensionless variables
\EQ
{
\rho=\sqrt{\left(r/\R\right)^2+1},\qquad\qquad\tau={t\over\R}.
} 
The conformal boundary of the space is approached as
$\rho\rightarrow\infty$ while  
the horizon is at $\rho=1$, where the coefficient of $d\tau^2$
vanishes. The slices with constant $\rho$ are manifestly $dS_3\times
S^1$. 

The scalar fields in this geometry have a natural expansion 
in terms of harmonics on the $S^2\times S^1$ spatial slices 
\EQ
{
\Phi(\rho,\chi,\tau,\Omega)
=\sum_{\ell,m,n}\;{\cal } 
A_{\ell \,m}\;Y_{\ell \,m}(\Omega)\;e^{in\chi}\int {d\nu\over
  2\pi}\;\Phi_{n}\,(\nu, \rho)\;{\cal T}_{\ell}(\nu,\tau).
\label{decomp}
}
The normal mode expansion above involves spherical harmonics on $S^2$,
the discrete Fourier modes on $S^1$ and ${\cal T}_{\ell}(\nu,\tau)$
which  solve the scalar wave equation on
$dS_3$
\EQ
{
{1\over \cosh^2 \tau}\partial_\tau \left(\cosh^2\tau \;\partial_\tau{\cal
    T}_{\ell}(\nu,\tau)\right)+{\ell(\ell+1)\over \cosh^2\tau}{\cal T}_{\ell}(\nu,\tau) 
= -(\nu^2+1){\cal
    T}_{\ell}(\nu,\tau).
}
For every $\ell \in {\mathbb Z}$, the equation has two kinds of
solutions that will be relevant for us:
\begin{itemize}
\item[] i) normalizable modes labelled by integers
$- i\nu = 1,2,\ldots \ell$; 
\item[] ii) delta-function normalizable modes
labelled by a continuous frequency variable $\nu \in {\mathbb R}$. 
\end{itemize}
We will return to this point when we discuss R-current correlators.

General solutions to this equation can be expressed in terms of
associated Legendre functions
\EQ
{
{\cal T}_{\ell}(\nu,\tau)=\;{1\over \cosh\tau}
\left(\;A_l\;P_\ell^{i\nu}(\tanh\tau)+\;B_l\;Q_\ell^{i\nu}(\tanh\tau)\right).
}
In the usual approach to quantizing free scalar fields in de Sitter
space, the integration constants $A_\ell$ and $B_\ell$ are determined by the 
choice of de Sitter vacuum \cite{Mottola:1984ar, Birrell:1982ix, Spradlin:2001pw}. However, in the present
context, the constants will be specified by picking out infalling wave
solutions at the horizon of the topological black hole. These are the
holographic boundary conditions relevant for real time response
functions in the strongly coupled field theory on $dS_3\times S^1$.

It is useful to see 
the scalar wave equation in this background recast as a
Schr\"odinger equation. We can do so by using Regge-Wheeler type variables  
\EQ
{
u={1\over 2}\ln\left({\rho+1\over \rho-1}\right)\qquad{\rm or} \qquad
\rho=\coth u,
\label{rg1}
}
and 
\EQ
{
\Psi_n =  \sqrt{\rho(\rho^2-1)}\;\Phi_n,
\label{rg2}
}
$\Phi$ being the scalar field in the bulk. 
\begin{figure}[h]
\begin{center}
\includegraphics[width=3.5in]{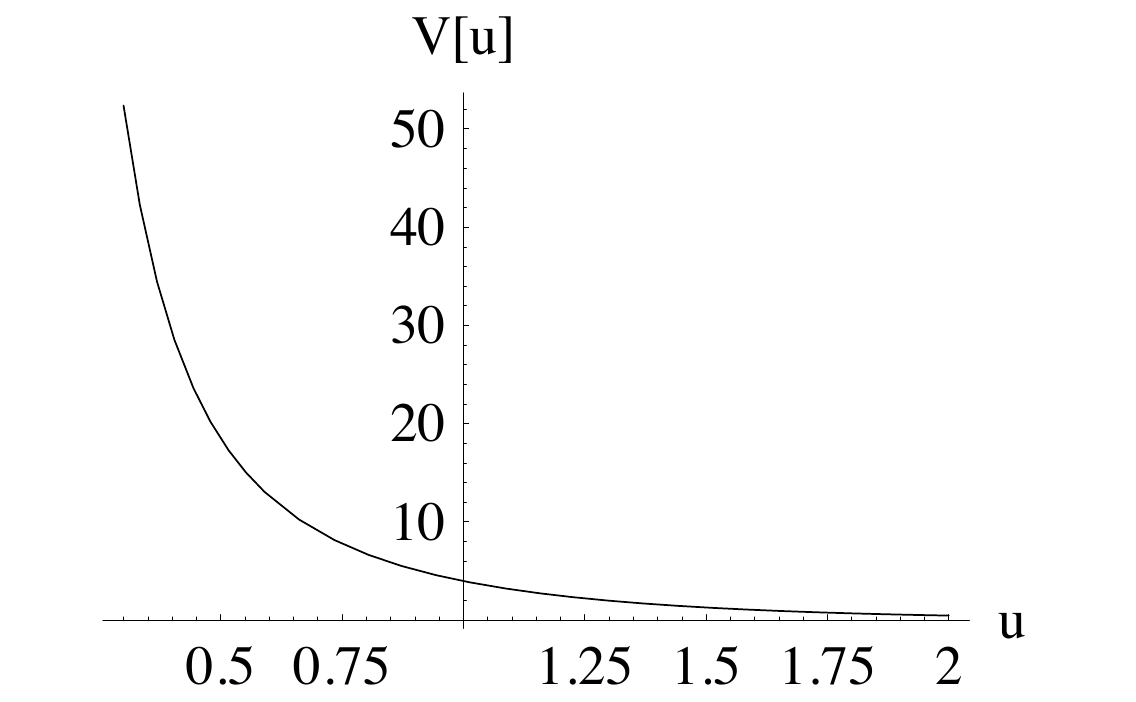} 
\end{center}
\caption{\footnotesize The Schr\"odinger potential for the
topological AdS black hole.}
\label{vofu}
\end{figure}
In these coordinates, the horizon is approached as $u\rightarrow
\infty$ while the conformal boundary is at $u=0$.
Following the above coordinate and field redefinitions, the
Schr\"odinger wave equation in the topological black hole geometry is given by
\EQ
{
-{d^2\over du^2}\;\Psi_{n}(\nu,u) 
+ V_{n}(u)\;\Psi_n(\nu, u)= \nu^2\;\Psi_n(\nu,u),
}
with the following potential (see Fig. \ref{vofu})
\EQ
{V_{n}(u)= ((M R_{\rm AdS})^2+\tfrac{15}{4}){1\over \sinh^2 u}+(\tfrac{1}{4}+
\tfrac{R^2_{\rm AdS}}{r_\x^2}\,n^2){1\over\cosh^2 u}\;. 
\label{schrodinger1}
}
Here, $\Psi_n = \Phi_n\sqrt{\rho(\rho^2-1)}$ and 
we have allowed for a generic non-zero mass $M$, since we will
eventually be interested both in the massless and massive cases.

As expected for AdS black holes, the potential decays exponentially near
the horizon $u\rightarrow \infty$, while blowing up near the boundary
at $u\rightarrow 0$. In the near horizon region where the potential
vanishes, the solutions with $\nu>0$ 
are travelling waves and there is a natural choice
of incoming and outgoing plane wave solutions. 
For any $n$ and $M$, the equation has analytically tractable 
solutions in terms of
hypergeometric functions. We will use these to calculate the retarded
Green's functions for the boundary gauge theory, {\it i.e.,} the ${\cal N}=4$
Super Yang-Mills at strong coupling on $dS_3\times S^1$.

Although analytical solutions exist for all non-zero $n$ and
$M$, we will restrict attention, for simplicity, to two special cases:
\begin{itemize}
\item[] i) $n\neq 0$ and 
$M R_{\rm AdS}=0$; 
\item[] ii) $n=0$ and $M R_{\rm AdS}\neq 0$. 
\end{itemize}
In each of these two cases 
the radial equation is solved by two linearly independent
hypergeometric functions. For the case of $M=0$ and $n\neq 0$, we have
\begin{eqnarray}
\Phi_n(\nu,\rho)\,&=&\,C_1\; \rho^{-i\,n\,\R/r_\x}\,\,(\rho^2-1)^{-i{\nu\over 2}-{1\over 2}} \,\,
{}_2
F_1\left(-\tfrac{1}{2}-\tfrac{i}{2}(\nu+n\,\tfrac{\R}{r_\x}),
\;\tfrac{3}{2}-\tfrac{i}{2}(\nu+n\,\tfrac{\R}{r_\x});
\;1-i n\,\tfrac{\R}{r_\x};\;\rho^2\right)\nonumber \\
& &+\,\,C_2 \; \rho^{i \,n\,\R/r_\x}\,\,\;(\rho^2-1)^{-i{\nu\over 2}-{1\over 2}}\;\,
{}_2
F_1\left(-\tfrac{1}{2}-\tfrac{i}{2}(\nu-n\,\tfrac{\R}{r_\x}),
\;\tfrac{3}{2}-\tfrac{i}{2}(\nu-n\,\tfrac{\R}{r_\x});
\;1+i n \,\tfrac{\R}{r_\x};\;\rho^2\right)\nonumber\\
\label{massless}
\end{eqnarray}

Similarly, for massive fields with $n=0$, we have
\begin{eqnarray}
\Phi_0(\nu,\rho)& =& \;C_1\;
(\rho^2-1)^{{-\tfrac{1}{2}}(4-\Delta)}\; \,
{}_2F_1\left(\tfrac{1}{2}(3-\Delta)-\tfrac{1}{2}i\nu,\;    
\tfrac{1}{2}(3-\Delta)+\tfrac{1}{2}i   
\nu;\;3-\Delta;\;-(\rho^2-1)^{-1}\right)\nonumber \\
& & +\,\,C_2\; (\rho^2-1)^{-\tfrac{1}{2}\Delta}\;
{}_2F_1\left(\tfrac{1}{2}(\Delta-1)-
\tfrac{1}{2}{i \nu},
\;\tfrac{1}{2}(\Delta-1)+\tfrac{1}{2}i{\nu};\;\Delta-1;\;\,
-(\rho^2-1)^{-1}\right).\nonumber\\
\end{eqnarray}
Here,
\begin{equation}
\Delta=2+\sqrt{4+(M\R)^2}.
\end{equation}

The correct linear combination, relevant for the holographic
computation of correlators, is  picked by applying the requirement
of purely infalling waves at the horizon of the topological black hole
at $\rho\approx1$.

\subsection{Scalar Glueball Correlators}

As argued in \cite{Balasubramanian:2002am, Balasubramanian:2005bg}, 
the topological black hole
in AdS space is automatically a solution to the Type IIB supergravity
equations of motion, since it can be obtained via a double Wick rotation
(and an identification) of $AdS_5\times S^5$. 
The ${\cal N}=4$ supersymmetric Yang-Mills theory on the $dS_3\times S^1$
conformal boundary of the topological black hole, has two $SO(6)$
singlet, scalar glueball fields 
\EQ
{
{\cal G}(\vec x,t)=\Tr F_{\mu\nu}F^{\mu\nu} \qquad {\rm and} \qquad
\tilde {\cal G}(\vec x,t)=\Tr F_{\mu\nu}\tilde F^{\mu\nu},
}
which are $\Delta=4$ operators of the boundary field theory. These are dual to the dilaton and the RR-scalar in the Type IIB theory
on the bulk spacetime and 
 both solve the massless  scalar wave equation in the topological
 black hole geometry. 
 
 The retarded propagators for the scalar glueball
 fields are known on ${\mathbb R}^{3,1}$ at weak coupling both at zero
 and finite temperature \cite{Hartnoll:2005ju}. 
Since the operators are chiral primary operators in the
 ${\cal N}=4$ theory, at zero temperature their propagators on ${\mathbb
   R}^{3,1}$ receive no quantum corrections and the strong coupling
 results from supergravity are in exact agreement with those of
 the free field theory. At finite temperature, however, when
 supersymmetry is broken, strong and
 weak coupling results on ${\mathbb R}^{3,1}$ differ \cite{Son:2002sd,
   Hartnoll:2005ju}. Computations of the glueball correlators also
 exist in the
 free ${\cal N}=4$ theory at finite temperature and on a spatial $S^3$, both
 in the confined and deconfined phases \cite{Hartnoll:2005ju}. Their
 strong coupling counterparts have not been determined.

The present case, with the field theory on $dS_3 \times S^1$, 
is intriguing for the following reasons. First of all, there is the lack of
supersymmetry. Secondly, the
boundary field theory sees a cosmological horizon on $dS_3$ 
accompanied by its
thermal bath. It would be interesting to observe the emergence of
the boundary Gibbons-Hawking temperature  
from a holographic calculation of its
correlators at strong coupling. Finally, when the radius of the boundary $S^1$
decreases below a critical value, $r_\x/\R \leq {1\over 2\sqrt{2}}$, the
topological black hole decays via a bounce to the small AdS bubble of
nothing. We would like to understand how 
boundary field theory correlators at strong coupling 
on $dS_3\times S^1$ change across this transition. The 
transition from the topological black hole to the Bubble of Nothing is
a ${\mathbb Z}_N$ breaking transition. This is understood precisely as
in the finite temperature situation: the ${\mathbb Z}_N$ breaking transition is due to a non-zero
expectation value for the Wilson loop around the spatial $S^1$.

Curiously, it is apparent that in the classical supergravity
approximation, the bulk scalar glueball correlators are insensitive to 
fermions and their boundary conditions around the spatial $S^1$. 
It would be interesting to understand whether 
this is related to large-$N$ volume independence 
\cite{Kovtun:2007py,Unsal:2008ch} in the ${\mathbb Z}_N$ symmetric phase.

We are primarily interested in real time response and for
the sake of simplicity, we will first 
study only the response functions for glueball fluctuations
that are homogeneous on the spatial $S^2$ slices at the boundary, {\it
  i.e.},
\EQ
{
G_R(\t,\t'\,;n\,;l=0)
= -i\int \frac{d\Omega}{4\pi}\,\frac{d\Omega'}{4\pi} 
\;\int{d\chi\over 2\pi} \;e^{-in \chi} 
\,\Theta(\t-\t')\;\langle\,\left[\;{\cal G}(\Omega,\chi,\t),\;{\cal
    G}(\Omega',0,\t')\;\right]\,\rangle.  
}
We will work with the dimensionless variables $\t= t/R_{\rm AdS}$,  
$\t' = t'/R_{\rm AdS}$ and restore appropriate dimensions when necessary. 
As we will see when we look at the R-current correlators, it is 
straightforward to generalize to 
the case of inhomogeoneous fluctuations on the spatial sphere. 

For the moment, let us focus our attention on the $s$-wave 
($\ell=0$) retarded correlation function of the
scalar glueball operator. For the $s$-waves, the correlator turns out to
be a function of $(\t-\t')$ so that it is natural to 
define the temporal Fourier transform of this as
\EQ
{
\tilde G_R(\nu\,;n) = \int _{-\infty}^\infty {d\t}
\, e^{-i\nu (\t-\t')} \, G_R(\t,\t'\,;n\,;\ell=0).
}
To calculate it at strong coupling and in the large radius regime
($r_\x\geq {\R\over2\sqrt 2}$), 
we solve the axion-dilaton wave equation
which is the equation for a massless, minimally coupled scalar field in the
background of the topological AdS black hole. 

\subsubsection{Spatially Homogeneous Case with $n=\ell=0$ }
We begin by looking at the spatially
homogeneous response functions, $\ell=n=0$, 
on the $dS_3\times S^1$ slices. The solutions to
the radial part of the Klein-Gordon equation in the massless limit 
are the hypergeometric functions
\EQ
{
{\Phi}_0^{(1)}(\nu,\rho)=\frac{\pi}{4}\;{1+\nu^2\over\cosh
\left(\tfrac{\pi \nu}{2}\right)}
\;\;{}_2F_1\left(\;-\tfrac{1+i\nu}{2}\;,\;-\tfrac{1-i\nu}{2}\;;\;1\;;
  \;\tfrac{\rho^2}{\rho^2-1}\right)     
}
and
\EQ
{
{\Phi}_0^{(2)}
(\nu,\rho)=(\rho^2-1)^{-2}\;{}_2F_1\left(\tfrac{3-i\nu}{2}\;,\;
  \tfrac{3+i\nu}{2}\; 
;\;3\;; \;-\tfrac{1}{\rho^2-1}\right).
}
For the $\ell=0$ modes, the temporal dependence is also particularly
simple and it has a natural interpretation in terms of positive and
negative frequency states
\EQ
{
{\cal T}_{0}^+(\nu,\tau)= {e^{-i \nu \tau}\over \cosh\tau} \qquad
{\rm and}\qquad
{\cal T}_{0}^-(\nu,\tau) ={e^{i \nu \tau}\over \cosh\tau}.
\label{modes}
}
Solving the Dirichlet problem and extracting correlation
functions holographically requires us to
first pick the correct linear combination of the
two solutions which is smooth near the horizon
$\rho\rightarrow 1$ and represents an incoming wave falling
into the horizon. 

In the near horizon region, the asymptotic forms of the solutions are:
\begin{eqnarray}
{\Phi}^{(1)}_0(\nu, \rho\rightarrow1) \rightarrow
i\;(2(\rho-1))^{-\frac{1-i \nu}{2}} \;e^{{\pi\over 2}\nu}\;
{\Gamma(-i \nu)\;\Gamma\left({3+i\nu\over2}\right)\over 
\Gamma\left(-{1+i\nu\over 2}\right)}
+i\;(2(\rho-1))^{-\frac{1+i\;\nu}{2}}
\;e^{-{\pi\over 2}
\nu}{\Gamma(i \nu)\;\Gamma\left({3-i\nu\over2}\right)
\over \Gamma\left(-{1-i\nu\over 2}\right)} \nonumber \\
\end{eqnarray}
and
\EQ
{
{\Phi}^{(2)}_0(\nu, \rho\rightarrow1)\rightarrow 
 \;(2(\rho-1))^{-\tfrac{1-i \;\nu}{2}} 
{2\; \Gamma(-i \nu)\over 
\Gamma\left({3-i\nu\over 2}\right)^2}
+\;(2(\rho-1))^{-\tfrac{1+i\;\nu}{2}} {2 \;\Gamma(i
\nu) 
\over \Gamma\left({3 + i\nu\over 2}\right)^2}.
}
We note that these modes diverge as $1/\sqrt{(\rho-1)}$ near the
horizon. However, employing the measure implied by the bulk metric
$\sqrt {-g} \sim (\rho^2-1)^{5/2}$, these are still normalizable in the vicinity
of the horizon. 

We can now 
pick a linear combination such that only the incoming positive frequency 
waves are allowed at the
horizon. 
This means, assuming that ${\cal T}_{0}^+$ are the 
positive frequency modes with ${\rm
  Re}(\nu)>0$, 
the solution to the radial wave equation should behave like
$(\rho-1)^{-\frac{1+i\nu}{2}}$ near the horizon. 

In conjunction with this, we have the properly normalized 
boundary behaviour as $\rho \rightarrow \infty$,
\EQ
{
\Phi^{(1)}_0(\nu,\rho\rightarrow \infty)\rightarrow 1 +\ldots, \qquad {\rm and} \qquad 
{\Phi}^{(2)}_0(\nu, \rho\rightarrow \infty)\rightarrow
{1\over\rho^4}
+ \ldots.}
The complete solution to the boundary value problem for a massless
scalar with $\ell=n=0$, in the topological AdS black hole is then
\EQ
{
{\Phi}_0(\nu,\rho)= {\Phi}^{(1)}_0(\nu,\rho) +
i {\pi\over 32}\;e^{{\pi\over
    2}\nu}\;{(\nu^2+1)^2\over\cosh{{\pi\over 2}\nu}}
\;{\Phi}^{(2)}_0(\nu,\rho).
\label{sol}
}
Following the holographic 
prescription \cite{Son:2002sd,Policastro:2002se} for computing real
time correlators, the Yang-Mills 
retarded correlation function is obtained by analyzing the
boundary terms from the on-shell scalar action
\SP
{
S
&= {N^2\over 16\pi^2}
\int d\tau\int d\Omega\int d\chi\;g^{\rho\rho}
\sqrt{-g}\; \;\Phi(\tau,\rho)\;\partial_\rho
\Phi(\tau,\rho)\;\big|_{\rho\rightarrow \infty}\,,
\\
}
where for the spatial $s$-wave we have defined
\EQ
{
\Phi(\tau,\rho)= \int_{-\infty}^{\infty} {d\nu\over 2\pi} \; {\cal
  T}_{0}^+(\nu,\tau)\;\Phi_0(\nu,\rho).
} 
Putting together the explicit expressions for ${\cal
  T}_{0}^+(\nu,\tau)$ and the 
boundary behaviour of the solution \eqref{sol}, we are immediately led
to the (unrenormalized)
$s$-wave retarded correlator in frequency space, which includes all contact
terms that are given by finite polynomials in the frequency $\nu$
\SP
{
\tilde G_R(\nu; 0)= &\, {N^2\over 16\pi^2}\left(-\tfrac{1}{8}(1+\nu^2)^2
\left[\psi\left({3-i\nu\over2}\right)+ 
\psi\left({3+i\nu\over2}\right)- i\pi
\coth\left(\pi\tfrac{1}{2}(\nu+i)\right)
\right]\right.\\
&\left.+\,\tfrac{1}{4}(1+\nu^2)^2\;\left(\ln \rho
  -\gamma_E+1\right)\big|_{\rho \rightarrow \infty}+ 
\tfrac{1}{2}(1+\nu^2){\rho^2}\big|_{\rho\rightarrow \infty}\right).
\label{ret}
}
We remark that unlike the case of the Poincare' patch description of 
AdS space, the
non-normalizable solution in the topological AdS black hole (akin to
global AdS), contains a term proportional to $1/\rho^2$ in its near-boundary
asymptotics, but it only
contributes a quadratically divergent contact term in the correlation function
above. 
\begin{figure}[h]
\begin{center}
\includegraphics[width=3.5in]{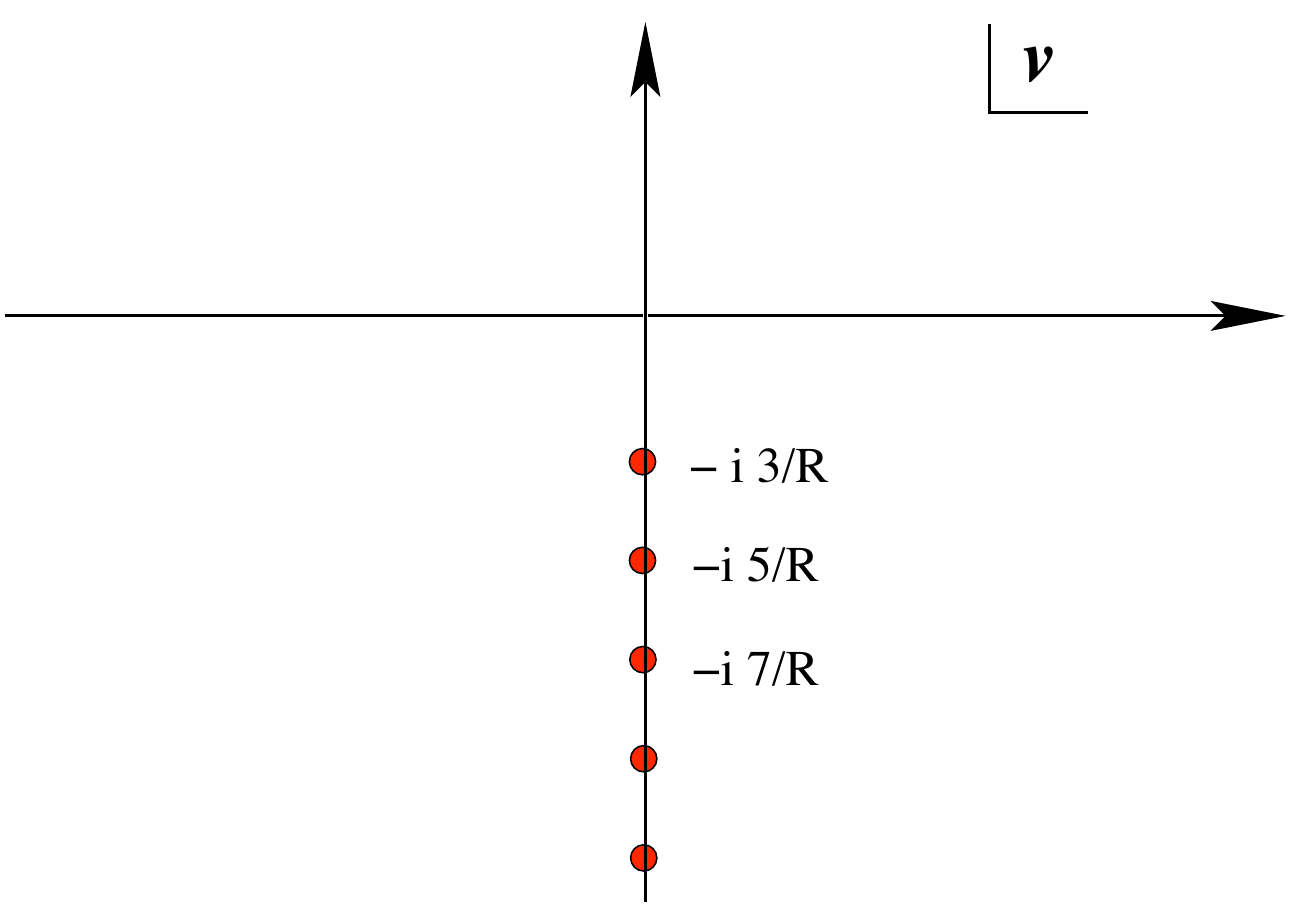} 
\end{center}
\caption{\footnotesize The analytic structure in complex
frequency plane, of the massless,
spatially homogeneous ($l=n=0$), retarded Green's function in the
${\mathbb Z}_N$ symmetric phase, corresponding to the topological
AdS black hole.}  
\label{retbon}
\end{figure}

The divergent and scheme-dependent contact terms can be minimally
subtracted away to yield the renormalized, retarded Green's function. 
Up to now, we have been working with dimensionless variables,
corresponding to a de Sitter boundary of unit curvature. Restoring
dimensionful constants with the replacement
\EQ
{
\nu \rightarrow
\nu R_{\rm dS},
}
$R_{\rm dS}$ being the radius of curvature of the boundary $dS_3$ or the
inverse Hubble constant\footnote{Obviously, this is identical to the radius of curvature of the bulk $R_{\rm AdS}$. We will use both interchangeably: $R_{\rm dS}$ when describing a quantity in the boundary field theory and $R_{\rm AdS}$ when describing a quantity in the bulk gravity theory.} , the renormalized retarded Green's function
continued into the complex frequency plane is given by
\EQ
{
\tilde G_{R_{\rm dS}}(\nu;0)=-\frac{N^2}{64\pi^2}(R_{\rm dS}^{-2}+\nu^2)^2 \;
\left[
\psi\left({3 -i\nu R_{\rm dS}\over2}\right)
-{2i\nu R_{\rm dS} \over(1+\nu^2 R_{\rm dS}^2)}\right].
\label{homog}
}
For ${\rm Im}(\nu)=0$, its real and imaginary parts match
\eqref{ret} and the function is analytic in the upper half plane,
with only isolated simple poles in the lower half plane at
\EQ
{
\nu_k = -i(3+ 2k){1\over R_{\rm dS}},\qquad {\rm with} \qquad k\in {\mathbb Z}.
}
This is depicted in Fig. \ref{retbon}.

As argued in \cite{Son:2002sd}, poles of the retarded correlator
in a black hole background coincide with the quasinormal
frequencies of the black hole. The quasinormal frequencies
and the retarded glueball 
correlator found for the topological black hole in $AdS_5$ are remarkably
similar to the corresponding objects in the BTZ black hole
\cite{Son:2002sd}.

\subsubsection{Non-zero Momentum along $S^1$ and $\ell=0$}

It is relatively easy to allow for a non-zero discrete momentum $n/r_\x$
along the spatial $S^1$. To solve the
boundary value problem in the topological black hole
background, this requires
the modes \eqref{massless}. Following the same steps as in the $s$-wave correlator,
we find that 
the retarded Green's function is
\begin{eqnarray}
\tilde G_R(\nu\,; n) & =&
-\frac{N^2}{128\pi^2}\left((\nu
  -\tfrac{n}{r_\x})^2+R_{\rm dS}^{-2}\right) 
\left((\nu +\tfrac{n}{r_\x})^2+ R_{\rm dS}^{-2}\right)\, \times\nonumber \\
&&\left[\psi\left(\frac{3}{2}- i\tfrac{R_{\rm dS}}{2}(\nu  - 
    \tfrac{n}{r_\x})\right)+\psi\left(\frac{3}{2}-
    i\tfrac{R_{\rm dS}}{2}(\nu  + \tfrac{n}{r_\x})\right)
- {2iR(\nu - \tfrac{n}{r_\x})
\over(\nu - \tfrac{n}{r_\x})^2R_{\rm dS}^2+1}-{2i R_{\rm dS}(\nu +\tfrac{n}{r_\x})
\over(\nu +\tfrac{n}{r_\x})^2R_{\rm dS}^2+1}\right]\,, \nonumber \\
\\
&& n\in {\mathbb Z}. \nonumber
\end{eqnarray}
Here,  we have restored the dimensionful constants. When $n=0$, this matches our expression for the $s$-wave correlator
\eqref{homog}.

\begin{figure}[h]
\begin{center}
\includegraphics[width=3.5in]{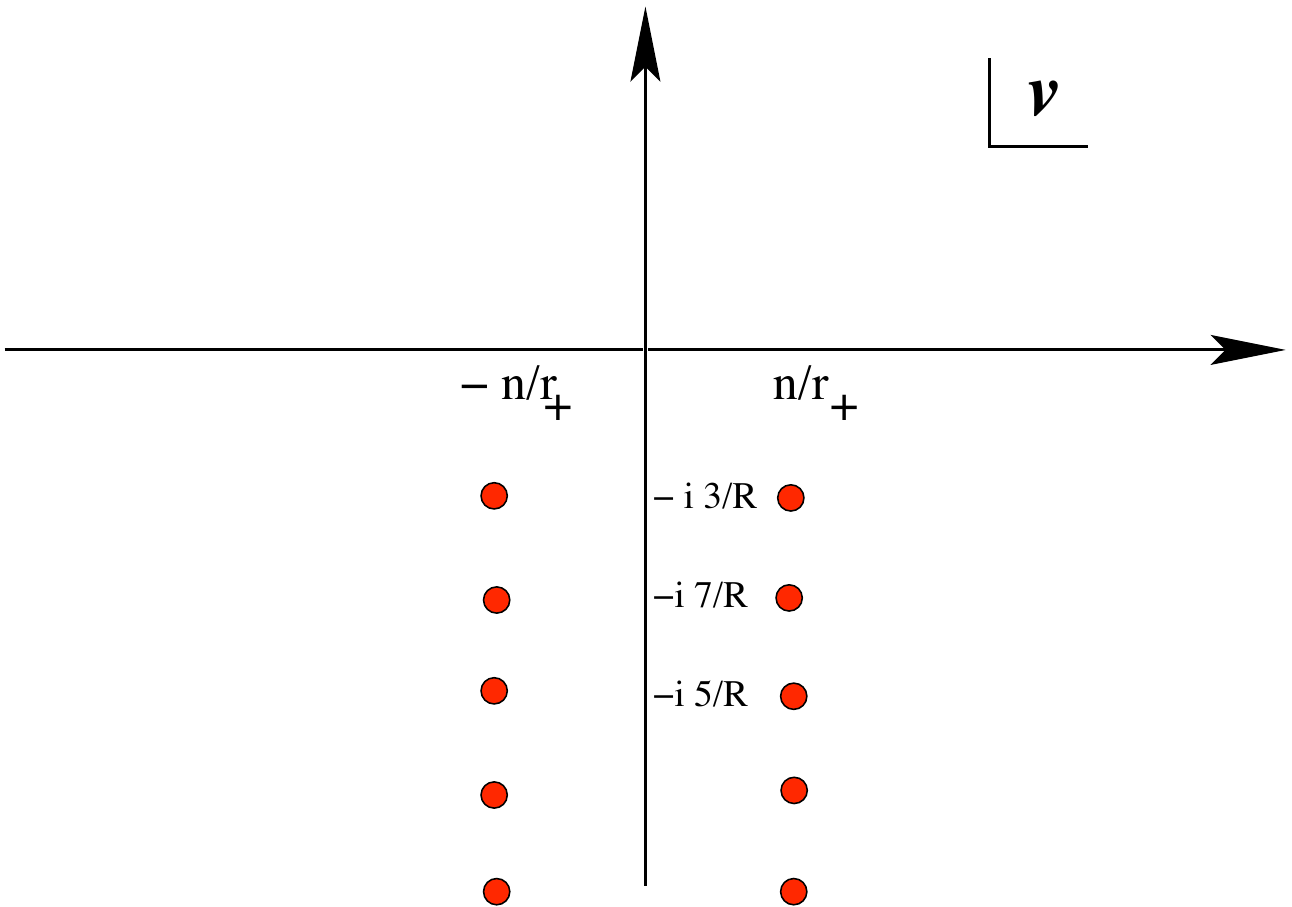} 
\end{center}
\caption{\footnotesize Simple poles in the
frequency plane, of the massless retarded Green's function with non-zero 
momentum along the spatial circle.}
\label{retdouble}
\end{figure}

The Green's function has nonanalyticities only in the lower half
plane, with simple poles at  
\EQ
{
\nu_k^\pm = - i(3+2k)R_{\rm dS}^{-1} \pm {n\over r_\x};\qquad k,n\in {\mathbb
  Z},
}
giving the quasinormal frequencies of the topological black hole, with
non-zero momentum along the spatial $S^1$.
Interestingly, each simple pole at $n=0$ ``splits'' into 
two simple poles at non-zero $n$ (see Fig. \ref{retdouble}).

We would like to also note that our expressions for the correlation functions on $dS_3\times S^1$ in
the ${\mathbb Z}_N$ symmetric phase satisfy the obvious
consistency check: in the high frequency/large momentum limit $\nu R, n
\gg 1$, they reproduce the flat space Green's function for the scalar glueball
operator
\EQ
{
\tilde G_R(\nu\,;n)\Big|_{\nu R,\,n\gg1}\longrightarrow -{N^2\over
  128\pi^2}(\nu^2-\tfrac{n^2}{r_\x^2} )^2\;\ln(\nu^2-\tfrac{n^2}{r_\x^2} )
.}
\subsection{Thermal Effects and the Gibbons-Hawking Temperature}
The de Sitter space has a cosmological horizon and an associated Gibbons-Hawking
temperature \cite{Gibbons:1977mu}. We therefore expect our boundary
($dS_3\times S^1$) field theory correlators to exhibit thermal
properties. For real frequencies $\nu$, the digamma functions have
an imaginary part so that
\SP
{
{\rm Im} \;\tilde G_R(\nu\,;n)\Big|_{{\rm Im}(\nu)=0}=&
\,{N^2\over 256\pi}\left((\nu
  -\tfrac{n}{r_\x})^2+R^{-2}\right) 
\left((\nu +\tfrac{n}{r_\x})^2+ R^{-2}\right)\times\\
&\left[\coth\left(\frac{\pi R}{2}(\nu+\tfrac{n}{r_\x}-iR^{-1})\right)
+\coth\left(\frac{\pi R}{2}(\nu+\tfrac{n}{r_\x}-iR^{-1})\right)
\right].
\label{imds}
}
Here we have used $\tanh(x)=\coth(x+i\tfrac{\pi}{2})$ to cast the
result in a form that will make the connection to thermal physics
explicit. 

In {\em flat space} and in free field theory at finite temperature
$T\neq 0$, the scalar glueball 
propagator with zero spatial momentum and frequency $\omega$
is proportional to the digamma
function \cite{Hartnoll:2005ju}
\EQ
{
\tilde G_R(\omega)\big|_{\rm Flat \;space}=-{N^2\over 2\pi^2
}\;\omega^4 \psi\left(- i 
\omega\over 4\pi T\right) + {\rm analytic}.
}
The imaginary part of the flat space glueball correlator is given by the
spectral function
\EQ
{
{\rm Im}\;\tilde G_R(\omega)\big|_{\rm Flat\;space}= -{N^2\over 2\pi^2}
\omega^4 \pi \coth\left( {\omega\over 4 T}\right).
\label{imflat}
}
In weakly coupled, or free field theories at finite temperature on
flat space, the one-loop spectral function reflects the physical effect of 
``Bose enhancement'', which follows from the stimulated emission of bosons into
the heat bath. In perturbative field theory on flat space, this
manifests itself as an enhancement of the decay rate
 of an unstable boson in a heat bath by a factor (relative to the vacuum
decay rate)
\EQ
{
\coth\left({\omega\over 4 T}\right)= 1+ 2n_B\left(\tfrac{\omega}{2}\right) = 
1+{2\over e^{\omega/2T}-1}.
}
Here, we have assumed that the unstable boson is at rest with energy $\omega$.

While there may not be an obvious way to define a spectral representation in de
Sitter space, the similarity between our strongly coupled de Sitter
space result \eqref{imds} and \eqref{imflat}
is obvious. In particular, it allows the identification of a temperature in 
de Sitter space
\EQ
{
T_H = {R_{\rm dS}^{-1}\over 2 \pi} 
}
which is precisely the value of the Gibbons-Hawking temperature. 

We note
that despite the similarity between the expressions for $dS_3$ and
thermal correlators in 
flat space, there is a crucial difference -- the frequency or
energy appearing in the Bose-Einstein-like 
distribution function in de Sitter
space, is not the real frequency $\nu$ 
\eqref{imds}, but in fact $\nu - i R_{\rm dS}^{-1}$. 
This difference can be traced back to
the definition of our positive and negative frequency modes
\eqref{modes}. For real $\nu$, the positive frequency modes are red-shifted
away in the far future. To get propagating modes in the future, we
would need to choose $\nu =\omega+ i R_{\rm dS}^{-1}$ with $\omega\in {\mathbb
  R}$.

It has been argued in \cite{Kovtun:2007py, Unsal:2008ch} that
correlation functions of large $N$ gauge theories in the 
${\mathbb Z}_N$ symmetric phase, with some or all spacetime
directions compactified, should be independent of the volume of the
compact directions. In the present situation, this would imply 
that on $dS_3\times S^1$ with antiperiodic boundary conditions for the
fermions, large $N$ correlators in the ${\mathbb Z}_N$
symmetric phase ($r_\x >
R_{\rm AdS}/2\sqrt 2  $) 
should be independent of the radius of the $S^1$.  In particular
then, for perturbations which are homogeneous along the circle, the
correlation functions should be independent of $r_\x$ and should match
up with the result on $dS_3\times {\mathbb R}$. The latter is obtained
by a (double) Wick rotation of $S^3\times {\mathbb R}$.
Since $\Tr F^2$ is a chiral primary in the ${\cal N}=4$ 
theory and its correlator on $S^3\times {\mathbb R}$ is not renormalized
by interactions, one would expect this to be true also on $dS_3\times
{\mathbb R}$. We would like to emphasize that this expectation is only for correlators evaluated on the vacuum obtained by analytic continuation. Therefore, we should not expect that the correlators in a general $\alpha$-vacuum to be renormalized. In particular, the transition amplitudes between two different vacua will not be renormalized: in free or weakly coupled theory, they are identical to zero, but our results from Chapter \ref{chapter:alpha} seem to suggest that they are not zero in the strong coupling regime.

\subsection{The Massive Case} \label{sec:massive}

The holographic calculation of correlation functions in the
topological AdS black hole can be easily extended to the case of massive
scalars. In the context of the Type IIB theory, such massive states are
stringy excitations with masses $M^2 \sim \alpha'^{-1}\gg R^{-2}_{\rm
  AdS}$. 
A scalar field of mass $M$ in the bulk is dual to a scalar operator
${\cal O}_\Delta$ in the field theory with conformal dimension
$\Delta=2+\sqrt{4+(MR_{\rm AdS})^2}$. We will study below the free massive
scalar in the bulk geometry to extract information on the analytic
structure of correlators of high dimension operators in the field
theory. 

There are two primary motivations for looking at high dimension operators
in the field theory: 
\begin{itemize}
\item[] i) The works of
\cite{Fidkowski:2003nf,Festuccia:2005pi} have demonstrated that
propagators of heavy fields, in the geodesic approximation, may be
used to probe the bulk geometry behind horizons and perhaps 
extract information on singularities behind such horizons. 
\item[] ii) One of our main goals is to look for signatures of the
transition between a ${\mathbb Z}_N$ symmetric phase and a ${\mathbb
  Z}_N$ broken phase. In the bulk theory, the latter phase is the
small bubble of nothing geometry. Correlators in this latter geometry
can only be computed using an eikonal (WKB) approximation involving high
frequencies and/or large masses. 
\end{itemize}

Extending the holographic analysis done above for massless fields, to
massive scalars in the topological black hole geometry,
we find that the  frequency space correlator is
\EQ
{
\tilde G_R(\nu)=  {\cal C}_\Delta\;
{\Gamma\left(\tfrac{1}{2}\left(\Delta-1-i\nu R_{\rm dS} \right)\right)^2 
\Gamma\left(3-\Delta\right)\over
\Gamma\left(\tfrac{1}{2}\left(3-\Delta-i\nu R_{\rm dS}\right)\right)^2
\Gamma(\Delta-1)},
\label{massive}
}
where the normalization constant ${\cal C}_\Delta = 2(\Delta -2)
\epsilon^{2(\Delta-4)}$, with $\epsilon \rightarrow 0$ as the boundary
is approached.
\begin{figure}[h]
\begin{center}
\includegraphics[width=3.5in]{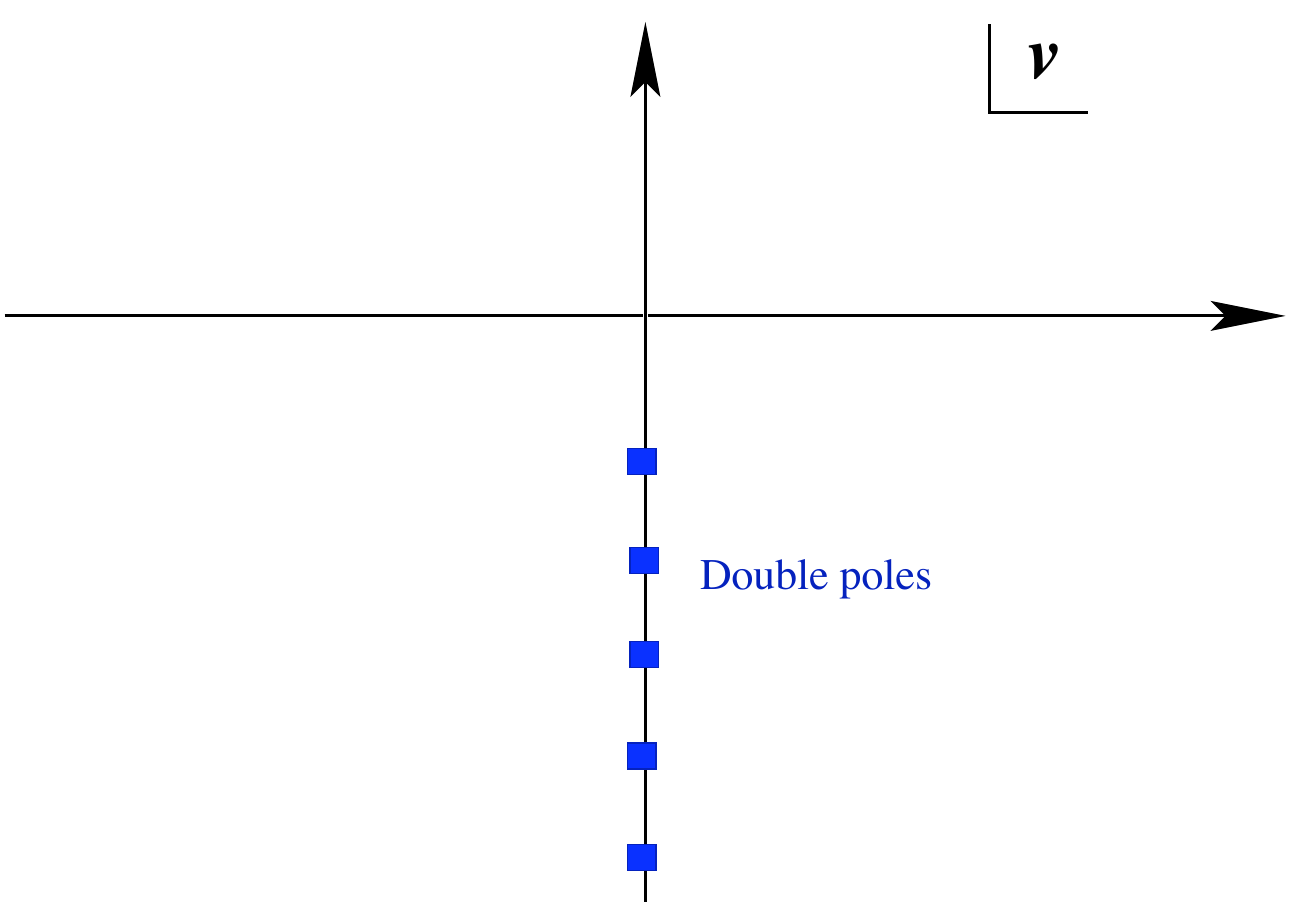} 
\end{center}
\caption{\footnotesize Double poles in the lower half plane
for the massive (retarded) propagator at zero spatial momentum.}
\label{masstbh}
\end{figure}

In the massless limit $MR_{\rm AdS}\rightarrow 0$, after subtracting an additional divergent contact
term, this reproduces the
expression \eqref{homog} found previously. The massive correlator has an analytic structure that is
qualitatively distinct from the massless case. In particular, the
retarded correlator has an infinite set of {\em double poles} and
{\em simple poles} at
\EQ
{
\nu_k = -i (\Delta-1+2k){1\over R_{\rm dS}}\,,\qquad k\in {\mathbb Z}.
}
See Fig. \ref{masstbh}.

The significance of the appearance of double poles in the massive
retarded propagator is not entirely clear. Such double poles have also 
appeared in 2d CFT correlators with non-integer conformal dimensions from
the BTZ black hole \cite{Son:2002sd}, at zero spatial momentum. At
finite spatial momentum, \textit{i.e.}, $n\neq 0$, we expect the double
poles to split into simple poles. 

For real frequencies, the massive correlator also has an imaginary
part
\EQ
{
{\rm Im}\;\tilde G_R (\nu)= - \tfrac{\pi^2}{2} {\cal C}_\Delta 
{\Gamma(3-\Delta)\over \Gamma(\Delta-1)}\;
{\sin\left(\pi\,\Delta\right)\;\sinh\left(\pi \;\nu R_{\rm dS} \right)\over
|\Gamma\left(\tfrac{1}{2}(3-\Delta-i\nu R_{\rm dS})\right)
\cos\left(\tfrac{\pi}{2}(\Delta-i\nu R_{\rm dS})\right)|^4}.
\label{immassive}
}
The de Sitter thermal origin of this result is not as explicit as
for the massless scalar. However, after identifying the de Sitter
Hawking temperature to be $T_H=R_{\rm dS}^{-1}/2\pi$, it is worth comparing
the above expression with the imaginary part of the propagator in two
dimensions for
large non-integer conformal dimension deduced from  
the non-extremal BTZ black hole \cite{Son:2002sd}. The similarities
between the two results, particularly the numerator of 
\eqref{immassive}, are striking.

\begin{figure}[h]
\begin{center}
\includegraphics[width=3.5in]{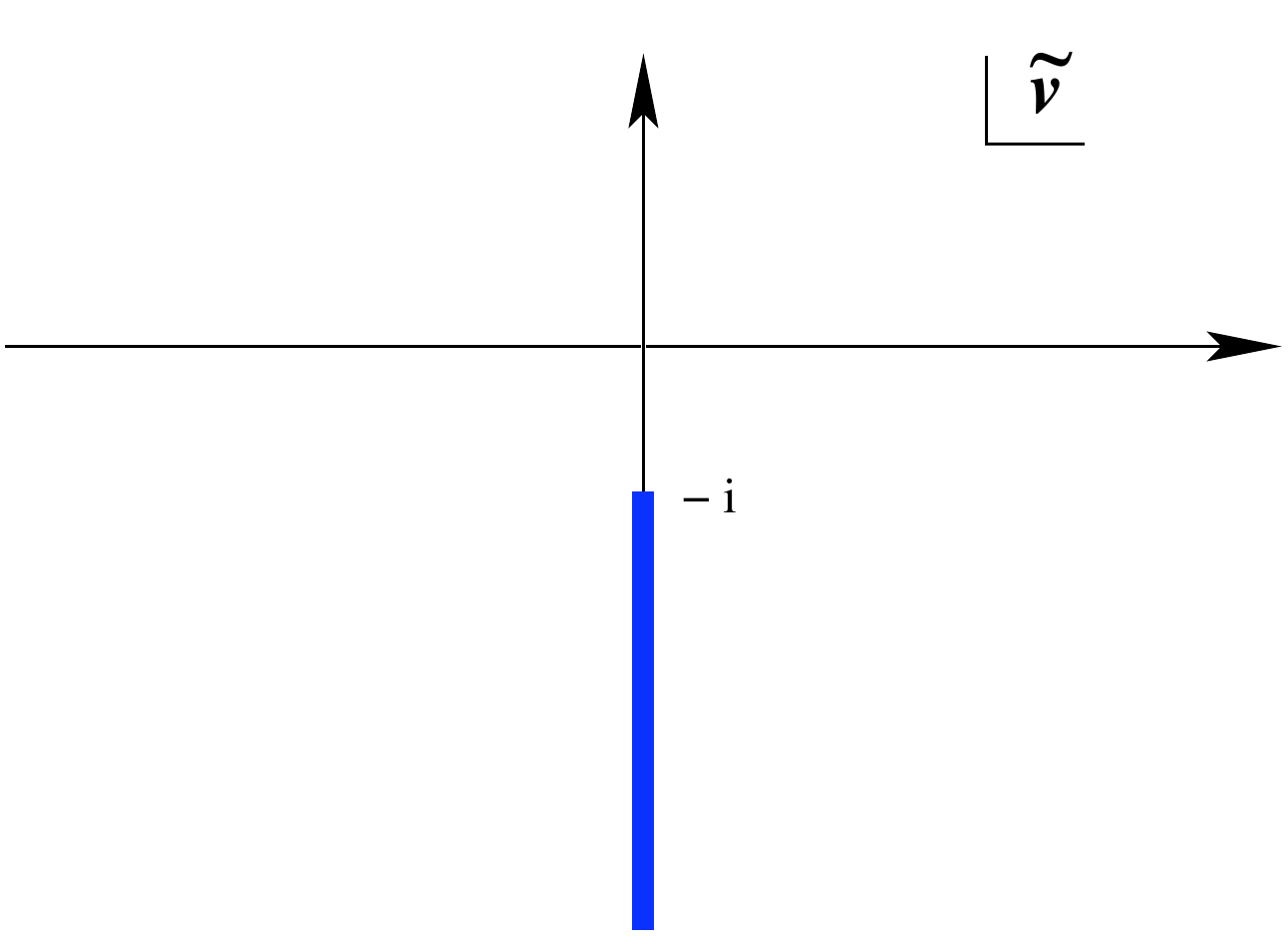} 
\end{center}
\caption{\footnotesize For large mass and frequency, the
propagator in the topological black hole phase, has a branch cut as
indicated in the rescaled frequency plane, $\tilde\nu=\nu/M$. This
results from the apparent merging 
of the infinite set of isolated poles of the exact
Green's function.} 
\label{largetbh}
\end{figure}

The large mass, high
frequency limit of this result is easily obtained, using Stirling's
approximation  
\EQ
{
\Gamma(z)\big|_{z\gg 1}\simeq \sqrt{2\pi} {1\over \sqrt z}\;e^{-z} \; z^z .
 }
When the masses are taken to be large so that $MR_{\rm AdS} \gg 1$, then
$\Delta \approx M R_{\rm AdS}$. In this high
frequency and large mass limit, it is useful to define a rescaled
frequency variable 
\EQ
{
\tilde\nu \equiv {\nu \over M},\qquad \nu R_{\rm dS},\; MR_{\rm AdS}\gg 1,
}
so that
\EQ
{
G_R(\tilde\nu)\sim {\cal C}_\Delta \left({1- i
  \tilde\nu\over 2}\right)^{MR_{\rm dS} (1-i \tilde\nu)}\;\left({-1-i\tilde\nu\over 2}\right)^{MR_{\rm dS}(1+i\tilde\nu)}.
\label{largemass}
}
Here, we have ignored an overall (real) phase due to the frequency
independent coefficients in the large mass limit. At first sight, a
potentially problematic feature of this 
approximation appears. It seems that there is a branch point singularity at $\tilde\nu=+i$ which
would imply a non-analyticity in the upper half plane, inconsistent
with the definition of a retarded propagator. We note that this 
feature is purely a result of the high frequency approximation
and the exact result \eqref{massive} has no singularities in the upper
half of the complex frequency plane. 
Indeed, closer
inspection reveals that the putative branch cut originating 
at $\tilde\nu = + i$ has a vanishing
discontinuity in the limit of large $MR_{\rm AdS}$. The spurious branch cut
originates from the equally spaced zeroes of $G_R(\nu
)$ appearing
to coalesce in the high frequency limit. The branch cut
discontinuity at $\tilde\nu=-i$ is, however, a genuine non-analyticity and
originates from the infinite set of poles merging into a continuum in
the high frequency approximation. See Fig. \ref{largetbh}

The easiest way to understand the singularities and discontinuities of
the function above is by examining the function $z^ {2 MR_{\rm dS} z}
\;(z-1)^{-2 MR_{\rm dS}(z-1)}$ and then make the replacement 
$z \rightarrow (1-i\tilde\nu)/2$. 

Along the branch cut ${\cal B} \equiv \left\{\tilde\nu \in [-i,
  -i\infty)\right\}$, we find that the retarded propagator has a
discontinuity
\SP
{
{\rm Disc}|_{\cal B} \;\tilde G_R(M\tilde\nu)= &2 i \;{\cal C}_\Delta
\;\sin(2\pi MR_{\rm dS})
\left({|\tilde\nu|+1\over 2}\right)^{MR_{\rm dS}(|\tilde\nu|+1)}
\left({|\tilde\nu|-1\over 2}\right)^{-MR_{\rm dS}(|\tilde\nu|-1)}\,,
\\
& \tilde\nu = -i |\tilde\nu|\,, \quad |\tilde\nu|\geq 1.
}
It is clear that the discontinuity is large in the large mass limit.

Now, let us look closely at the putative branch cut along the imaginary
axis ${\cal B}' \equiv \left\{\tilde\nu \in [-i, i]\right\}$. Computing the
discontinuity across this, we have
\SP
{
{\rm Disc}|_{{\cal B}'} \;\tilde G_R(M\tilde\nu)&= \,2 i\; {\cal C}_\Delta
\sin\left(MR_{\rm dS}\pi(1- x)\right) \left({1+x\over 2}\right)^{MR_{\rm dS}(1+x)}
\left({1-x\over 2}\right)^{MR_{\rm dS}(1-x)},\\
&\tilde\nu = i x,\,\quad -1\leq x\leq 1.
}
This vanishes when $MR_{\rm dS} \rightarrow \infty$, for two reasons: 
the rapid sinusoidal oscillations will give vanishing contribution
to any contour integral along ${\cal B}'$ and furthermore, the amplitude
of the oscillations vanishes exponentially as evident from the
expression above.

In the leading high frequency approximation , for real frequencies, the
imaginary and real parts of the Green's function are given by
\SP
{
\tilde G_R(\nu)
&\approx {\cal C}_\Delta\;\left({\nu^2+M^2\over 4M^2}\right)^{MR_{\rm dS}}
e^{-\tfrac{1}{\pi} \tfrac{\nu}{T_H}\tan^{-1}\tfrac{\nu}{M}}\; 
e^{-\tfrac{|\nu|}{2T_H}}\left(\cos(\tfrac{M}{2T_H})- i\; 
{\rm sgn}(\nu)\sin(\tfrac{M}{2T_H})\right),\\
&\qquad T_H={R_{\rm dS}^{-1}\over 2\pi}\,,\quad\nu\in{\mathbb R}.
}
The high frequency, large mass limit thus appears to retain features
of the thermal effects of de Sitter space.  
This result can be deduced from \eqref{largemass} after choosing an
appropriate branch of the function and also directly follows from 
\eqref{immassive}. We will see subsequently that these high frequency
expressions can also be derived by solving the wave equations using 
a WKB approximation, providing a consistency check.

\section{WKB for the AdS Bubble of Nothing}

An exact holographic computation of correlation functions in the AdS bubble of
nothing background appears difficult as analytical solutions to the
wave equation in this background are not known. Despite this, we may
obtain the boundary Green's function following a systematic
approximation. In particular, we will employ the WKB approximation to
determine boundary correlation functions at high frequency and large mass.
The WKB approximation has been used successfully to find high frequency Green's functions in the Big AdS-Schwarzschild
black hole \cite{Festuccia:2005pi}. In this approximation, lines of isolated singularities (poles)
get replaced by branch cuts since in the high frequency regime, the
separation between poles effectively goes to zero, as seen in our
example above in Section \ref{sec:massive}. 

We first take the massive scalar wave equation 
\EQ
{
{1\over \sqrt{-g}}\;\partial_\mu (g^{\mu\nu}\sqrt{- g}\;\partial_\nu \Phi)-
M^2 \Phi=0, 
}
in the bubble background 
and expand the scalar field in harmonics on
$dS_3\times S^1$ as in \eqref{decomp}. The harmonics $\Phi_n(\nu,
\rho )$ then satisfy a radial differential equation which can be put
in the form
\SP
{
{1\over
4(x-1)(x+\tilde r_h^2)(x-\tilde r_h^2-1)}\left({\nu^2+1}
-{n^2}{R^2_{\rm AdS}\over r_\x^2}{(x-1)^2
\over(x+\tilde r_h^2)(x-\tilde
r_h^2-1)} 
-{M^2R_{\rm AdS}^2} (x-1)\right)\Phi_n &\\\\
+ {d^2\Phi_n\over dx^2}+\left({1\over x-1}+{1\over x+\tilde r_h^2}+
{1\over x-\tilde r_h^2-1}\right){d\Phi_n\over dx} &=0,}
where $x=\rho^2$ and $n$ labels the momentum along the $S^1$. 

This is an ordinary differential
equation with four regular or nonessential 
singular points at $x=1, \,-\tilde r_h^2,\,
\tilde r_h^2+1$ and $\infty$. Analytical solutions for this type of
equation are unknown. In fact, a similar differential equation was
encountered in the computation of glueball masses at strong coupling 
in the three dimensional effective theory obtained from
thermal ${\cal N}=4$ SYM \cite{Csaki:1998qr} on ${\mathbb R^3}\times
S^1$ with SUSY-breaking boundary conditions. In that case, the 
dual bulk geometry is the Euclidean black brane solution in
AdS space where the thermal circle shrinks to zero size smoothly. 

It will be convenient to rewrite the differential equation in the Schr\"odinger form by introducing the variables
\EQ
{\Phi={\Psi\over \sqrt{\rho(\rho^2-1)}},}
\EQ
{u\;=\;
{\tilde r_h\over 1+2\tilde r_h^2}\; 
\cot^{-1}\left({\rho\over \tilde
    r_h}\right)+{\sqrt{1+\tilde r_h^2}\over{1+2\tilde r_h^2}}\;\coth^{-1}
\left({\rho\over\sqrt {1+\tilde r_h^2}}\right).
\label{newvar}
}
These are the natural generalizations of \eqref{rg1} and \eqref{rg2}
to the bubble of nothing geometry. The cigar in the geometry gets
smoothly capped off at $\rho=\sqrt{\tilde r_h^2+1}$, where the
spacetime ends. In terms of
the  $u$ coordinate, this occurs as  $u\rightarrow\infty$. 
In terms of the harmonics of $\Psi$ 
on the $dS^3\times S^1$ slices, as in \eqref{decomp}, we obtain the
Schr\"odinger equation 
obeyed by the harmonics $\Psi_n(\nu,u)$ in the small AdS
bubble of nothing. It is given by
\EQ
{-{d^2\over du^2}\;\Psi_{n}(\nu,u) 
+ \tilde V_{n}(\nu, u)\;\Psi_n(\nu, u)= 0, 
\label{schr2}}
with the potential
\begin{eqnarray}
\tilde V_{n}(\nu, u) &=& 
\left( (MR_{\rm AdS})^2- {\nu^2+1\over \rho^2-1}\right)
\left(\rho^2-1-{1\over \rho^2}\;\tilde r_h^2\;(\tilde r_h^2+1)\right) \nonumber \\
& & +\,n^2\, {R^2_{\rm AdS}\over r_\x^2}\;{\rho^2-1\over
  \rho^2}  + {1\over 
  4\rho^2}\;
(15\rho^4-10\rho^2-1).\label{schr3}
\end{eqnarray}
Here, $\rho$ is implicitly a function of $u$, determined by the solution to 
\eqref{newvar}.
In addition to the fact that the Schr\"odinger potential is far more
complicated than \eqref{schrodinger1}, one crucial difference to the case of the topological black hole is that the potential cannot be defined independent of the
frequency itself. This is due to the term proportional to $\tilde r_h$
in Eq. \eqref{schr3}. This situation is also in contrast to the
case of the big AdS-Schwarzschild black hole in
\cite{Festuccia:2005pi}. Here, for a
given frequency and mass, we need to find the ``zero energy'' eigenstate
of the Schr\"odinger problem  \eqref{schr2}.

\begin{figure}[h]
\begin{center}
\includegraphics[width=3.5in]{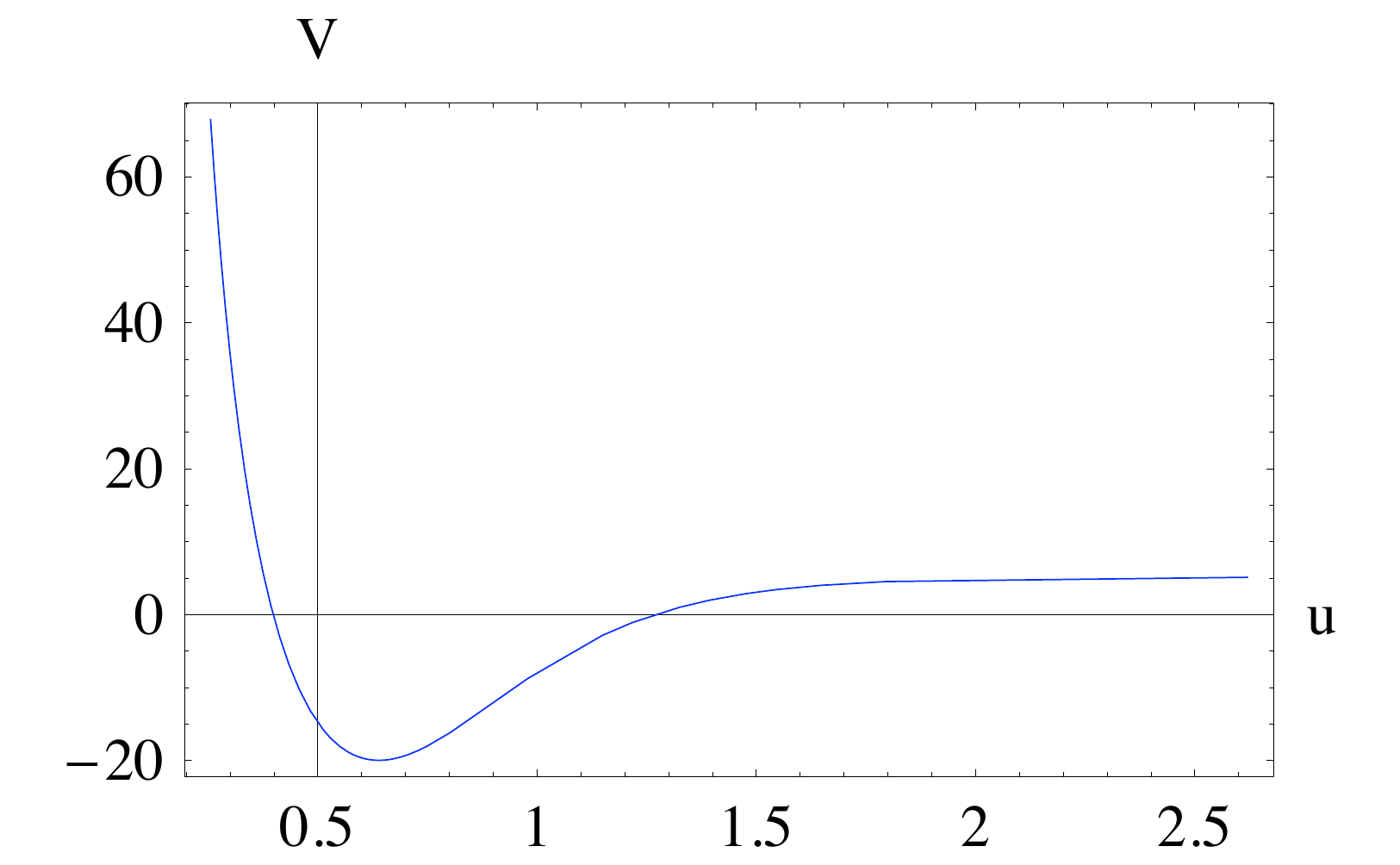} 
\end{center}
\caption{\footnotesize The Schr\"odinger potential for a
massive scalar in the AdS bubble of nothing. In the above plot the
dimensionless frequency $\nu=7$, the mass $MR_{\rm AdS}= 2$ and $\tilde r_h = 
r_h/R_{\rm AdS} =1$.}
\label{schrpot1} 
\end{figure}

\begin{figure}[h]
\begin{center}
\includegraphics[width=3.5in]{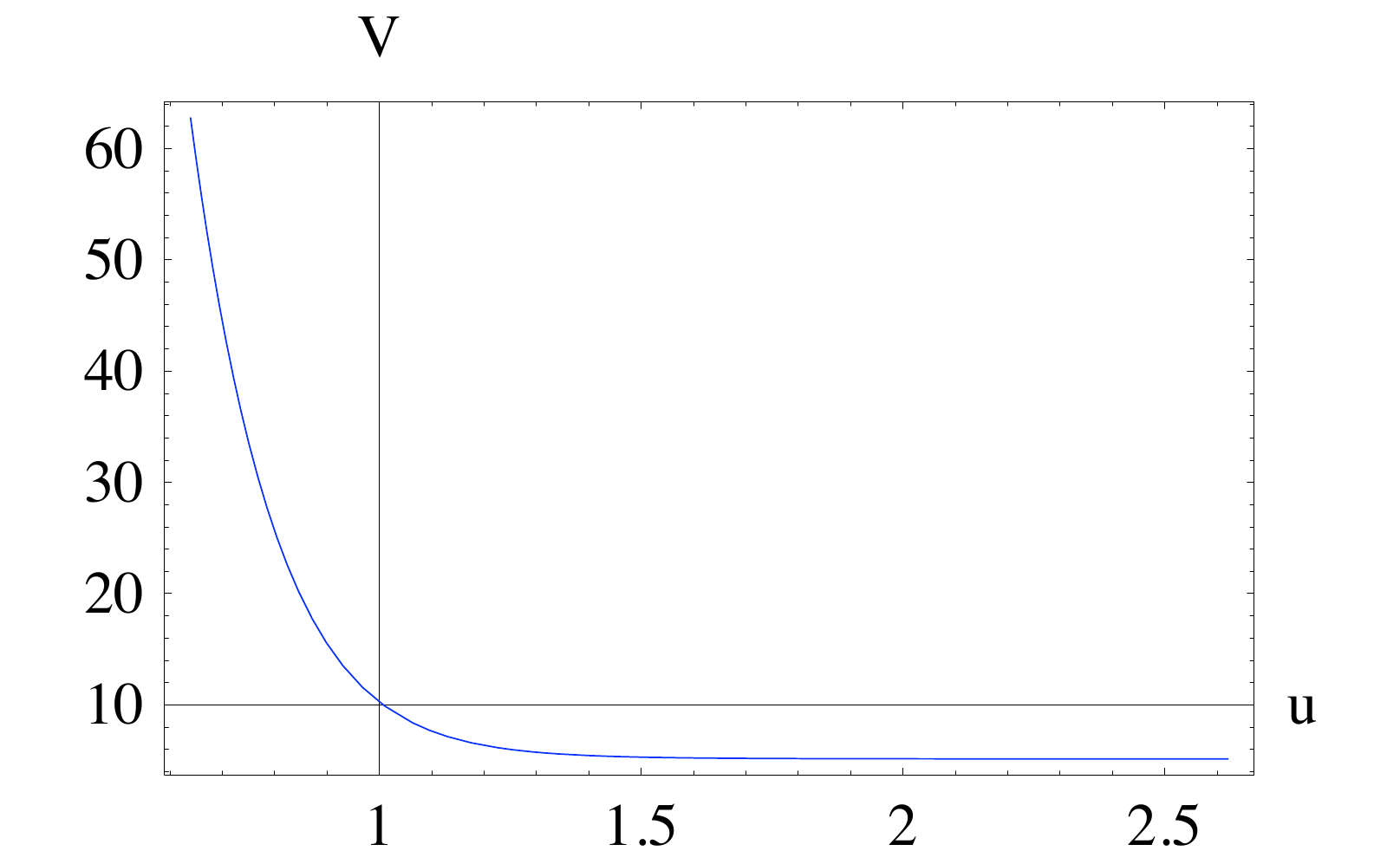} 
\end{center}
\caption{\footnotesize The Schr\"odinger potential for 
$\nu=7$, the mass $MR= 7$ and $\tilde r_h = 
r_h/R_{\rm AdS} =1$. }
\label{schrpot2}
\end{figure}

Interestingly, 
the qualitative behaviour of the potential $\tilde V_n$ changes,
depending on the relative values of the mass and the frequency. This
is illustrated in Figures (\ref{schrpot1}) and (\ref{schrpot2}). 
The qualitative nature of the potential is easy to grasp in the high
frequency limit, wherein we take 
\EQ
{
\nu\rightarrow\infty \qquad {\rm and} \qquad MR_{\rm AdS} \rightarrow\infty,
}
while keeping
\EQ
{\tilde \nu={\nu\over MR_{\rm AdS} }}
fixed. For simplicity, we also set $n=0$, focussing attention on the
spatially homogeneous fields.
In this approximation, the potential $\tilde V_0(\nu, u)$ becomes
\EQ
{
{\cal V}(\tilde\nu,u)=(MR_{\rm AdS} )^2
\left(1-{\tilde\nu^2\over \rho^2-1}\right) 
\left(\rho^2-1 - {1\over \rho^2 }\;\tilde r_h^2(\tilde r_h^2+1)\right).
\label{wkbpot}
}
Having thrown away a subleading constant
in the large frequency limit, the potential is vanishing  at the tip of the cigar $\rho =
\sqrt{\tilde r_h^2+1}$. Using $u\sim -\ln(\rho-\sqrt{\tilde
  r_h^2+1})$ near this point, it follows that  
${\cal V}(\tilde \nu,u)$ decays
exponentially as a function of $u$. 

The potential has another zero  
at $\rho=\sqrt{{\tilde\nu}^2+1}$. Now, since the spacetime ends at
$\rho=\sqrt{1+\tilde r_h^2}$, the number of zeroes of the potential 
depends on whether
$|\tilde\nu|$ is greater than or less than $\tilde r_h$. In particular, when
$|\tilde\nu| > \tilde r_h$, the potential energy has two zeroes or turning points, 
and qualitatively
resembles Fig. \ref{schrpot1}, while for $|\tilde\nu|< \tilde r_h$, it behaves 
as in Fig. (\ref{schrpot2}). In the latter case, the potential asymptotes to a constant
different from zero in the figures, however, this constant becomes
negligible in the high frequency limit.

Zero energy Schr\"odinger wave functions in the potential
\eqref{wkbpot} are marginally bound states for 
$|\tilde\nu| > \tilde r_h$, since the
potential has two turning points. For $|\tilde\nu| <\tilde r_h$, there is only
one turning point and the wave function should exhibit a qualitative
change in its behaviour at $|\tilde\nu|=\tilde r_h$. This also
strongly suggests that 
in the complex frequency plane, $\tilde\nu=\pm \tilde r_h$ should be
singularities of 
boundary correlation functions.

The zero energy wave function for the Schr\"odinger equation can be
obtained in the WKB approximation, where care needs to be taken in
applying the matching conditions at each of the turning points in the
potential. We treat the two cases $|\tilde\nu|> \tilde r_h$ and $|\tilde\nu|<
\tilde r_h$ separately.

\subsection{WKB Approximation for $\nu > \tilde r_h$}

There are two distinct regions in the potential: 
\begin{itemize}
\item the ``quantum
tunnelling region'',  which corresponds to the regime
\begin{equation}
\sqrt{1+\tilde\nu^2} < \rho <\infty,
\end{equation}
 which we label  as Region I; 
 \item the ``propagating region'', which corresponds to the regime
 \begin{equation}
 \sqrt{1+\tilde
  r_h^2} < \rho < 
\sqrt{1+\tilde\nu^2},
\end{equation}
which we will call Region II.
\end{itemize}

In Region I, we write the WKB solutions as
\EQ
{
\Psi_{\rm WKB}(\tilde\nu, u) = 
\;{1\over {\cal V}^{1/4}}\;\left(A_+
\exp\left({\int_{u_c}^u \sqrt {\cal V}
  \;du}\right) + A_-\;
\exp\left({-\int_{u_c}^u \sqrt {\cal V}
  \;du}\right)\right),
\label{wkb1}
}
where the classical turning point $u_c$ is defined by
\EQ
{
\rho(u_c) \equiv \sqrt{\tilde\nu^2+1}.
}
These represent the growing and decaying modes in the near boundary
region of the bulk geometry. This can be understood easily as
follows. Since
\EQ
{
{du\over d\rho}\bigg|_{\rho\to \infty} \approx - {1\over \rho^2}
}
in this regime, we have
\EQ
{{\cal V} \approx (MR_{\rm AdS})^2\;\rho^2,}
which then immediately yields the near boundary WKB solution \eqref{wkb1}
for $\Psi$. 
This, together with its relation \eqref{newvar}
to the massive bulk scalar field $\Phi$, implies that
\EQ
{
\Phi_{\rm WKB}\big|_{\rm \rho\rightarrow\infty}\sim 
A_+\;(\cdots)\rho^{-2-MR_{\rm AdS}}+A_- \;(\cdots)\;\rho^{-2+MR_{\rm AdS}},
\label{asympwkb}
}
where the ellipses denote unspecified normalization constants.
The two power laws appearing in this solution are
precisely the 
normalizable and non-normalizable modes of the massive scalar
field, in the limit of large mass, 
in an asymptotically locally AdS spacetime. Following the
prescription for computing the retarded Green's functions, we
normalize $\Psi_{\rm WKB}$ \eqref{asympwkb} so that it approaches unity
near the boundary.

In the interior, however, for $\rho\leq
\sqrt{1+\tilde\nu^2}$ the solutions enter Region
II and become oscillatory. In Region II, we have
\EQ
{
\Psi_{\rm WKB}(\tilde\nu, u)=\;{1\over |{\cal V}|^{1/4}}\left(B_+\;
\exp\left(i{\int_{u_c}^u \sqrt {|{\cal V}|}
  \;d\rho}\right) + B_-\;
\exp\left({-i\int_{u_c}^u \sqrt {|{\cal V}|}
  \;du}\right)\right).
\label{wkb2}
}
The constants are uniquely determined by the WKB matching conditions at
the classical turning points of the potential ${\cal V}(\tilde\nu, u)$ and the
one normalization condition on $A_-$ near the boundary. 

Near the turning point $u=u_c$ corresponding to $\rho = \sqrt{1+
  \nu^2}$, since we are well away from any extrema, we can assume
  that
\EQ
{
{\cal V}(\nu, u)=\kappa (u_c-u)+\ldots\,, \qquad {u \to u_c}.
}
In this region, where the potential is basically linear, the exact
  solution in terms of Airy functions is given by
\EQ
{
\Psi\big|_{u\to u_c}=
A_+\; {2\sqrt\pi\over \kappa^{1/6}} \;{\rm Ai}\left(\kappa^{1/3}(u_c-u)\right)
+A_-\; {\sqrt\pi \over \kappa^{1/6}}\;{\rm Bi}\left(\kappa^{1/3}(u_c-u)\right)
.
\label{airy}}
The normalizations and constants of integration have been chosen
  carefully so that near the
  turning point, the exact solution, which is given in terms of the Airy functions, matches the WKB solution
  \eqref{wkb1} 
  in Region I ($u< u_c$) away from the turning point. 
  
  Now, we can
  continue the solution \eqref{airy} into Region II ($u> u_c$). For
  $u> u_c$, where the WKB solution \eqref{wkb2} should be valid, the
  Airy functions have the asymptotic form
\EQ
{
\Psi\big|_{u > u_c}\approx 
2 A_+\;{\sin\left(\tfrac{2}{3}\sqrt{\kappa}(u-u_c)^{3/2}+\tfrac{\pi}{4}\right)
\over \kappa^{1/4}(u-u_c)^{1/4}}+A_-\;
{\cos\left(\tfrac{2}{3}\sqrt{\kappa}(u-u_c)^{3/2}+\tfrac{\pi}{4}\right)
\over \kappa^{1/4}(u-u_c)^{1/4}}.
}
Comparison with \eqref{wkb2} near the turning point then implies
\EQ
{
B_+ = e^{i\tfrac{\pi}{4}}\left(\tfrac{1}{2}A_- - i A_+\right)\,, \qquad {\rm and} \qquad
  B_-= e^{-i\tfrac{\pi}{4}}\left(\tfrac{1}{2}A_-+i A_+\right).
\label{cond1}}

There is yet another condition that emerges from the behaviour of the
  solutions near the second ``turning point'', $\rho\rightarrow\sqrt{1+\tilde
  r_h^2}$ or $u\rightarrow \infty$  
where the space ends. In this region, we have
\EQ
{
{du\over d\rho}\big|_{\rho\to\sqrt{1+\tilde r_h^2}} \approx -
  {\sqrt{1+\tilde r_h^2}\over 2(2\tilde r_h^2+1)(\rho - \sqrt{1+\tilde r_h^2})}
}
so that
\EQ
{
\rho-\sqrt{1+\tilde r_h^2}\;\approx \;2\sqrt{1+\tilde r_h^2}
\exp\left(- 2 \tfrac{(1+ 2\tilde
  r_h^2)}{\sqrt{1+\tilde r_h^2}}\;u +\tfrac{\tilde r_h}{\sqrt{1+\tilde r_h^2}} \cot^{-1}
\left(\tfrac{\sqrt{1+\tilde r_h^2}}{\tilde r_h}\right)\right).
}
It follows then that, as a function of $u$, the high
frequency potential decays exponentially,
\EQ
{
{\cal V}(\nu, u)\big|_{u\rightarrow\infty}\approx (MR_{\rm AdS})^2
  \left(1-{\nu^2\over \tilde r_h^2}\right) 
\;\exp\left(- 2 \tfrac{(1+ 2\tilde
  r_h^2)}{\sqrt{1+\tilde r_h^2}}\; u+{\rm constants}\right).
}
We note that for $\nu>\tilde r_h$, this potential approaches zero from
below. Let us define constants $A$ and $B$, in terms of which the potential
  is simply
\EQ
{
\tilde V_0(\nu, u)\big|_{u\rightarrow\infty}\approx - A e^{-B u},
}
where $A$ and $B$ can be read off easily from the expressions
  above. 
  
The Schr\"odinger equation with an exponentially decaying
  potential
\begin{equation}
-\Psi''(u)-A e^{-B u}\Psi(u)=0
\end{equation}
is solved exactly by Bessel functions
\EQ
{
\Psi= C_1 \;J_0\left(
2\tfrac{\sqrt{A}}{B} \,e^{-Bu/2}\right)+C_2\; Y_0\left(
2\tfrac{\sqrt{A}}{B} \,e^{-Bu/2}\right).
}
Since we are looking for a zero energy eigenfunction of the
  Schr\"odinger problem, this means that for a potential that
 vanishes at infinity, the corresponding (normalizable)
wavefunction can only be zero. This is an important difference to the
  black hole case where the wave functions are infalling plane waves
  at the horizon. In the bubble geometry, however, since spacetime
ends smoothly in the interior where the cigar caps off, there is no
freedom in choosing the boundary condition at the tip of the cigar --
we must require
regularity (normalizability) of solutions in the interior. This means
that the 
solution to the Schr\"odinger equation \eqref{schr2} must approach a constant 
exponentially as $u\to \infty$.
  
Requiring that the wave function $\Psi$ vanish or approach a constant
  as $u\to \infty$ then eliminates the term proportional to $Y_0$. Hence,
  in the exponentially decaying region 
\EQ
{
\Psi(u)\propto J_0\left(
2\tfrac{\sqrt{A}}{B} \,e^{-Bu/2}\right).
}
The WKB approximation should then match onto the Bessel function for large
values of the argument of the Bessel function. Using the standard
asymptotic expansion for Bessel functions 
\EQ
{
J_0(x)\big|_{x\gg 1}\simeq {\cos(x)\over \sqrt{\pi x}}+
  {\sin(x)\over\sqrt{\pi x}},
\label{asymj}
}
we can deduce a relationship between the constants $B_+$ and
  $B_-$ in \eqref{wkb2}. To make this precise we define 
  the integral
\EQ
{{\cal S}_{II}(u) = \int^{u}_\infty \sqrt{|{\cal V}(\nu,u)|}\;du
=- MR_{\rm AdS}\int_{\tilde r_h}^{\rho^2-1} dx \sqrt{x^2-\nu^2\over{(x^2-\tilde r_h^2)
(x^2+1+\tilde r_h^2)}} ,
}
which in radial coordinate $\rho$ is given by
\begin{eqnarray}
{\cal S}_{II}(u(\rho))&=& i\;\frac{MR_{\rm AdS}}{\sqrt{1+2\tilde r_h^2}} 
\left[ \nu \left( F 
\left(\sin^{-1}\sqrt{\tfrac{1}{a}\tfrac{\rho^2+\tilde r_h^2}
          {\rho^2-1}}\;\bigg|\;k\right)
-\frac{1}{\sqrt k} \;K\left(\tfrac{1}{k}\right)\right)+\right. \nonumber
\\ \nonumber
\\
&&
\left.
\tfrac{1}{\nu}(1+\tilde r_h^2)
\left(\Pi\left(a;\,\,\sin^{-1} \sqrt{\tfrac{1}{a}
\tfrac{\rho^2+\tilde r_h^2}{\rho^2-1}} 
\,\bigg| \,k\right)-\frac{1}{\sqrt k}\;
\Pi\left(b\big|\tfrac{1}{k}\right)\right)\right].
\end{eqnarray}
Here, $F$ and $K$ are the elliptic integrals of the first kind, $\Pi$ is the elliptic integral of the third kind and
\EQ
{
 a= \frac{\nu^2+1+\tilde r_h^2}{\nu^2}\;,\qquad 
b=\frac{1+2\tilde r_h^2}{\tilde r_h^2}\;,\qquad k= \frac{a}{b}.
}

Then, the WKB solution in Region II is 
\EQ
{
\Psi_{\rm WKB}= {1\over |{\cal V}|^{1/4}}\left(B_+\;e^{i {\cal S}_{II}(u)-i
  {\cal S}_{II}(u_c)}+B_-\, e^{-i {\cal {S}}_{II}(u)+ i {\cal
    S}_{II}(u_c)}\right).
\label{wkblarge}
}
For large $u$, 
\EQ
{
{\cal S}_{II}(u)\big|_{u\gg 1} \approx \int_{\infty}^u \sqrt A e^{-B u/2}=
- 2{\sqrt A\over B}\; e^{-Bu/2}.
}
Using this result and comparing \eqref{wkblarge} to the asymptotics of
the Bessel function \eqref{asymj}, we find
\EQ
{
{B_+\over B_-}= \; i e^{2i\; {\cal S}_{II}(u_c)}.
\label{cond2}
} 
The final ingredient consists in determining $A_+$ and $A_-$. To this end,
we first define 
\EQ
{{\cal S}_I(u)=\int^u_{\infty}\sqrt{{\cal V}(\nu,u)}\;du,}
which then gives us
\EQ
{{\cal S}_{I}(u(\rho))=\frac{MR_{\rm AdS}}{\sqrt{1+2\tilde r_h^2}} \left[\nu 
F
\left(\sin^{-1}\left(\sqrt{\tfrac{1}{a}\tfrac{\rho^2+\tilde r_h^2}
          {\rho^2-1}}\right),\;k\right)+
\tfrac{1}{\nu}(1+\tilde r_h^2)\,\Pi\left(a;\sin^{-1}\left(\sqrt{
\tfrac{1}{a}\tfrac{\rho^2+\tilde r_h^2}
{\rho^2-1}}\right) \bigg | \,k\right)
\right],}
with
\EQ
{ a= \tfrac{\nu^2+1+\tilde r_h^2}{\nu^2}\;,\qquad 
b=\tfrac{1+2\tilde r_h^2}{\tilde r_h^2}\;,\qquad k= \tfrac{a}{b}.
}  
Near the boundary $u\to 0$ or equivalently $\rho\to \infty$, we find
\SP
{
\Psi_{\rm WKB}\approx &\,A_-\tfrac{1}{\sqrt {MR_{\rm AdS}}} \;\rho^{MR_{\rm AdS}-\tfrac{1}{2}} \;e^{i\pi
  MR_{\rm AdS}/2}\;\left({4\over 1+\tilde r_h^2}\right)^{MR_{\rm AdS}/4}\;\nu^{-MR_{\rm AdS}/2}\,\times\\
&\exp\left[\tfrac{MR_{\rm AdS}}{\sqrt{1+2\tilde r_h^2}}\;
\left(\nu (F
\left(\csc^{-1}\sqrt{a},\;k\right)- K(k))-\tfrac{1}{\nu }(1+\tilde
r_h^2)
\;\Pi(a|k)\right)\right]\\
&+\; A_+\,\tfrac{1}{\sqrt {MR_{\rm AdS}}} \;\rho^{-MR_{\rm AdS}-\tfrac{1}{2}} \;e^{-i\pi
  MR_{\rm AdS}/2}\;\left({4\over 1+\tilde r_h^2}\right)^{-MR_{\rm AdS}/4}\nu^{MR_{\rm AdS}/2}\,\times\\
&\exp\left[-\tfrac{MR_{\rm AdS}}{\sqrt{1+2\tilde r_h^2}}\;
\left(\nu (F
\left(\csc^{-1}\sqrt{a},\;k\right)- K(k))-{1\over \nu }(1+\tilde
r_h^2)
\;\Pi(a|k)\right)\right].
}
Combining \eqref{cond1} and \eqref{cond2}, we obtain
\EQ
{
A_+ = - A_-{1\over 2} \; \tan\left({\cal S}_{II}(u_c)\right).
}

\begin{figure}[h]
\begin{center}
\includegraphics[width=3.5in]{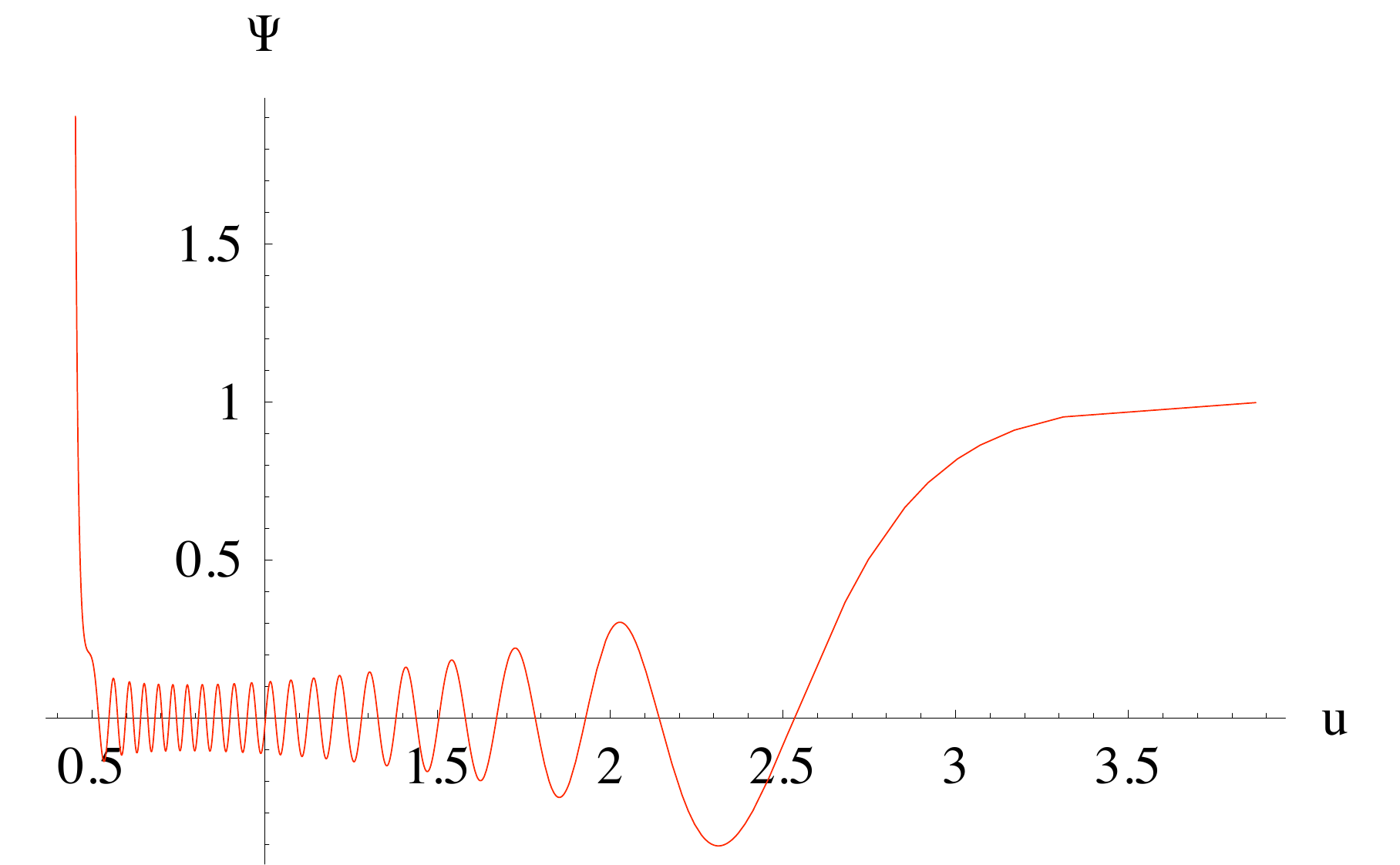} 
\end{center}
\caption{\footnotesize The exact numerical solution to the
Schr\"odinger problem for the AdS bubble of nothing. Here,
$\nu R_{\rm AdS}=200$, $MR_{\rm AdS}=100$ and $r_h/R_{\rm AdS}=1$. The solution approaches a
constant for $u\gg 1$, while the classically forbidden region is $0< u
< u_c \approx 0.49$.}
\label{wkbhi}
\end{figure}

The Green's function at high frequency is completely
determined (up to contact terms) by the ratio $A_+/A_-$. After substituting the
solution \eqref{asympwkb} into the boundary action we find that
\EQ
{
\tilde G_R(\tilde\nu) \approx \lim _{\epsilon\rightarrow 0}\;
{MR_{\rm AdS}} \;\exp\left(2 \int_{u_c}^\epsilon \sqrt{{\cal V}(\tilde\nu,
    u)}\;du\right)  
\;\tan\left(\int_{\infty}^{u_c}
\sqrt{|{\cal V}(\tilde\nu, u)|} 
\;du\right).
\label{gfhi}
}
The exponential prefactor in this expression is the 
WKB transmission coefficient into the classically forbidden Region I.
The physically relevant contribution to the transmission coefficient
is the constant, $\epsilon$-independent term in an expansion around
the boundary $\epsilon \to 0$. The leading $\epsilon$ dependence, which
is an overall multiplicative constant proportional to
$\epsilon^{2MR_{\rm AdS}}$, can be absorbed into the normalization of the
correlation function.

The first observation we can make without actually evaluating
\eqref{gfhi} is that it has an infinite set of poles as a function of
$\tilde\nu$ on the real axis for $\tilde\nu^2> r_h^2$. These occur whenever
\EQ
{{\cal S}_{II}(\tilde\nu, \;MR_{\rm AdS})= \int_{\infty}^{u_c} du
\sqrt{|{\cal V}(\tilde\nu, u)|}=(n+ \tfrac{1}{2}){\pi}\,,
\quad n\in {\mathbb Z}.
}
That is, the semiclassical action in the propagating region is a
half-integral multiple of $\pi$. This is of course the condition for
the existence of a bound state wave function and the poles in the
Green's function reflect the appearance of these bound states at
certain values of $\tilde\nu$ on the real axis. This is unlike the
corresponding correlators in the topological black hole phase, which do not
have any poles on the real axis.

At first sight, the poles on the real frequency axis might appear
somewhat surprising. However, they have the following natural physical
interpretation. The low energy physics of the gauge theory on the
boundary of the AdS bubble of nothing is that of supersymmetric
Yang-Mills theory 
at large $N$ and strong 't Hooft coupling on $dS_3$. The
antiperiodic boundary conditions on the spatial $S^1$ make all
fermionic excitations massive and the broken supersymmetry leads to
large radiative corrections to the scalar masses. In the strongly
coupled theory, the dynamical scale of the three dimensional effective
theory is expected to be set by $r_\x^{-1}$, the scale of the 
compact $S^1$ direction. When the 
Gibbons-Hawking temperature $T_H= 1/(2\pi
R)$ in $dS_3$ is smaller than $r_\x^{-1}$, we expect the gauge theory to be in a
confined phase where the degrees of freedom are gauge singlet
glueballs. The appearance of the isolated 
poles in the high frequency correlators is consistent with this
physical picture.

The WKB integral in Region II, 
in the high frequency approximation to the Green's function \eqref{gfhi},
can be expressed in terms
of complete elliptic integrals as
\SP
{{\cal S}_{II}(\tilde\nu,\; MR_{\rm AdS})=  &
\,\tfrac{1}{2}MR_{\rm AdS} \int_{\tilde\nu^2+1}^{\tilde
    r_h^2+1}dx\;
{\sqrt{\tfrac{\tilde\nu^2}{x-1}-1}\over \sqrt{(x+\tilde
    r_h^2)(x-\tilde r_h^2-1)}}\\\\ 
= &\,i\,\frac{MR_{\rm AdS}}{\sqrt{1+2\tilde r_h^2}}\,
\left[|\tilde\nu|\left(K(\tfrac{a}{b})-{\sqrt {\tfrac{b}{a}}}\;
K\left(\tfrac{b}{a}\right)\right)+
\tfrac{1}{|\tilde\nu|}(1+\tilde r_h^2)
\left(\Pi(a \big|\tfrac{a}{b})-{\sqrt {\tfrac{b}{a}}}
\;\Pi\left(b\big|{\tfrac{b}{a}}\right)\right)\right],}
where
\EQ
{
a= \tfrac{\tilde\nu^2+1+\tilde r_h^2}{\tilde\nu^2}\;,\qquad 
b=\tfrac{1+ 2\tilde r_h^2}{\tilde r_h^2}\;, \qquad \tilde\nu^2 > r_h^2. 
}  
From the general characteristics of these elliptic functions and their
singularities \cite{elliptic}, it  can be checked that ${\cal S}_{II}(\tilde\nu,
\;MR_{\rm AdS})$ has no singularities on the real axis for 
$\tilde\nu^2 > \tilde r_h^2$. Potential logarithmic branch points 
at $\tilde\nu^2=\tilde r_h^2$ and at $\tilde\nu=0$, cancel out
between the individual terms above. In fact, for any fixed value of
$\tilde r_h$, it  also follows that, for large $\tilde\nu$, the WKB integral
increases linearly with $\tilde\nu$
\EQ
{
{\cal S}_{II} \propto |\tilde\nu|\,, \qquad |\tilde\nu|\gg 1.
} 
Hence, for $|\tilde\nu|\gg \tilde r_h$, the propagator \eqref{gfhi} 
has approximately equally spaced simple poles on the real axis,
whenever ${\cal S}_{II}= (n+\tfrac{1}{2})\pi$. 

Although there are no other sources of singularities from $S_{II}$, 
the WKB transmission coefficient in Region I, which also enters the
Green's function \eqref{gfhi}, can have branch point singularities on
the real axis
\SP
{
{\cal S}_I(\tilde\nu,\;MR_{\rm AdS})&= - \tfrac{1}{2}MR_{\rm AdS}
\int_{1+\tilde\nu^2}^{1/\epsilon^{2}}  
 \;dx\frac{\sqrt{1-\tfrac{\tilde\nu^2}{x-1}}}
{\sqrt{(x+\tilde r_h^2)(x-\tilde r_h^2-1)}}
\\\\
&= - \frac{MR_{\rm AdS}}{\sqrt{1+2 \tilde r_h^2}}\;
\left(|\tilde\nu| \left(F\left(\csc^{-1}\sqrt{a},\;\tfrac{a}{b}\right)- 
K\left(\tfrac{a}{b}\right)\right)
-\tfrac{1}{|\tilde\nu| }(1+\tilde r_h^2)
\;\Pi\left(a|\tfrac{a}{b}\right)\right)\\
&\quad-\tfrac{1}{2}MR_{\rm AdS}\left(\ln\left(\tfrac{2\epsilon^{-2}}{|\tilde\nu|
\sqrt{1+\tilde r_h^2}}\right) + i\pi\right).
}
This function is also free of any branch cuts at $\tilde\nu=\pm \tilde r_h$,
as can be checked by directly evaluating the integral at this
point. However, the logarithmic growth at large $\tilde\nu$ implies a branch
point at infinity.  

As an aside, we would like to mention that the high frequency limit in the
topological black hole phase \eqref{largemass} can be rederived by
formally setting $\tilde r_h=0$ in the WKB integrals and 
$\tilde G_R(\tilde \nu) \sim \exp( {2\,\cal S}_I)$.

\subsection{WKB Approximation for $|\nu|< \tilde r_h$}

For low real frequencies $\tilde\nu^2 < \tilde r_h^2$, the nature of the 
WKB potential changes (see Fig. (\ref{schrpot2})). When $|\nu|< \tilde r_h$, the potential energy ${\cal V}(\nu, u)$ is a
monotonic function of $u$ which exponentially vanishes as
$u\rightarrow\infty$. Now, the potential has effectively only one
turning point and the wave function has no region where it
propagates. The WKB solution \eqref{wkb1} in Region I should smoothly
match onto the exact solution of 
\EQ{-\psi''(u)+ A e^{-Bu}\psi(u)=0\,, }
where $A, B>0$.
The solutions to these are the modified Bessel's functions. Enforcing regular
behaviour as $u\to\infty$ picks out
\EQ
{
\psi(u)\propto I_{0}\left(2\tfrac{\sqrt A}{B}e^{-Bu/2}\right).
} 
The WKB approximation for $I_0(x)$ is valid when $x\gg 1$, and here
\EQ
{
I_0(x)\big|_{x \gg1}\simeq {1\over 2\sqrt {2\pi x}}\left(e^x+ i
  e^{-x}\right). 
}
We write the WKB solution to the wave equation in the bubble of
nothing background as
\EQ
{
\Psi_{\rm WKB}(\nu,u)= A_+\, {1\over {\cal V}^{1/4}}\exp\left(
\int_{\infty}^u\sqrt{{\cal V}} \,du\right)+A_- \,{1\over {\cal V}^{1/4}}
\exp\left(
- \int_{\infty}^u\sqrt{{\cal V}} \,du\right).
}
Comparison with the modified Bessel function implies
\EQ
{
A_+=i A_-.
}
Near the boundary $u\to 0$, which is equivalent to $\rho\to \infty$,
we have
\SP
{
\Psi_{\rm WKB}\approx &
\,A_-\frac{1}{\sqrt {MR_{\rm AdS}}} \;\rho^{MR_{\rm AdS}-\tfrac{1}{2}} \;e^{i\pi
  MR_{\rm AdS}2}\;\left({4\over 1+\tilde r_h^2}\right)^{MR_{\rm AdS}/4}\;\nu^{-MR_{\rm AdS}/2}\,\times\\
&\exp\left[\tfrac{MR_{\rm AdS}}{\sqrt{1+2\tilde r_h^2}}\;
\left(\nu (F
\left(\csc^{-1}\sqrt{a},\;k\right)- 
\tfrac{1}{\sqrt k}\,K\left(\tfrac{1}{k}\right))
-\tfrac{1}{\nu}(1+\tilde
r_h^2)
\;\tfrac{1}{\sqrt k}\Pi\left(a\big |\tfrac{1}{k}\right)\right)\right]\\
&+\; A_+\frac{1}{\sqrt {MR_{\rm AdS}}} \;\rho^{-MR_{\rm AdS}-\tfrac{1}{2}} \;e^{-i\pi
  MR_{\rm AdS}/2}\;\left({4\over 1+\tilde r_h^2}\right)^{-MR_{\rm AdS}/4}\,\nu^{MR_{\rm AdS}/2}\,\times\\
&\exp\left[-\tfrac{MR_{\rm AdS}}{\sqrt{1+2\tilde r_h^2}}\;
\left(\nu (F
\left(\csc^{-1}\sqrt{a},\;k\right)- 
\tfrac{1}{\sqrt k}K\left(\tfrac{1}{k}\right))
-\tfrac{1}{\nu }(1+\tilde
r_h^2)
\;\tfrac{1}{\sqrt k}\Pi\left(a\big |\tfrac{1}{k}\right)\right)\right].
\label{asympwkb2}
}
This allows us to compute the boundary action and the Green's function
\EQ
{
\tilde G_R(\tilde\nu) \approx \lim_{\epsilon\to 0} 
i\,MR_{\rm AdS}\,\exp({\cal S}_I ) = i \,MR_{\rm AdS}\,
\exp\left(2 \int_\infty^\epsilon \sqrt{{\cal V}(\tilde\nu,u)}\, du\right).
}
\begin{figure}[h]
\begin{center}
\includegraphics[width=3.5in]{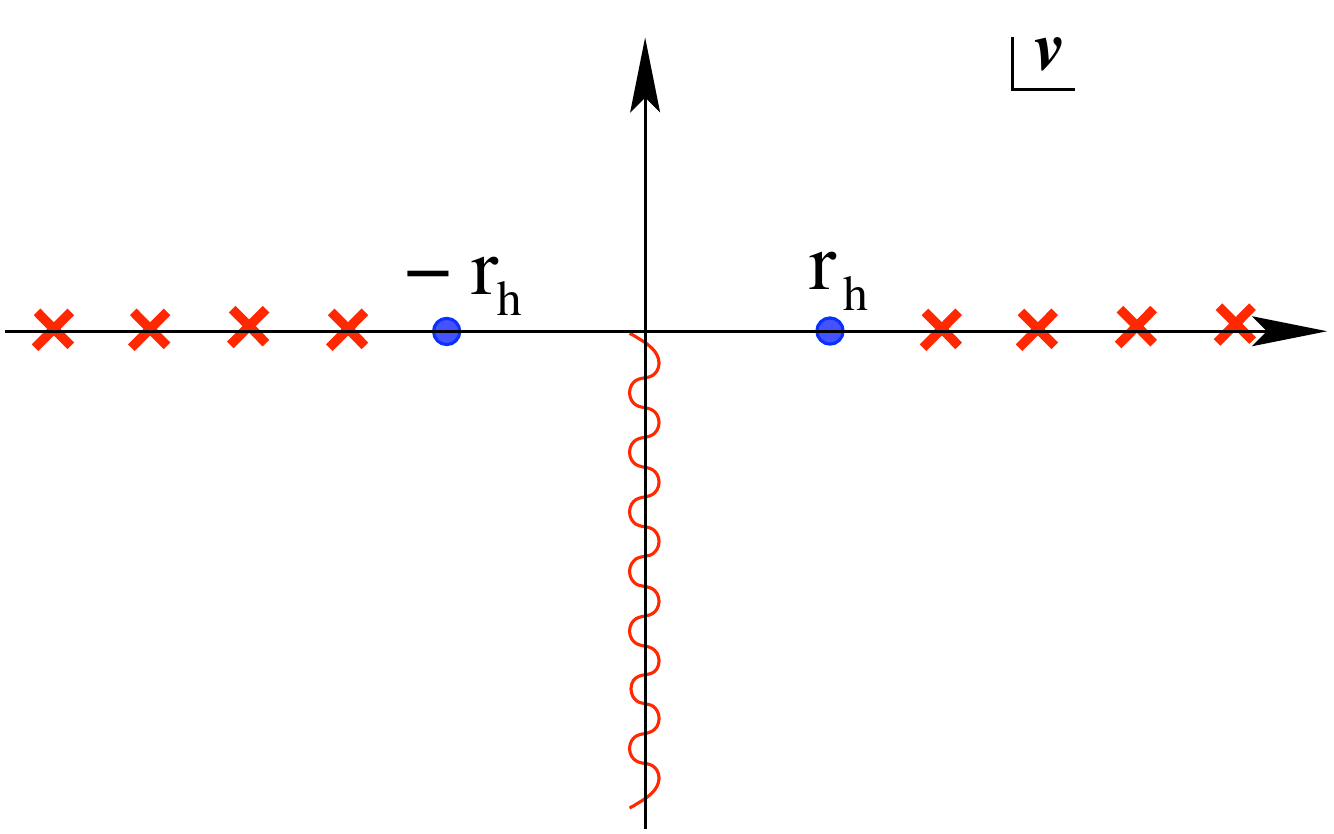} 
\end{center}
\caption{\footnotesize The analytic structure of the boundary 
Green's function in the `` bubble of nothing'' phase in the WKB
approximation $\nu \to \infty$, $MR_{\rm AdS}\to \infty$ with $\tilde\nu =
\nu/M R_{\rm AdS}$ fixed. Approximately equally spaced simple poles on the
real  
axis for $\tilde\nu^2> \tilde r_h^2$, are accompanied by branch points at
$\tilde\nu=0$ and infinity.  
}
\label{wkbgf}
\end{figure}
In terms of elliptic functions, the explicit form for the WKB action is
\SP
{
{\cal S}_I = &-\frac{MR_{\rm AdS}}{\sqrt{1+2\tilde r_h^2}}\;
\left(|\tilde\nu| (F
\left(\csc^{-1}\sqrt{a},\;\tfrac{a}{b}\right)- 
\sqrt{\tfrac{b}{a}}K\left(\tfrac{b}{a}\right))
-\tfrac{1}{|\tilde\nu| }(1+\tilde
r_h^2)
\;\sqrt{\tfrac{b}{a}}\Pi\left(a\big |\tfrac{b}{a}\right)\right)\\
&-\tfrac{1}{2}MR_{\rm AdS}\left(\ln\left(\tfrac{2\epsilon^{-2}}{|\tilde\nu|\,\sqrt{1+\tilde
        r_h^2}}\right)+ i \,\pi\right),
\label{wkbsing}
}
and the only singularity of this expression is the logarithmic singularity
at $\tilde\nu=0$ which complements the singularity at $\tilde\nu=\infty$, found
above.

\subsection{Combining the Results}

The resulting analytic structure of the Green's function is summarized
in Fig. (\ref{wkbgf}), which is to be contrasted with corresponding
Green's function (at large mass and frequency) in the topological
black hole phase in Fig. (\ref{largetbh}).
As already dicussed earlier, the isolated poles on the
real axis indicate glueball states and 
that the gauge theory is in the confined phase,
wherein the radius of the spatial $S^1$ is much smaller than the $dS_3$
radius of curvature. This may be interpreted as a hadronized phase, where the de Sitter temperature being too low for the degrees of freedom to
be deconfined. 

In this context, we should point out that the high
frequency WKB analysis, where the frequencies are much larger than the
de Sitter cosmological constant, is basically a flat space limit and
thus, the singularities of the Green's functions may be interpreted in
the standard way as in flat space. One feature of the propagator in
the bubble of nothing phase whose origin is not entirely clear is 
the branch point singularity at $\tilde\nu=0$. A similar branch point at
$\tilde \nu = -i$ was encountered in the high frequency limit in the
topological black hole phase. The associated branch cut was a
consequence of the apparent merger of the infinite set of quasinormal poles of
the topological black hole, in the high frequency limit. We do not
know of a similar interpretation of the branch cuts of \eqref{wkbsing}
and Fig. (\ref{wkbgf}).

\begin{savequote}[15pc]
\sffamily
I am a being of Heaven and Earth, of thunder and lightning, of rain and wind, of the galaxies.\qauthor{Eden Ahbez (1908 - 1995)}
\end{savequote}
\chapter{Strongly Coupled Plasma in de Sitter Space} \label{chapter:plasma}

In the previous chapter, we saw that the AdS topological black hole geometry corresponds to the deconfined or plasma phase of the strongly coupled ${\cal N} = 4$ supersymmetric Yang-Mills theory of $dS_3 \times S^1$. In this chapter, we will study further the properties of the strongly coupled plasma.

\section{R-current correlation functions}
Real time response to perturbations of a conserved charge density can
reveal interesting late time physics, such as hydrodynamic or
diffusive relaxation. It is now well understood and 
established via holographic
calculations in AdS 
black hole backgrounds that correlators of conserved global currents
exhibit hydrodynamic and diffusion poles \cite{Policastro:2002se, Son:2007vk}. R-charge diffusion in the
strongly coupled, high temperature ${\cal N}=4$ plasma was first
discovered in \cite{Policastro:2002se} and we have reviewed it in Section \ref{revplasma}. The universal features of
strongly coupled plasmas follow from the properties of the stretched
horizon of AdS black holes \cite{Son:2007vk}. 
It is therefore natural to ask whether the
horizon in the topological AdS black hole geometry implies 
hydrodynamic behaviour of correlation functions in the dual field
theory. The answer to this question will depend on the relevant time scales
involved since the boundary field theory is formulated on an expanding
background, namely $dS_3\times S^1$.

The strong coupling correlators for $SO(6)$ R-currents of ${\cal N} = 4$ SUSY
Yang-Mills theory on $dS_3 \times S^1$ will be obtained holographically from
the on-shell action for the $U(1)$ gauge fields in the topological
$AdS_5$ black hole background, by
following the prescription of \cite{Policastro:2002se}. The Maxwell
action for the gauge field is 
\begin{equation}
S = -\frac{1}{4 g^2_\mathrm{SG}} \int d^5 x \sqrt{-g} g^{\m \a} g^{\n
  \b} F_{\m \n} F_{\a \b}, \label{action} 
\end{equation}
varying which gives the following equations of motion
\begin{equation}
\frac{1}{\sqrt{-g}} \partial_\n \left( \sqrt{-g} g^{\m \a} g^{\n \b}
  F_{\a \b} \right) =0 \label{eom}. 
\end{equation}
Here, we have $g_\mathrm{SG}^2 = 16 \p^2 R_{\rm AdS} \slash N^2$. We
find it convenient to use the following form for the metric in the
region exterior to the horizon,
\begin{equation}
ds^2 = \frac{R^2_{\rm AdS}}{4 z^2 \left(1-z\right)} dz^2 + \frac{R^2_{\rm AdS}
  (1-z)}{z}\left[-d \t^2 + \cosh^2 \t \;d \W_2^2 \right]+\frac{r_\x^2}{z} \;d
\x^2.
\end{equation}
Substituting
the solution to the equation of motion that corresponds to
the boundary value $A_{\a}(z)\big|_{z=0} = A_{\a}^0$ back into the
gauge field action, we will get a generating functional for the
R-charge correlators of the field theory.   

Since the field theory lives on the $dS_3\times S^1$
boundary with spatial $S^2\times S^1$ spatial slices,  
it is natural to consider the late time behaviour of
long-wavelength fluctuations in the following two cases:
\begin{enumerate}
\item The R-charge perturbation is inhomogeneous on the $S^1$, but
  homogeneous on the spatial section of $dS_3$, 
\item The fluctuation is homogeneous on the circle, but inhomogeneous
  on the $S^2$. 
\end{enumerate}
Of particular interest is the presence or the lack of late time hydrodynamic
relaxation of this system.

\subsection{Inhomogeneous Perturbation on the $S^1$ \label{S1}}
In the first case, we will assume for simplicity 
that the R-charge perturbation carries no momentum around the 
$S^2$. Furthermore, we use gauge freedom to set the radial component
of the gauge potential $A_z = 0$. Hence, from the symmetries of the
configuration, there are two
remaining bulk gauge fields that are non zero:
\EQ
{
A_\tau=A_\tau(z,\tau,\x) \qquad {\rm and} \qquad A_\x = A_\x(z,\tau, \x).
}
Since the $\x$-direction is a spatial circle, the two components of
the vector potential,
$A_\tau$ and $A_\x$, can be conveniently expanded in Fourier modes 
on the circle. The time dependence can also be re-expressed in terms
of a mode expansion. There is a subtlety involved in this however.
For $A_\x$, which is a scalar in $dS_3$, the mode decomposition 
is straightforward
\EQ
{
A_\x(z,\t,\x)= \sum _n \frac{e^{i n \x}}{2\pi} 
\int_{-\infty}^\infty 
\frac{d\nu}{2\pi} \,\,{{\cal T}_\x (\nu,\t)}\;\;{\cal G}_n(\nu,z),
}
where $n \in \mathbb{Z}$. On the other hand, $A_\t$, which is a gauge field
in $dS_3$, has its complete time dependence captured by 
normal modes of two kinds: normalizable and delta-function
normalizable. In fact, below we will show that 
there is a single normalizable mode and a
continuum of delta-function normalizable states obtained as solutions
to a Schr\"odinger problem. Anticipating this, we write
\EQ
{
 A_\tau(z,\t,\x)= \sum _n \frac{e^{i n \x}}{2\pi} 
\left(\int_{-\infty}^\infty \frac{d\nu}{2\pi} \,{\cal T}_\tau (\nu,\t)\;  
{\cal F}_n(\nu,z) + {\cal T}_\t^{\rm N}(\t) \,{\cal F}^{{\rm N}}_n(z)\right), \label{eq:subt1}
}
where ${\cal T}_\t(\nu,\t)$ and ${\cal T}^{\rm N}_\t(\t)$ will be the
delta-normalizable and normalizable modes, respectively.
The mode functions ${\cal T}_{\x,\t}$ are solutions to 
\AL
{\dd{}{\t}\left({\cosh^{-2} \t} \,\dd{}{\t} \left(\cosh^2 \t \, {\cal T}_\t
    (\nu,\t) \right) \right)
&= - (\nu^2 +1) \;{\cal T}_\t (\nu,\t), \\
{\cosh^{-2}\t}{d\over d\t}\left(\cosh^2\tau\,{\cal T}_\tau\right)&= -i
(\nu+i){\cal T}_\chi.
}
The first of these can be put in the form of a Schr\"odinger equation,
by defining ${\cal T}_\t = \tilde{\cal T}/\cosh\t$, 
\EQ
{
-\frac{d^2}{d\tau^2}\tilde {\cal T} -{2\over \cosh^2\tau}\tilde{\cal T}
=\nu^2\,\tilde {\cal T}.
}
It is solved by the associated Legendre
function ${P}_1^{-i\nu}(\tanh \tau)$. 
For $\nu^2\geq 0$, there is a
continuous infinity of delta-function normalizable states, which yield
\EQ
{
{\cal T}_\t (\nu,\t) = 
\, \frac{e^{- i \nu \tau} }
{\cosh \t}\left(\frac{\nu-i\tanh\t}{\nu-i}\right) \qquad {\rm and} \qquad
{\cal T}_\x(\nu,\t) =  \frac{e^{- i \nu \tau} }{\cosh \t},\label{modes}
}
where $\nu^2
\geq 0.$

The Schr\"odinger potential above also has bound states for $\nu^2 <
0$. In fact, there is precisely one normalizable bound state with $\nu^2=-1$,
corresponding to the solution
\EQ
{
{\cal T}_\t^{\rm N}(\tau)= {1\over\sqrt 2}\,{1\over \cosh^2\tau}.
}
This solution can be directly obtained by evaluating $P_1^{-i\nu}(\tanh\t)$ at
$\nu^2= -1$  or can be systematically inferred 
from the Maxwell equations.

The continuum modes for $A_\x$ and $A_\t$
are orthonormal with respect to the inner product 
\begin{equation}
\langle {\cal T}_{a},{\cal T}_{a} \rangle  \equiv
\int_{-\infty}^\infty d\t \cosh^2 \t \; {\cal T}_{a}(\nu,\t) {\cal
  T}_{a}(\nu',\t) = 2 \pi \,  \d(\nu + \nu'), 
\end{equation}
for $a \in \{ \x,\t \}$, while the bound state is normalized so that
\EQ
{
\int_{-\infty}^\infty d\t \,\cosh^2\t\,\left({\cal T}^{\rm N}_\t\right)^2=1.
}

In the far future, the continuum modes are both given by ${\cal T}_a \sim
e^{-i\nu\tau} e^{-2\tau}$. Upon analytically continuing to 
the complex $\nu$ plane, for $\nu = i +\w$, with $\w \in {\mathbb R}$, they
are propagating, \textit{i.e.}, purely oscillatory, excitations with frequency $\w$ in the far
future. On the other hand the normalizable bound state decays
exponentially in the far past and future. Furthermore, being real, it does not
contribute to the flux at the horizon of the topological black hole.

Using these modes to eliminate the $\t$-dependence, we find
three equations that depend only on ${\cal F}_n(\nu,z)$ and ${\cal
  G}_n(\nu,z)$  
\AL
{-\tfrac{1}{R^2_{\rm AdS}}\,(\nu+i) \;{\cal F}_n'+ \frac{n (1-z)}{r_\x^2}
  \,{\cal G}_n' &= 0,   
\label{cons}\\
\tfrac{1}{R^2_{\rm AdS}}\dd{}{z}\left((1-z) {\cal F}_n' \right) + 
\frac{n}{r_\x^2 
}\,(\nu-i) \,{\cal G}_n - \frac{n^2}{r_\x^2} \,{\cal F}_n &=0, \label{eq3}\\  
4 z\dd{}{z}\left((1-z)^2 {\cal G}_n' \right) - n(\nu+i) {\cal F}_n
+(\nu^2+1) {\cal G}_n &= 0.
\label{eq4} 
}
Here, prime denotes a derivative with respect to $z$. The radial or
$z$-dependence of the bound state solution, ${\cal F}^{\rm N}(z)$ 
is found by analytically
continuing the profile for generic $\nu$ to $\nu=\pm i$.
We note that there
are three equations for two unknowns. Therefore, to ensure a non-trivial
solution, any two equations must imply the third and it is
straightforward to check that this is indeed the case. 

We  can then
use these equations of motion to derive two independent ones, 
each containing only one of the  unknown functions:
\AL
{
4z (1-z) {\cal F}_n''' -4 (3z-1) {\cal F}_n''-4 {\cal F}_n' &= \left[
  \bar{n}^2-\frac{\nu^2+1}{1-z} \right] {\cal F}_n',
\label{gaugea}\\
4 z (1-z) {\cal G}_n''' - 4(5z-1) {\cal G}_n''-
\frac{8(1-2z)}{1-z}{\cal G}_n' &= \left[\bar{n}^2 -
  \frac{(\nu^2+1)}{1-z}\right]{\cal G}_n', \label{gaugeb}
}  
where we have defined $\bar{n} \equiv \frac{n R_{\rm AdS}}{r_\x}$.

Unfortunately, we cannot use the perturbative method as described in Section \ref{revplasma} in this case, as the smoothness condition used in the perturbative method is not enough to fully determine the solution. Fortunately, these equations of motion can be solved
in terms of hypergeometric functions
\EQ
{
{\cal F}_n' = C_\t(\nu,\bar n)\;(1-z)^{-i(\nu-i)/2}\;{}_2F_1\left(
\tfrac{1}{2}+\tfrac{i}{2}(\bar n-\nu)\,,
\tfrac{1}{2}-\tfrac{i}{2}(\bar n+\nu)\,; 1-i\nu \,;1-z\right).
}
Here, $C_\t$ is a frequency dependent constant, to be determined by
the boundary values of the fields.
Near the boundary of AdS space, the solution approaches
\SP
{
{\cal F}_n'\big|_{z\to 0} =-\;C_\t\;\tfrac{\Gamma(1-i\nu)} 
{\Gamma\left(\tfrac{1}{2}+
\tfrac{i}{2}(\bar n-\nu)\right)
\Gamma\left(\tfrac{1}{2}-\tfrac{i}{2}(\bar n+\nu)\right)}\, \left(2
\gamma_E + \ln z +\psi\left(\tfrac{1}{2}+\tfrac{i}{2}(\bar
  n-\nu)\right) +
\psi\left(\tfrac{1}{2}-\tfrac{i}{2}(\bar n+\nu)\right)\right).
}
Imposing the boundary conditions at $z \to 0$
\EQ
{
\lim_{\e \to 0} {\cal F}_n(\nu,\epsilon)= {\cal F}^0_n(\nu) \qquad {\rm and} \qquad
\lim_{\e \to 0} {\cal G}_n(\nu,\epsilon)={\cal G}_n^0 (\nu),
}
from \eqref{eq3}, we find that
\EQ
{C_\tau(\nu,\bar n) = - \tfrac 
{\Gamma\left(\tfrac{1}{2}+
\tfrac{i}{2}(\bar n-\nu)\right)
\Gamma\left(\tfrac{1}{2}-\tfrac{i}{2}(\bar
  n+\nu)\right)}{4\Gamma(1-i\nu)}\;\left(\bar n^2 {\cal F}_n^0(\nu) -
\tfrac{R_{\rm Ads}}
  {r_\x}\,\bar n (\nu-i) \,{\cal G}^0_n(\nu)
\right).
}
With the normalization fixed in terms of the boundary values of the
relevant fields, we have
\SP
{
{\cal F}_n'(\nu,\epsilon)=\tfrac{1}{4}\left(\bar n^2 {\cal F}_n^0 -
  \tfrac{R_{\rm AdS}}{r_\x}\,\bar n (\nu-i) 
{\cal G}_n^0
\right)\;\left(2
\gamma_E + \ln \epsilon +\psi\left(\tfrac{1}{2}+\tfrac{i}{2}(\bar
  n-\nu)\right) + 
\psi\left(\tfrac{1}{2}-\tfrac{i}{2}(\bar n+\nu)\right)\right).
}
From \eqref{cons} we also obtain
\EQ
{
{\cal G}_n'(\nu,\epsilon) = \tfrac{r_\x}{R_{\rm AdS}}{\nu+i\over \bar n}
{\cal F}_n'(\nu, \epsilon).
}

Following identical steps for the
normalizable temporal modes, we have
\EQ
{
{\cal F}_n^{{\rm N}\prime}(\epsilon)=\tfrac{1}{4}\bar n^2{\cal
  F}_n^{{\rm N}0}\left(2
\gamma_E + \ln \epsilon +\psi\left(1+\tfrac{i}{2}\bar n\right) + 
\psi\left(1 -\tfrac{i}{2}\bar n \right)\right).
}
We can now plug these solutions into the boundary action to obtain the 
retarded R-current correlators.

The induced boundary action for bulk Maxwell fields has the form
\SP
{S=
{4 \pi \over g^2_{SG}} \sum_{n} \frac{1}{2 \pi}\,\left(\int_{-\infty}^\infty 
{d\nu\over
  2\pi}\left( r_\x {\cal F}_{-n}^0(-\nu)\;{\cal
    F}_n'(\nu)\big|_{z=\epsilon} -  
\tfrac{R^2_{\rm AdS}}{r_\x}\, 
{\cal G}_{-n}^0(-\nu) \;{\cal G}_n'( \nu)\big|_{z=\epsilon}
\right) + r_\x{\cal F}_{-n}^{{\rm N}0}{\cal F}_{n}^{{\rm
    N}\prime}\big|_{z=\epsilon}\right), 
\label{bulkmaxwell1}
}
where $g^2_{SG}= 16\pi^2 R_{\rm AdS}/N^2$. The complete real time
retarded correlation functions for the R-currents, can now be accessed readily.
First, we define the boundary values of our gauge fields as
\EQ
{
A_{\x,\t}(\epsilon,\t,\x)\equiv
\sum_n {e^{i n\x}\over 2\pi}A_{\x,\t}^n(\t)\,,
}
so that 
\EQ
{
{\cal G}_n^0(-\nu)=\int_{-\infty}^\infty\,A_\x^n(\t)\,{\cal T}_\x(\nu,\t)\cosh^2\t.
}
Similarly,
\EQ
{
{\cal F}^0_n(-\nu)=\int_{-\infty}^\infty\,
A_\t^n(\t)\,{\cal T}_\t(\nu,\t)\cosh^2\t\, \qquad {\rm and} \qquad
{\cal F}^{{\rm N}0}_n=\int_{-\infty}^\infty\,
A_\t^n(\t)\,{\cal T}_\t^{\rm N}(\t)\cosh^2\t\,.
}
Putting these ingredients together, we find that the boundary action
is 
\SP
{
S=&{N^2\over 4\pi R_{\rm
    AdS}}\;\sum_n \tfrac{1}{2\pi}\int_{-\infty}^\infty d\t\,\cosh^2\t
\int_{-\infty}^\infty d\t^\prime\,\cosh^2\t'\\
&\left[\int_{-\infty}^\infty{d\nu\over 2\pi} \left(2 \gamma_E+ \ln
    \epsilon + \psi\left(\tfrac{1}{2}+\tfrac{i}{2}(\bar n-\nu)\right) + 
\psi\left(\tfrac{1}{2} -\tfrac{i}{2}(\bar n+\nu) \right)\right)\times\right.\\
&\left.\times\left(
  \,A_\t^{-n}(\t)A_\t^{n}(\t')\,\tfrac{{\bar n}^2}{4}\,r_\x\,\,{\cal
    T}_\tau(\nu,\tau){\cal T}_\t(-\nu,\t')\,- \,A_\t^{-n}(\t)A_\x^n(\t')\,
{R_{\rm AdS}}
\tfrac{\bar n}{4}(\nu-i)\,\times \right.\right.\\\\
&\left.\left.\times {\cal T}_\t(\nu,\t){\cal T}_\x(-\nu,\t')\,-\, 
A_\x^{-n}(\t)A_\t^n(\t')\,R_{\rm AdS}\tfrac{\bar n}{4}(\nu+i)\,{\cal
  T}_\x(\nu,\t){\cal T}_\t(-\nu,\t')\,+\right.\right.\\\\
&\left.\left.+ A_\x^{-n}(\t)A_\x^{n}(\t')\tfrac{R^2_{\rm AdS}}{r_\x}
\tfrac{1}{4}(\nu^2+1){\cal T}_\x(\nu,\t){\cal T}_\x(-\nu,\t')\right)
+ r_\x A^{-n}_\t(\t)A^n_\t(\t')\times\right.\\\\
&\left.\times \tfrac{{\bar n}^2}{4}{\cal T}^{\rm N}(\t){\cal T}^{\rm N}(\t')
\left(2 \gamma_E+ \ln
    \epsilon + \psi\left(1+\tfrac{i}{2}\bar n\right)+ 
\psi\left(1 -\tfrac{i}{2}\bar n \right)\right)
\right].
\label{bulkmaxwell2}
}

The R-current correlators can be read off
from the finite and non-analytic pieces of this boundary action
\eqref{bulkmaxwell1}, \eqref{bulkmaxwell2}. We will denote the Fourier harmonics of the
R-current along the spatial circle as
\EQ
{
j_n^\mu(\tau) = \int_0^{2\pi}d\x \, e^{-i n\x}\, j^\mu(\x,\t),
}
where we have already restricted attention to the s-wave sector on the
spatial two-sphere. For perturbations with non-vanishing momentum
along the spatial circle, we define the retarded, real time, Green's
functions as  
\EQ
{G^{\mu\nu}(\t,\t'; n)=\langle[\,j_n^\mu(\tau),\,j_{-n}^\nu(\tau')]\rangle\,
\Theta(\tau-\tau')\,,\qquad \mu,\nu \in \{\x,\t\}.
}
The conserved global currents are in one-to-one correspondence with
the boundary values of the 5-d gauge fields. Functionally
differentiating the induced boundary action \eqref{bulkmaxwell2} 
with respect to the boundary values of the gauge fields, we find
\AL
{G^{\t \t}(\t,\t';n) = &\,
\cosh^2\t\cosh^2\t'\,\times 
\\\nonumber
&\qquad n^2\,\left[\int_{-\infty}^\infty {d\nu\over 2\pi}\,{\cal T}_\tau(\nu,\tau){\cal
  T}_\tau(-\nu,\t') \,\,{\bf \Xi}(\nu,n) 
+ {\cal T}_\t^{\rm N}(\t){\cal T}_\t^{\rm
  N}(\t')\,\,{\bf \Xi}(i, n)\right], \label{gtt}\\
G^{\x \x}(\t,\t';n) =&
\,\cosh^2\t\cosh^2\t'
\int_{-\infty}^\infty {d\nu\over 2\pi}\,{\cal T}_\x(\nu,\t)
{\cal T}_\x(-\nu,\t')\left(\nu^2 + 1\right)\,{\bf \Xi}(\nu,\,n), \\\nonumber\\
G^{\t \x}(\t,\t';n) = &\,G^{\x \t\,*}(\t,\t';n) 
=\, \cosh^2\t\cosh^2\t'\,n\int_{-\infty}^\infty{d\nu\over 2\pi}\,{\cal T}_\t(\nu,\t)
{\cal T}_\x(-\nu,\t')\left(\nu -i\right)
{\bf \Xi}(\nu, n),
}
where
\EQ
{{\bf \Xi}(\nu,\,n)= \tfrac{N^2 R_{\rm AdS}}{32 \pi^2 r_\x}\left(\psi\left(\tfrac{1}{2}+\tfrac{i}{2}(\bar  
  n-\nu)\right) +
\psi\left(\tfrac{1}{2}-\tfrac{i}{2}(\bar
  n+\nu)\right)\right).
}

It is easily established that the above expressions are indeed
retarded Green's functions and are non-vanishing only when $\t >
\tau'$. Two essential features ensure that
this is the case: 
\begin{enumerate}
\item The function ${\bf \Xi}(\nu, n)$ appearing
universally in the $\nu$-integrals has only simple poles in the lower half
complex plane at
\EQ
{
\nu= -i(2k +1) \pm \bar n\,,\qquad k\in {\mathbb Z}.
}
\item There is a second source of non-analyticities in the $\nu$-plane. This
lies in the $\nu$-dependent normalization of the mode functions ${\cal
T}_\t(\nu,\t)$ \eqref{modes}. The potentially worrisome aspect of this
is the appearance of a pole in the {\em upper} half plane at $\nu=+i$, which
gives a non-vanishing contribution for $\t < \t'$. However, the
potential problem is eliminated by the term dependent on the discrete,
normalizable mode
${\cal T}^N_\t$ in \eqref{gtt} which exactly cancels against the
contributions from the poles at $\nu =+i$, ensuring that our
Green's function is causal. 
\end{enumerate}
The remaining Green's functions are
manifestly free of singularities in the upper half plane. We thus see that the Son-Starinets recipe for determining real time
response functions works in the case of the topological black hole, 
provided we carefully account for the
contributions from both the continuum and discrete mode functions in $dS_3$.

It is also clear in the above expressions that there are no diffusion
poles. Instead, from the properties of the digamma function which we
have encountered before, the frequency space Green's function, which is
effectively ${\bf \Xi} (\nu,n)$\,, has only simple poles in the complex
$\nu$-plane, the lowest of these being at 
\EQ
{
\nu= -i\pm \,\bar n.
} 
Excitations with complex $\nu = \w - i$, and $\w\in {\mathbb R}$ 
will propagate at late times, where the perturbations simply
evolve as left- and right-moving excitations on the
$S^1$ without dissipating. 

The absence of any diffusion or transport like behaviour may be
intuitively explained by noticing the similarity of the topological
AdS black hole to the BTZ black hole. 
  Setting up an excitation with momentum only on the $S^1$ is
equivalent to stating that the variation of the fields along the $S^2$ is
zero. Therefore, aside from the time dependent factors associated to the de
Sitter expansion, the metric that is ``seen'' by the bulk fields is not
the full five dimensional metric, but its effective $2+1$
dimensional portion
\begin{equation}
ds^2 = -R^2_{\rm AdS}
(\rho^2-1) d \t^2 +\frac{R^2_{\rm AdS}}{\rho^2-1} d\rho^2  + \rho^2 r_\x^2 d \x^2.
\end{equation}
Comparing this with the metric for a $2+1$ dimensional BTZ black hole
with zero angular momentum 
\begin{equation}
ds^2 = -\left(\frac{r^2}{R^2} - M \right) dt^2 + \left(\frac{r^2}{R^2}
  - M \right)^{-1} dr^2 + r^2 d \phi^2, 
\end{equation}
we note the obvious similarity. Therefore, we expect the behaviour of
fields in the 
topological black hole background with an inhomogeneous excitation
around the $S^1$ to be similar to the behaviour of the fields in a BTZ
black hole background.  In other words, they should behave as in a
(1+1)-dimensional CFT \cite{Son:2002sd}, just as we see from our
results above. 

\subsection{Inhomogeneous Perturbation on the $S^2$ \label{dS3}}

We will now examine the real time response to fluctuations carrying
momentum along the spatial sections of the three dimensional de Sitter space.
Each spatial section of $dS_3$ is a two-sphere which undergoes
exponential expansion at late times. For this case, we will 
focus on a situation where an inhomogeneous R-charge perturbation is
set up on the two-sphere 
with only a dependence on the polar angle $\theta$ and time
$\tau$. For this configuration, the dual bulk gauge fields are
\EQ
{
A_\tau=A_\tau(z,\tau, \o)\,\qquad {\rm and} \qquad A_\theta = A_\theta(z,\tau, \o)\,, 
}
where $A_z=A_\chi=0$. Here, $A_z$ is set to zero by the gauge freedom and $A_\chi$
vanishes due to $\chi$-independence of the configuration. 

By spherical
symmetry, the scalar potential $A_\tau$ and the vector 
potential $A_\theta$, each can be expanded in terms of scalar
and vector spherical harmonics, respectively:
\EQ
{
A_\tau = \sum _{\ell=0}^\infty
{\cal F}_\ell(z,\tau)\;Y_\ell^0(\theta), \qquad
A_\theta=\sum_{\ell=1}^\infty {\cal G}_\ell(z,\tau)\;\partial_\theta
Y_\ell^0(\theta). 
}
Substituting these into the bulk Maxwell equations, we find
\AL
{
4 z(1-z){\partial_z}\left ((1-z){\partial_z
  {\cal G}_\ell}\right) -{\partial^2_\tau {\cal G}_\ell}
+{\partial_\tau {\cal F}_\ell}=&\,0,
\label{atheta}
\\\nonumber\\
 4 z(1-z)\partial_z \left ((1-z){\partial_z
  {\cal F}_\ell}\right) - {\ell(\ell+1)\over \cosh^2 \tau}
\left({\cal F}_\ell -\partial_\tau {\cal G}_\ell\right)=&\,0,
\label{atau}
\\\nonumber\\
\partial_\tau \left(\cosh^2\tau\partial_z {\cal F}_\ell\right) +
\ell(\ell+1)\partial_z {\cal G}_\ell=&\,0.
\label{conserve}
}
From equations \eqref{atau} and \eqref{conserve} we obtain a 
differential equation for ${\cal F}'_\ell \equiv \partial_z {\cal F}_\ell$
\EQ
{
4 \partial_z \left(z(1-z)\partial_z\left((1-z){\cal F}'_\ell\right)\right)
-{\ell(\ell+1)\over \cosh^2\tau} {\cal F}'_\ell
-{1\over\cosh^2\tau}\partial_\tau^2 \left(\cosh^2\tau {\cal F}'_\ell\right)=0.
\label{radial}
}
Now, we will separate out the explicit temporal dependence, keeping in
mind, as before, the possibility of contributions from 
both discrete and continuous modes
\EQ
{
{\cal F}'_\ell(z,\tau)=\int_{-\infty}^\infty{d\nu}
\; {\cal T}_\ell(\nu,\t) \;F_\ell(\nu,z) + 
\sum_{m}{\cal T}_{\ell\,m}^{\rm N}(\t)\, F_{\ell \,m}^{\rm N}(z). \label{eq:subt2}
}
The mode functions ${\cal T}_\ell$ satisfy
\EQ
{
\left[- \partial_\tau^2 -
{\ell(\ell+1)\over \cosh^2\tau} \right]
\left(\cosh^2\tau\,{\cal T}_\ell\right) = \nu^2 \;
\left(\cosh^2\tau\,{\cal T}_\ell\right) ,
\label{pt}
}
which is a Schr\"odinger equation whose potential clearly will have
both bound states and scattering or continuum states. The full set of
solutions form an orthonormal, complete set. Indeed, the
delta-normalized eigenstates are the Legendre functions 
\EQ
{
{\cal T}_\ell(\nu,\t)= 
\Gamma(1+i \nu)\,{P_\ell^{-i \nu}(\tanh\tau)\over \cosh^2\tau}.
}
Those with $\nu^2>0$ are scattering states with continuous values of 
$\nu \in {\mathbb R}$, while the
discrete, ``bound states'' have $-i \nu =1,2,\ldots \ell$,
\EQ
{
{\cal T}_{\ell\,m}^{\rm N} 
= \sqrt{m(\ell-m)!\over (\ell+m)!}
\,\,{P_\ell^m(\tanh\tau)\over \cosh^2\tau}\,,\qquad m=1,2,\ldots \ell.
}

For $\nu^2> 0$, the late time $\tau \to\infty$ behaviour
of the modes will be significant, 
\EQ
{
{\cal T}_\ell(\nu,\t)\big|_{\tau\gg 1} \to e^{-i \nu\tau}\; e^{-2\tau},
}
as these modes are oscillatory.  Applying infalling boundary
conditions on these modes at the
horizon of the topological AdS black hole, \eqref{radial} yields
\EQ
{
F_\ell(\nu,z)= C_\ell(\nu)\; (1-z)^{-1-i\nu/2}
\;{}_2 F_1\;\left(-i\tfrac{\nu}{2},1-i\tfrac{\nu}{2}\,;1-i\nu\,;1-z\right).
\label{gaugesol}
}
For the discrete series, the radial profile in the bulk $F_{\ell \,m}^{\rm N}(z)$ is obtained by evaluating $F_\ell(\nu,z)$ at
$\nu= -i m$. 

The asymptotic form of the radial dependence of the bulk potential
${\cal F}'_\ell $, can be determined from \eqref{gaugesol}
\EQ
{
F_\ell\big|_{z\to 0} = C_\ell(\nu)\left(C_1(\nu) + C_2(\nu)\;\ln z
\right),
}
where
\AL
{
&C_1= - \frac{2\gamma_E
  +\psi(1-\tfrac{i\nu}{2})+\psi(-\tfrac{i\nu}{2})}{
\Gamma(1-\tfrac{i\nu}{2})\Gamma(-\tfrac{i\nu}{2})},\\
&C_2 = -{1\over \Gamma(1-\tfrac{i\nu}{2})\Gamma(-\tfrac{i\nu}{2})}.
}
We can solve for $C_\ell$ in terms of the boundary values of the gauge
potentials, using the bulk equation of motion \eqref{atau} for
${\cal F}_\ell'$ near the boundary which yields
\EQ
{
4 \int_{-\infty}^\infty {d\nu\over 2\pi}
\; \Gamma(1+i \nu)P_\ell^{-i\nu}(\tanh\tau)\;\ C_\ell(\nu) \;C_2(\nu)
= \ell(\ell+1)\left({\cal F}_\ell^0(\t)-\partial_\tau
  {\cal G}_\ell^0(\t)\right).
}
We note that the other equation of motion \eqref{atheta} for ${\cal G}_\ell'$,
just yields
the time derivative of this condition, so that we have only one
equation to determine the coefficient $C_\ell$. This equation can
be solved if we recall that the associated Legendre functions are
mutually orthogonal\footnote{The orthogonality of these functions for  
  purely imaginary order follows from the fact that they are
  eigenfunctions of the Schr\"odinger equation in the $\rm sech^2$
  potential, $\left(-\tfrac{d^2}{d\tau^2}- \ell(\ell+1)/\cosh^2\tau
  \right)P_\ell^{i\nu}(\tanh \tau)= \nu^2 P_\ell^{i\nu}(\tanh\tau)
$. In particular for $\nu\in {\mathbb R}$, these are scattering
states and are delta-function normalizable, and the eigenfunctions
corresponding to two different eigenvalues are orthogonal as usual.}
\EQ
{
\int_{-\infty}^\infty d\tau
\;P_\ell^{i\nu}(\tanh\tau)P_\ell^{-i\nu'}(\tanh\tau)=
{\delta(\nu-\nu') \over \Gamma(1-i\nu)\Gamma(1+i \nu)}.
}
Thus, we have
\EQ
{
C_\ell = {\ell(\ell+1)\over 4 C_2}\Gamma(1-i \nu)\int_{-\infty}
^\infty
d\tau'
\left({\cal F}_\ell^0(\t')-\partial_{\tau'} {\cal G}_\ell^0(\t')\right)
\; P_\ell^{i\nu}(\tanh\tau'),
}
from which we obtain the solution 
\SP
{
\cosh^2\tau\;{\cal F}'_\ell(\epsilon,\tau) =& \; {\ell(\ell+1)\over 4}
\, 
\int_{-\infty}^\infty
d\tau'\left({\cal F}_\ell^0(\t')-\partial_{\tau'}
    {\cal G}_\ell^0(\t')\right)\\
&\bigg[\int_{-\infty}^\infty {d\nu\over 2\pi}\, {\pi \nu\over \sinh\pi\nu}\;
P_\ell^{-i\nu}(\tanh\tau)P_\ell^{i\nu}(\tanh\tau')\,\left(\ln \epsilon
  +2 \gamma_E+ 
\psi(-\tfrac{i\nu}{2})+\psi(1-\tfrac{i\nu}{2})\right) \\
& 
+ \sum_{m=1}^\ell m \tfrac{(\ell-m)!}{(\ell+m)!}\,
P_\ell^m(\tanh\t)P_\ell^m(\tanh\t')
\left(\ln\epsilon + 2\gamma_E+\psi\left( \tfrac{m}{2}\right)+
  \psi(1+\tfrac{m}{2})\right)\bigg]. \label{derivatives}
}
Plugging these back into the expression for the boundary action
\eqref{bdryaction}, we obtain the generating functional for two point
correlators of R-currents.

Putting the above ingredients together, 
the general form of the electric field along the
radial direction in the bulk is 
\EQ
{
A_\tau'(z,\tau,\o)=
\sum_{\ell=0}^\infty Y_\ell^0(\theta)\left(\int 
_{-\infty}^\infty\;{d\nu\over 2\pi }\, {\Gamma(1+ i \nu)}\,
{P_\ell^{-i \nu}(\tanh\tau)\over \cosh^2\tau}\;
F_\ell(\nu,z)+\sum_{m=1}^\ell
\sqrt{m\tfrac{(\ell-m)!}{(\ell+m)!}} \, {P_\ell^m(\tanh\t)\over
  \cosh^2\t}\, F_{\ell \,m}^{\rm N}(z)
\right).
}
This also allows us to automatically solve for $A_\theta'$
using \eqref{conserve} and we get
\EQ
{
A_\theta'(z,\tau,\theta)= - \sum_{\ell=1}^\infty
{\partial_\theta Y_\ell^0(\theta)\over
  \ell(\ell+1)}\,\left(\int_{-\infty}^\infty {d\nu\over 2\pi}\; 
\Gamma(1+i\nu)\,\partial_\tau P_\ell^{-i \nu}(\tanh\tau)\;
F_\ell(\nu,z)+ \sum_{m=1}^\ell
  \sqrt{m\tfrac{(\ell-m)!}{(\ell+m)!}}
\,\,\partial_\t P_\ell^m(\tanh\t)\,F_{\ell\, m}^{\rm N}(z)
\right).
}
Now, the bulk gauge field action can be shown to induce a boundary term which
will be the generating functional for the boundary R-current
correlators. Using the explicit solutions above, the induced boundary
action becomes
\EQ
{
S=  {1\over 2 g^2_{SG}}\int d\tau \;
r_\x\left[\sum _{\ell=0}^\infty 
{\cal F}_\ell(z,\tau)\, {\cal F}'_\ell(z,\tau) 
\cosh^2\tau + \sum_{\ell=1}^\infty 
\ell(\ell+1)\,{\cal G}_\ell(z,\tau)\, {\cal G}'_\ell(z,\tau)
\right]_{z=\epsilon\to 0}.
\label{bdryaction}
}
The next step is to express this completely in terms of the boundary values of
the gauge potentials 
${\cal F}_\ell^0(\t) \equiv {\cal F}_\ell(\epsilon,\tau)$ and ${\cal
  G}_\ell^0(\tau) \equiv {\cal G}_\ell(\e,\t)$.  Their radial derivatives 
${\cal F_\ell}'$ and ${\cal G}_\ell'$, equivalently $A_\tau'$ and $A_\theta'$, at the boundary
$z=\epsilon$, are also determined completely by the 
boundary values of the gauge potentials, ${\cal F}_\ell^0(\t)$ and
${\cal G}_\ell^0(\tau)$ as in \eqref{derivatives}.

From the boundary action above, we thus find that the real time, 
retarded Green's functions for the R-charge currents $j^{\mu}$, in the
gauge theory are 
\AL
{
G^{\tau\tau}(\tau,\t'; \ell)=&  
\,\ell(\ell+1)
\left[\int_{-\infty}^\infty  
{d\nu\over 2\pi} \; \,{\pi \nu\over \sinh\pi\nu}\; 
P_\ell^{-i\nu}(\tanh\tau)P_\ell^{i\nu}(\tanh\tau')\,{\bf \Upsilon}(\nu)+
\right.\\\nonumber
&\left.\qquad\quad+ \sum_{m=1}^\ell \, (-1)^m\,m \,
P_\ell^m(\tanh\t)P_\ell^{-m}(\tanh\t')\,{\bf
    \Upsilon}(i\,m)\right], \\\nonumber\\
G^{\theta\theta}(\t,\t';\ell)=&\,\ell(\ell+1) 
\left[\int_{-\infty}^\infty  {d\nu\over 2\pi}\;{\pi \nu\over \sinh\pi\nu}\; 
\partial_\t P_\ell^{-i\nu}(\tanh\tau)\, \partial_{\t'}P_\ell^{
  i\nu}(\tanh\tau') {\bf \Upsilon}(\nu)+ \right.\\\nonumber
&\left.\qquad\quad+ \sum_{m=1}^\ell \, (-1)^m\,m\,
\partial_\t P_\ell^m(\tanh\t)\partial_{\t'}P_\ell^{-m}(\tanh\t')\,
{\bf
    \Upsilon}(i\,m)\right], \\\nonumber\\
G^{\tau\theta}(\tau,\t';\ell)=& \ell(\ell+1)\left[\int_{-\infty}^\infty 
{d\nu\over 2\pi}\;{\pi \nu\over \sinh\pi\nu}\; 
P_\ell^{-i\nu}(\tanh\tau)\, \partial_{\t'}P_\ell^{i\nu}(\tanh\tau')
\,{\bf \Upsilon}(\nu)+\right.\\\nonumber
&\left.  \qquad\quad+ \sum_{m=1}^\ell (-1)^m\, m\,
P_\ell^m(\tanh\t)\partial_{\t'}P_\ell^{-m}(\tanh\t')\,
{\bf \Upsilon}(i\,m)\right],\\\nonumber\\
}
where
\EQ
{{\bf \Upsilon}(\nu)=\tfrac{N^2 r_\x}{64\pi^2 R_{\rm AdS}}\,
\left(\psi(-\tfrac{i\nu}{2}) - \tfrac{1}{i \nu})\right).}

We need to first confirm that these Green functions satisfy basic
consistency checks. Specifically, the retarded functions must vanish
for $\t < \t'$.  As in the previous case, this property is not
manifest, but follows from the nature of the non-analyticities of
${\bf \Upsilon}(\nu)$, and the normalized mode functions 
$\Gamma(1+i\nu)P_\ell^{-i\nu}(\tanh\tau)$, in the complex $\nu$-plane. 
For real values of $\nu$, the associated Legendre function is
\cite{elliptic, Abramowitz:1964} 
\EQ
{
\Gamma(1+i\nu)\,P_\ell^{-i \nu}(\tanh\t)= e^{-i\nu\t}\,
{}_2F_1(-\ell,\,\ell+1;\, 1+ i\nu,\, (1-\tanh\t)/2).
}
For integer $\ell$, the hypergeometric function is a finite polynomial in 
$\tanh\t$ and a ratio of degree $\ell$ polynomials of $\nu$.
Therefore, the exponential frequency dependence means that, for $\t-\t' <  0$,
the integrals over the frequency $\nu$ can be  
evaluated by closing the contour in the upper half plane. The function 
${\bf \Upsilon}(\nu)$ has no poles in the upper half complex plane. It
has simple poles at 
\EQ
{
\nu= 0, \,-2i, \,-4i \ldots
}
The normalized modes, $\Gamma(1+i \nu) P_\ell^{-i\nu}(\tanh\t)$ have
exactly $\ell$ simple poles in the upper half plane at
$\nu=i,\,2i\ldots \ell i$. The contributions from these are, however,
cancelled by the inclusion of the discrete modes in the retarded
Green's function above. Hence, our correlators are zero for $\t < \t'$.

\subsection{Late Time Behavior}

The real time response functions in general contain important
information on the long time relaxation of perturbations away from the
equilibrium or ground state. In thermal field theory on flat space,
the relaxation of such fluctuations of conserved charges proceeds
via hydrodynamic or diffusion modes. The response functions at strong
coupling then exhibit diffusion poles in frequency space, $G^{\t\t}
\propto (i\omega-  D k^2)^{-1}$, where $\omega$ is the frequency and
$k$, the soft spatial momentum. Due to the explicit time dependence of
the background metric, we cannot do a similar frequency space study of
the full Green's functions in de Sitter space. Instead, we could analyze
their behaviour as functions of time.

In de Sitter space, perturbations labelled by wave number
$\ell$, get red-shifted so that given sufficient time, 
their physical wavelengths become super-horizon sized. This happens when 
\EQ
{
{ \ell \, e^{-\tau}\over R} \sim {1\over R}.
}
At late enough times, even very high harmonics on the sphere get
stretched  and eventually exit the horizon. To zoom in on the time
evolution of such modes, it is useful to think of $\ell e^{-\tau}$,
the physical wave number, as being fixed as $\t \to
\infty$. For example, in this late time approximation,
we neglect terms like $\ell \,e^{-2\tau}$
in comparison to powers of $\ell \,e^{-\t}$.
This is practically equivalent to going to planar coordinates
for de Sitter space and the mode functions behave as 
\EQ
{
P_\ell^{-i \nu}(\tanh\t)\big|_{\ell e^{-\t}=\,{\rm fixed}} \longrightarrow 
\ell^{-i \nu}\,J_{i\nu}\left(2\ell e^{-\t}\right).
\label{bessel}
}
It is possible to derive this by replacing the potential
$\ell(\ell+1){\rm sech}^2\t$ in the mode
equation \eqref{pt},  with $4\,\ell^2\,e^{-2\tau}$.
We note that, instead of a fixed physical wavelength, if we focus
attention on fixed comoving wavenumber, which is given by $\ell$, all modes
simply approach the $s$-wave at late times
\EQ
{
\lim_{\t \to \infty}\,\Gamma(1+i\nu)\, 
P_\ell^{-i\nu}(\tanh\tau)\big|_{\ell\,\,{\rm fixed}} 
\longrightarrow 
e^{-i\nu \t}.
}

\begin{figure}[h]
\renewcommand{\figurename}{Fig.}
\begin{center}
\includegraphics[width=2.5in]{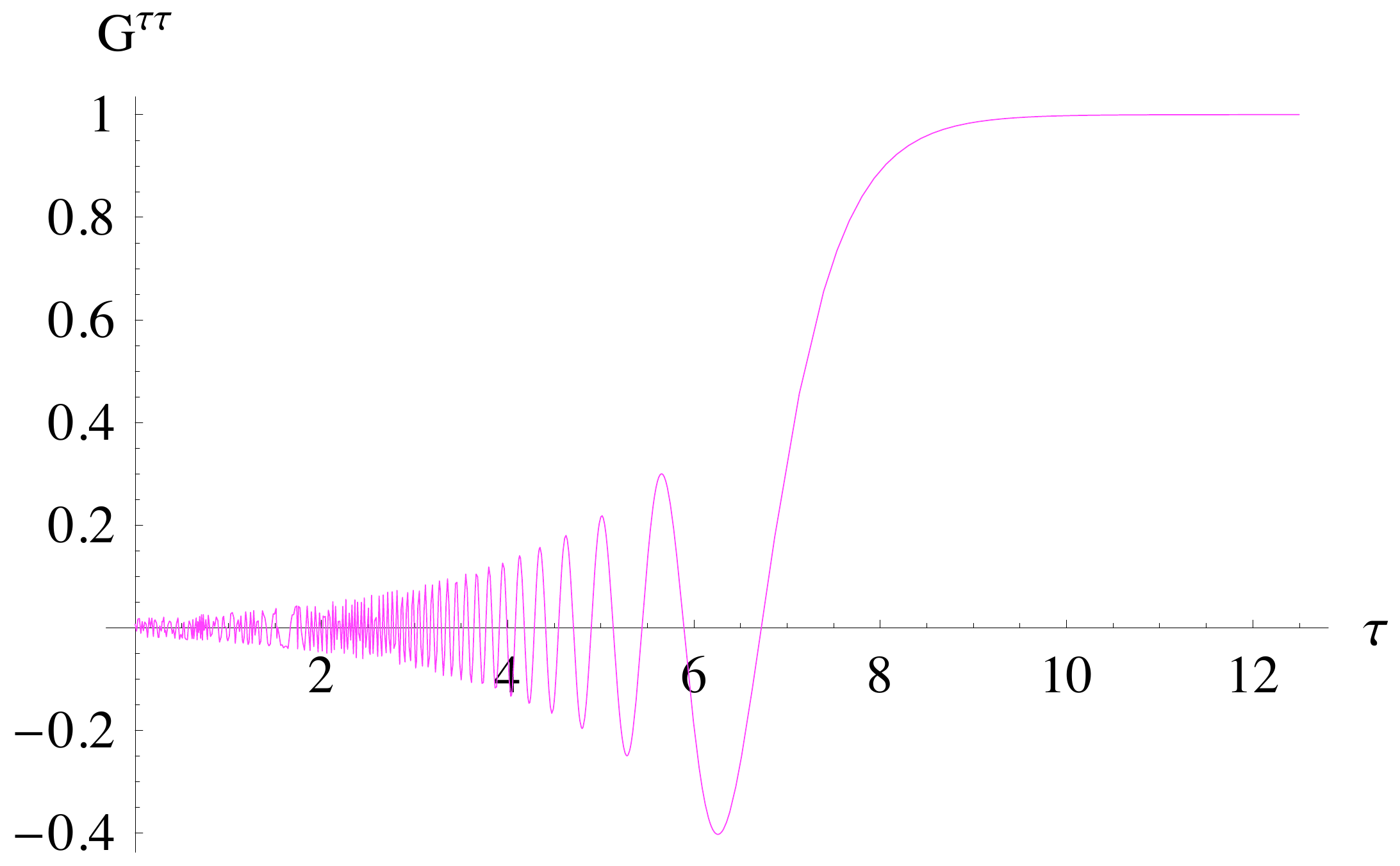} \hspace{0.3in}
\includegraphics[width=2.5in]{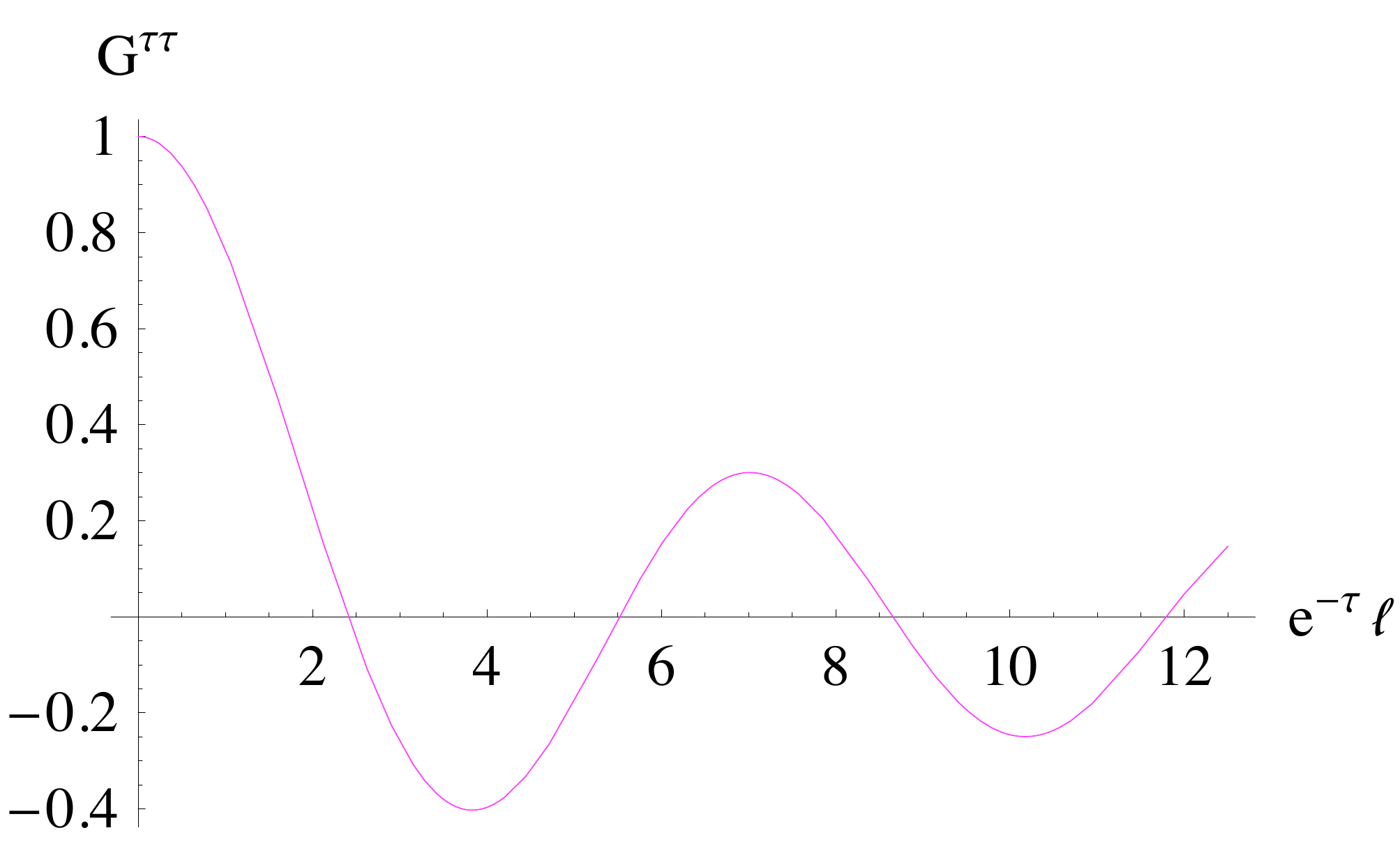}
\end{center}
\caption{\footnotesize The leading behaviour of correlators
$G_{\t\t}$ (and $G_{\t\o}$), up to normalization
constants: as a function of $\t$ on the left with $\ell=1000$ and, on
the right, as a function of $\ell e^{-\t}$ }\label{1}
\end{figure}

\begin{figure}[h]
\renewcommand{\figurename}{Fig.}
\begin{center}
\includegraphics[width=2.5in]{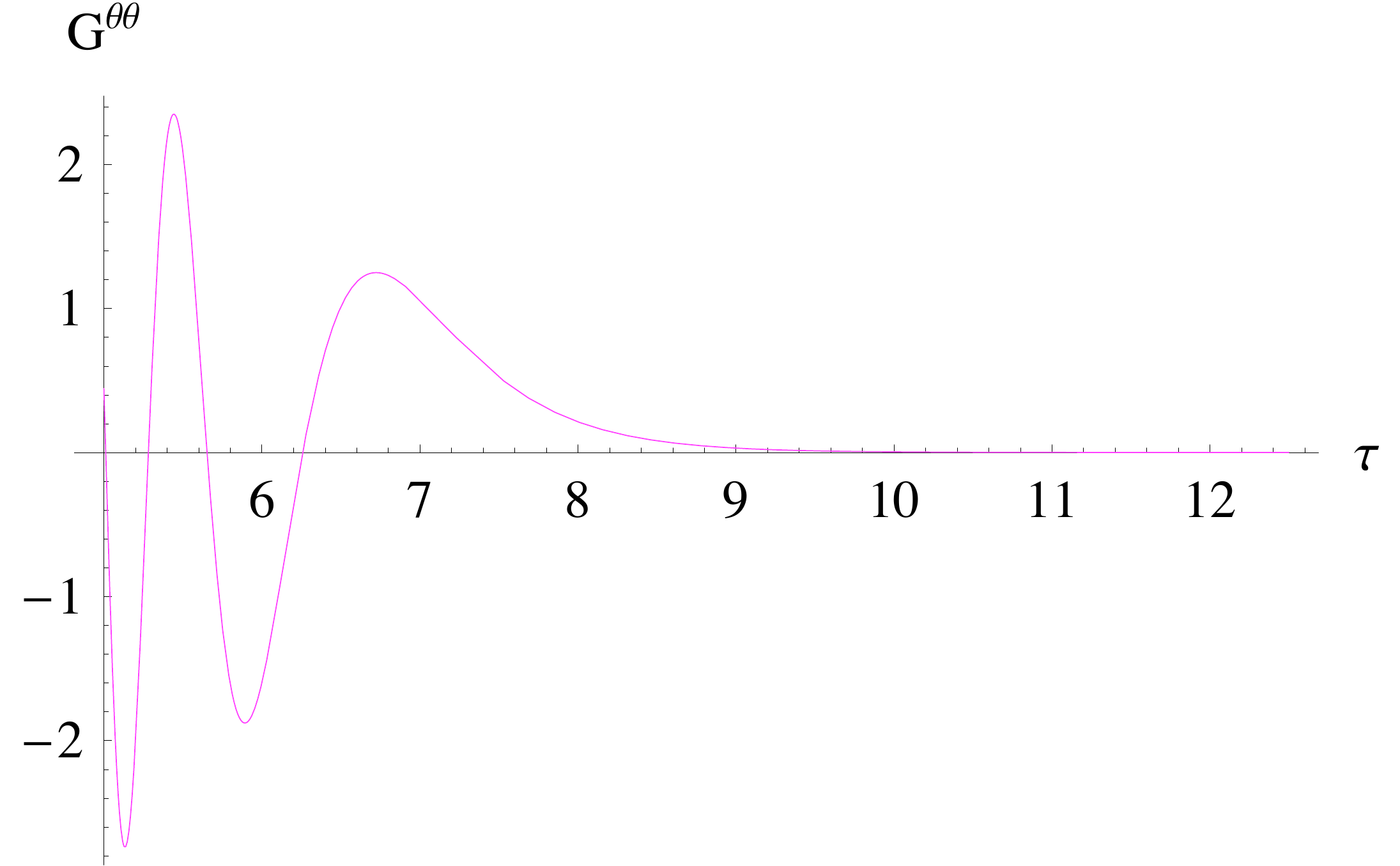}\hspace{0.3in}
\includegraphics[width=2.5in]{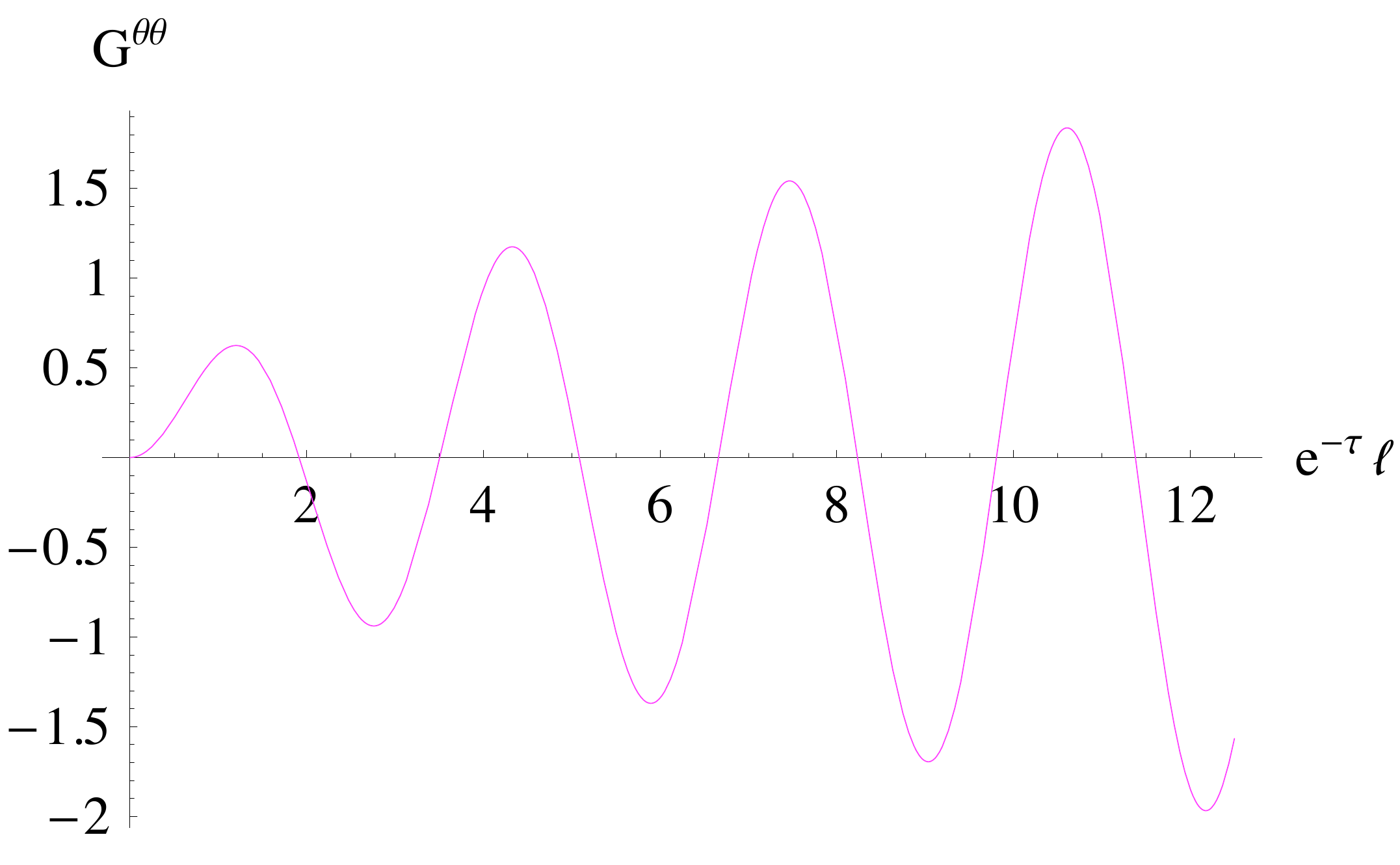}
\end{center}
\caption{\footnotesize The leading behaviour of 
$G_{\o\o}$ (and $G_{\o\t}$): The left figure is plotted as a function
of time for $\ell=1000$, while the right hand side figure is a
function of $\ell e^{-\t}$.\label{2}}  
\end{figure}

For fixed physical wavelengths, $\ell\, e^{-\tau}$, or equivalently, at
the time when a harmonic $\ell$ crosses the horizon, the 
real time correlators are given by the exact results with the replacement 
\eqref{bessel}. The integral over $\nu$ can be easily evaluated using
the method of residues, and it turns out that the leading contribution
at late times is from the residue at $\nu=0$. Thus,
\EQ
{
G^{\t\t}(\t,\,0;\ell),\,\,\, G^{\t\theta}(\t,\,0;\ell)\,\, 
\sim \,\,J_0(2\,\ell e^{-\tau} )
\label{late1}
}
and
\EQ
{
G^{\theta \t}(\t,0;\,\ell),\,\,\,G^{\theta\theta}(\t,0;\,\ell)\,\,\sim
\,\,\partial_\t J_0(2\,\ell e^{-\t}).
\label{late2}
}
This  late time behaviour is depicted in
Figs. (\ref{1}) and (\ref{2}). 

The late time behaviour deduced above is not characteristic of
diffusion in de Sitter space. Suppose that the R-charge fluctuation
relaxed via diffusion modes, then the covariant conservation of the
R-current together with Fick's law would lead to the diffusion equation
\EQ
{
\partial_\t j^\tau  =  D \nabla_\theta\nabla^\theta \,j^\tau,
}
in $dS_3$, $D$ being the diffusion constant. 
The spherical harmonics of $j^\tau$ on the expanding spatial
spherical sections would then obey
\EQ
{
\partial_\tau j^\t_\ell = - D\,{\ell(\ell+1)\over \cosh^2\t}\, j^\t_\ell.
}
At late times $\tau \to \infty$ and large enough $\ell$, this is
solved by
\EQ
{
j^\t_\ell \sim \exp\left(\tfrac{1}{2} D\, \ell^2\,e^{-2\tau}\right).
\label{diffusive}
}

The large time behaviour of the Green's functions \eqref{late1} and
\eqref{late2} do not match up with expected diffusive relaxation
\eqref{diffusive} on $dS_3$. A natural reason for this is that the
rate of exponential expansion of the spatial section and the
Gibbons-Hawking temperature are both set by $R_{\rm dS}^{-1}$. Thus, the mean
free path or the mean free time between collisions is comparable to
the expansion time scales so that the system never enters a
diffusive regime.

\begin{savequote}[15pc]
\sffamily
Give me ambiguity or give me something else.
\qauthor{Anonymous}
\end{savequote}
\chapter{Vacuum Ambiguity in de Sitter Space at Strong Coupling} \label{chapter:alpha}
 
In previous chapters, we saw that the Son-Starinets prescription automatically chose the boundary field theory to be in the Euclidean or Bunch-Davies vacuum. In this chapter, we will calculate the symmetric two-point functions at strong coupling and address the issue of vacuum ambiguity in de Sitter space at strong coupling.

This chapter  is organized as follows.  In Section \ref{alpha}, we review the issue of vacuum ambiguity in de Sitter space at the weak coupling regime. In Section \ref{calc}, we perform the holographic computation of the symmetric two-point functions as functions of the geodesic distance. 

\section{$\a$-vacua in de Sitter Space \label{alpha}}
As we reviewed in Section \ref{desitterrev}, the three-dimensional de Sitter space $dS_3$ can be realized as the hypersurface described by 
\EQ
{-x_0^2+x_1^2+x_2^2+x_3^2 = R_{\rm dS}^2,
}
with $R_{\rm dS}$ being the $dS_3$ radius. The metric for the global patch is given by
\EQ
{ds^2 = - dt^2 +R_{\rm dS}^2 \cosh^2 \left({t\over R_{\rm dS}}\right)  \;d\Omega_2^2 \label{dsmetric}.
}
For a more detailed discussions on the classical geometry of de Sitter space and its other aspects see for example \cite{Spradlin:2001pw}. 

This spacetime does not contain a globally time-like Killing vector and thus does not have a unique vacuum state. To see this in a more formal manner, let us consider the symmetric two-point function in a de Sitter invariant state $|\a\rangle$. For simplicity, we consider only a free massive scalar field theory here. For the treatment of weakly interacting theories, see \cite{Collins:2003mj} and references therein.

$|\a\rangle$ being de Sitter invariant implies that the symmetric two-point function $G_{\a}(x,y)$ can only depend on the two spacetime points $x$ and $y$ via the geodesic distance $d(x,y)$. The symmetric two-point function obeys the massive scalar field equation
\EQ
{\left(-\Box_x + m^2 \right) G(x,y) = 0,
}
which can be expressed as
\EQ
{\left(Z^2-1\right)^{-1/2} \partial_Z\left[\left(Z^2-1\right)^{3/2} \, \partial_Z G\left(Z(x,y)\right) \right] + m^2\, R_{\rm dS}^2\, G\left(Z(x,y)\right)=0, \label{dsode}
}
where we have introduced the real function
\EQ
{Z(x,y) = \cos{d(x,y) \over R_{\rm dS}}\, .
}
Here, the mass term may include the contribution that comes from coupling to the curvature.

A solution is given by
\EQ
{G_E\left(Z(x,y)\right) = {\G(h_+) \, \G(h_-) \over (4 \p)^{3/2} \, \G(3/2)} \,\, _2F_1\left(h_+,h_-;{3 \over 2};{1+Z(x,y) \over 2}\right),
}
with
\EQ
{h_{\pm} = 1 \pm \sqrt{1-m^2 R_{\rm dS}^2} \,\, .
}
Here, the normalization is chosen such that the short distance singularity matches to that of Minkowski space.

We note that if $G_E(Z)$ is a solution to Eq. (\ref{dsode}), then $G_E(-Z)$ is also a solution. Since for $m^2>0$ these two solutions are linearly independent, then the general solution to Eq. (\ref{dsode}) can be expressed as
\EQ
{G_{\a}(x,y)= {1+e^{\a+\a^*} \over 1-e^{\a+\a^*}} \, G_E\left(Z(x,y)\right) + {e^{\a}+e^{\a^*} \over 1-e^{\a+\a^*}} \, G_E\left(-Z(x,y)\right). \label{alphaweak}
}
The constants in the $\a$-vacuum propagator (\ref{alphaweak}) are chosen such that the modes of the free massive scalar field $\Phi$ in the $\a$-vacuum can be related to those in the Bunch-Davies vacuum by the Bogoliubov transformation
\EQ
{\Phi_n^{\a} = {\Phi_n^E + e^{\a}\, \left(\Phi_n^E\right)^* \over \sqrt{1 - e^{\a+\a^*}}}\,.
}
The short-distance singularity of the $\a$-vacuum propagator (\ref{alphaweak})  is related to the singularity of the Bunch-Davies vacuum propagator by a factor $\left(1+e^{\a+\a^*}\right)/\left(1-e^{\a+\a^*}\right)$. 

Let us also mention a peculiar, but widely used, interpretation of the $\a$-vacuum propagator (\ref{alphaweak}). For a point $x=(t,\theta,\phi)$, we can introduce its antipodal point $\bar{x}=(-t,\theta+\p,\phi+\p)$. It is easy to show that $Z(x,\bar{y})=-Z(x,y)$. Therefore, the $\a$-vacuum propagator (\ref{alphaweak}) can be rewritten as
\SP
{G_{\a}(x,y) = &\,\,  {1 \over 1-e^{\a+\a^*}} \, G_E\left(Z(x,y)\right) + {e^{\a+\a^*} \over 1-e^{\a+\a^*}} \, G_E\left(Z(\bar{x},\bar{y})\right) \\
& + {e^{\a} \over 1-e^{\a+\a^*}} \, G_E\left(Z(\bar{x},y)\right)+ {e^{\a^*} \over 1-e^{\a+\a^*}} \, G_E\left(Z(x,\bar{y})\right). \label{antipode}
}
The interpretation is then that the extra contributions in the $\a$-vacuum propagator (\ref{alphaweak}) can be thought of as arising from an image source at the antipodal point. We would like to emphasize that even though this is an interesting interpretation, the existence of vacuum ambiguity in de Sitter space does not depend on this interpretation concerning image source at the antipodal point. 

Finally, let us remark that there is no de Sitter invariant vacuum in the case of massless minimally coupled scalar fields \cite{Allen:1985ux}. In the case of interest, since the ${\cal N}=4$ Super Yang-Mills is conformally coupled to the de Sitter background ({\it cf.} \cite{Hollowood:2006xb}), there is an infinite class of de Sitter invariant vacua at weak coupling and, as we will see in Section \ref{calc}, this vacuum ambiguity persists at strong coupling.

\section{The Symmetric Two-point Functions at Strong Coupling \label{calc}}

We would like to calculate the symmetric two-point function of a scalar operator ${\cal O}$ of strongly coupled ${\cal N}=4$ Super Yang-Mills in the ${\mathbb Z}_N$-invariant phase, living in $dS_3\times S^1$. This operator is dual to a scalar field $\Phi$ living in the topological $AdS_5$ black hole, with the mass $M$ of $\Phi$ is related to the dimension $\D$ of the operator ${\cal O}$ by
\EQ
{\D = 2 + \sqrt{4 + M^2 \, R_{\rm AdS}^2}.
}
To simplify notation, we will henceforth set $R_{\rm AdS}=1$. This also means that $R_{\rm dS}=1$, as the de Sitter radius of the boundary spacetime is equal to the AdS radius of the bulk. We will calculate the correlators, by first solving the equation of motion of $\Phi$ in the background (\ref{in z}), with the relevant boundary condition given by $\Phi(z,x)\big|_{\rm boundary}=\Phi_0(x)$. This equation of motion is given by
\EQ
{z^5 \partial_z \left({1+z^2 \over z^3} \, \partial_z \Phi \right) + {z^2 \over 1+ z^2} \, {1 \over r_{\chi}^2} \, \partial_{\chi}^2 \Phi + z^2 \, \Box_{dS_3} \Phi - M^2 \, \Phi = 0.
}
Since we are mainly interested in the vacuum structure of the field theory in de Sitter space, we will consider no $\chi$-dependence in $\Phi$. This is done for simplicity reason, but the generalization is straightforward. 

The solution to the equation of motion will then be of the form
\EQ
{\Phi(z,x) = \int dy \int dk^2 \, a_{k^2} \, R_{k^2}(z) \, F_{k^2}(x,y) \, \Phi_0(y), \label{ansatz}
}
such that 
\EQ
{\Box_{x \in dS_3} \, F_{k^2}(x,y) = k^2 F_{k^2}(x,y), \label{dssplit}
}
and
\EQ
{z^5 \partial_z \left({1+z^2 \over z^3} \, \partial_z R_{k^2} \right) + z^2 \, k^2 \, R_{k^2} - M^2 \,R_{k^2} = 0. \label{radialsplit}
}
The radial equation of motion has two solutions
\EQ
{R_{k^2}^- = z^{4-\D} \, _2F_1\left({3-\D \over 2}-{\sqrt{1-k^2} \over 2}, {3-\D \over 2}+{\sqrt{1-k^2} \over 2}; 3-\D; -z^2\right),
}
and
\EQ
{R_{k^2}^+ = z^{\D} \, _2F_1\left({\D-1 \over 2}-{\sqrt{1-k^2} \over 2}, {\D-1 \over 2}+{\sqrt{1-k^2} \over 2}; \D-1; -z^2\right).
}
As one goes to the boundary $z=\e \ll 1$, the two solutions behave as $R_{k^2}^- \rightarrow \e^{4-\D}$ and $R_{k^2}^+ \rightarrow \e^{\D}$. Therefore, the most general solution is given by
\EQ
{\Phi(z,x) = \int dy \int dk^2 \, a_{k^2} \, {R_{k^2}^-(z) + \b_{k^2} \, R_{k^2}^+(z) \over R_{k^2}^- (\e)+ \b_{k^2} \, R_{k^2}^+(\e)}  \, F_{k^2}(x,y) \, \Phi_0(y), \label{sol1}
}
with the condition
\EQ
{\Phi_0(x) = \int dy \int dk^2 \, a_{k^2} \, F_{k^2}(x,y) \, \Phi_0(y).
}
We note that we could also add the term
\EQ
{\d \Phi(z,x) =  \int dy \int dk^2 \,  \gamma_{k^2} \, R_{k^2}(z)^+   \, F_{k^2}(x,y)
}
to the solution (\ref{sol1}) while still obeying the boundary conditions. However, as this term corresponds only to giving the boundary operator ${\cal O}$ non-trivial expectation value \cite{Balasubramanian:1998de}, we can omit it by requiring $\langle {\cal O} \rangle = 0$.

For explicitness, in the following calculation we will restrict ourselves to the scalar glueball operators 
\EQ
{{\cal O}= \Tr \, F_{\m\n} F^{\m\n} \qquad {\rm and} \qquad {\cal O} = \Tr \, F_{\m\n} \tilde{F}^{\m\n}. 
}
As we mentioned in earlier chapters, these operators are dual to the dilaton and the RR-axion, respectively, in the Type IIB Superstring theory on the bulk. The scalar glueball operators are of dimension 4, while both of the bulk fields are massless. The calculations for other operators will be similar.

In this case, the general solution to the radial equation of motion becomes
\EQ
{R_{k^2}(z) = R_{k^2}^- + \b_{k^2} \, R_{k^2}^+\, ,
}
with 
\EQ
{R_{k^2}^- = {1 \over \G({1 - \sqrt{1-k^2} \over 2}) \G({3 + \sqrt{1-k^2} \over 2})}\, _2F_1\left(-{1 \over 2}-{\sqrt{1-k^2} \over 2}, -{1 \over 2}+{\sqrt{1-k^2} \over 2}; 1; z^2+1\right),
}
and
\EQ
{R_{k^2}^+ = z^{4} \,\, _2F_1\left({3 \over 2}-{\sqrt{1-k^2} \over 2}, {3 \over 2}+{\sqrt{1-k^2} \over 2}; 3; -z^2\right).
}
Here, we have normalized this solution such that $R_{k^2}(0) = 1$.

Let us now turn to the part of the equation of motion that concerns the de Sitter factor of bulk spacetime (\ref{dssplit}). Since our goal is to obtain symmetric two-point functions in de Sitter invariant vacuum, $F_{k^2}(x,y)$ must be a function of $Z(x,y)$ and satisfies 
\EQ
{\left(Z^2-1\right)^{-1/2} \partial_Z\left[\left(Z^2-1\right)^{3/2} \, \partial_Z F_{k^2}\left(Z(x,y)\right)  \right] + k^2\, F_{k^2}\left(Z(x,y)\right) =0. \label{ds}
}
For $k^2\leq 1$, we can rewrite $\sqrt{1 - k^2} = \ell$ and we have a solution
\EQ
{F_{\ell}\left(Z\right) =  \left(Z^2-1\right)^{-{1 \over 4}} \, P_{\ell-{1 \over 2}}^{1 \over 2}\left(Z\right),
}
which, for non-negative $\ell \in {\mathbb Z}$, forms a set of complete orthonormal functions in the interval $Z \in [-1,1]$. This interval $Z \in [-1,1]$ corresponds to the point $x$ and $y$ having space-like separation. The orthogonality condition is given by
\EQ
{\int_{-1}^1 dZ \, P_{\ell-{1 \over 2}}^{1 \over 2}\left(Z\right) \, \left(P_{m-{1 \over 2}}^{1 \over 2}\left(Z\right) \right)^*= \d_{\ell,m}.
}

For $k^2\geq 1$, we can rewrite $\sqrt{1 - k^2} = i \n$ and the solution
\EQ
{F_{\n}\left(Z\right) =  \left(Z^2-1\right)^{-{1 \over 4}} \, P_{-{1 \over 2}+i \n}^{1 \over 2}\left(Z\right)
}
forms a set of complete orthonormal functions in the interval $Z \in [1,\infty)$ for non-negative $\n \in {\mathbb R}$. This interval $Z \in [1,\infty)$ corresponds to time-like separation between $x$ and $y$. The orthogonality condition is given by
\EQ
{\int_{1}^{\infty} dZ \, P_{-{1 \over 2}+i \n}^{1 \over 2}\left(Z\right) \, \left(P_{-{1 \over 2}+i \m}^{1 \over 2}\left(Z\right) \right)^*= \d(\n-\m).
}
Similarly, for $Z \in (-\infty,-1]$, the solution
\EQ
{F_{\n}\left(Z\right) =  \left(Z^2-1\right)^{-{1 \over 4}} \, P_{-{1 \over 2}+i \n}^{1 \over 2}\left(-Z\right)
}
forms a set of complete orthonormal functions for non-negative $\n \in {\mathbb R}$. This interval corresponds to the antipodal point of $y$ being time-like separated from $x$.

Hence, we find the solution to the equation of motion for the massless fields in the bulk to be
\SP
{\Phi(z,x) = & \int \limits_{\tiny \begin{array}{cc} {\rm spacelike}\\{\rm separation} \end{array}}dy \, \sum_{\ell \in {\mathbb Z}^+}  \bigg(R_{1-\ell^2}^-(z) + \b_{\ell} \, R_{1-\ell^2}^+ (z)\bigg) \, \sqrt{2 \over \pi} \,\,{ P_{\ell-{1 \over 2}}^{1 \over 2}\big(Z(x,y)\big) \over \sqrt[4]{{Z(x,y)}^2-1} } \,\,\, \Phi_0(y) \\
\\
& + \int \limits_{\tiny \begin{array}{cc} {\rm timelike}\\{\rm separation} \end{array}}dy \,\int_{0}^{\infty}  d\n\bigg(R_{1+\n^2}^- (z)+ \b_{\n} \, R_{1+\n^2}^+ (z)\bigg) \, \sqrt{2 \over \pi} \,\,{ P_{-{1 \over 2}+i \n}^{1 \over 2}\big(Z(x,y)\big) \over \sqrt[4]{{Z(x,y)}^2-1} } \,\,\, \Phi_0(y). 
}
Substituting this to the action, we can read the symmetric two-point function.

For points $x$ and $y$ that have a time-like separation, the two-point function is given by
\EQ
{G_{{\bar{\b}}}(x,y) = {N^2 \over 2^{7/2} \, \pi^{5/2}} \,\int_{0}^{\infty}  d\n \,\, \Bigg[{z^2+1\over z^3} \,\, \partial_z \bigg(R_{1+\n^2}^- (z)+  \b_{\n} \, R_{1+\n^2}^+ (z)\bigg)\Bigg]_{z=\e} {P_{-{1 \over 2}+i \n}^{1 \over 2}\big(Z(x,y)\big) \over \sqrt[4]{{Z(x,y)}^2-1}} \, .
}
There is the ambiguity $\bar{\b} = \{\b_{\n}\}$ as for any given $\b_{\n}$, the radial wave function $R_{1+\n^2}=R_{1+\n^2}^- + \b_{\n} \, R_{1+\n^2}^+$ is normalizable at the horizon. This ambiguity implies that the vacuum ambiguity in de Sitter space persists at strong coupling. As the set $\{\b_{\n}\}$ parametrizes the two-point function and thus the vacua at strong coupling, we will dub this infinite class of vacua as $\{\b_{\n}\}$-vacua.

As in earlier chapters, the propagator for the Bunch-Davies vacuum can be obtained by requiring the radial wave function to oscillate as $R_{1+\n^2}\propto z^{-i \n}$ near the horizon, {\it i.e.}, no term that oscillates as $z^{i \n}$. Such radial wave function is given by
\EQ
{R^E_{1+\n^2}=R_{1+\n^2}^- + \b^E_{\n} \, R_{1+\n^2}^+\, ,
}
with
\EQ
{\b^E_{\n} = i {\pi\over 32}\;e^{{\pi\over 2}\nu}\;{(\nu^2+1)^2\over\cosh{{\pi\over 2}\nu}} \, . \label{be}
}
Therefore, the symmetric two-point function of the Bunch-Davies vacuum is
\SP
{G_{E}(x,y) = \, & {N^2 \over 2^{7/2} \, \pi^{5/2}} \,\int_{0}^{\infty}  d\n \,\, \Bigg[-\frac{{(1+\nu^2)}^2}{8} \Bigg(\psi\left({3-i\nu\over2}\right)+ \psi\left({3+i\nu\over2}\right)- i\pi \coth {\pi (\nu+i) \over 2} \Bigg) \\
& +\frac{{(1+\nu^2)}^2}{4}\;\left(\ln z -\gamma_E+1\right)\bigg|_{z \rightarrow \infty}+ \frac{(1+\nu^2){z^2}}{2}\bigg|_{z \rightarrow \infty}\,\, \Bigg] \, \,\, {P_{-{1 \over 2}+i \n}^{1 \over 2}\big(Z(x,y)\big) \over \sqrt[4]{{Z(x,y)}^2-1}} \, .
}
The divergent and scheme-dependent contact terms can be minimally subtracted away to yield the renormalized two-point function. We get
\EQ
{G_{E}(x,y) =  {N^2 \over 2^{11/2} \, \pi^{5/2}} \,\int_{0}^{\infty}  d\n \, {(1+\nu^2)}^2 \;\left[\psi\left({3 -i\nu \over2}\right)-{2i\nu  \over 1+\nu^2 }\right] \,\, {P_{-{1 \over 2}+i \n}^{1 \over 2}\big(Z(x,y)\big) \over \sqrt[4]{{Z(x,y)}^2-1}} \, .
}

Similarly, for two points $x$ and $y$ that have space-like separation, the symmetric two-point function is given by 
\EQ
{G(x,y) =  {N^2 \over 2^{11/2} \, \pi^{5/2}} \, \sum_{\ell} \, {(1-\ell^2)}^2 \;\left[\psi\left({3 + \ell \over2}\right)+{2 \ell \over1- \ell^2}\right] \,\, {P_{\ell-{1 \over 2}}^{1 \over 2}\big(Z(x,y)\big) \over \sqrt[4]{{Z(x,y)}^2-1}} \, .
}
We note that
\EQ
{G_{E}(x,y; \n) =  {N^2 \over 2^{11/2} \, \pi^{5/2}} \, {(1+\nu^2)}^2 \;\left[\psi\left({3 -i\nu \over2}\right)-{2i\nu  \over 1+\nu^2 }\right] \,\, {P_{-{1 \over 2}+i \n}^{1 \over 2}\big(Z(x,y)\big) \over \sqrt[4]{{Z(x,y)}^2-1}}  \label{BDmode}
}
is none other than the analytic continuation of 
\EQ
{G(x,y;\ell) =  {N^2 \over 2^{11/2} \, \pi^{5/2}}  \, {(1-\ell^2)}^2 \;\left[\psi\left({3 + \ell \over2}\right)+{2 \ell \over1- \ell^2}\right] \,\, {P_{\ell-{1 \over 2}}^{1 \over 2}\big(Z(x,y)\big) \over \sqrt[4]{{Z(x,y)}^2-1}} \, ,
}
where we have analytically continued the ``frequency" $\n = i \ell$ and used the fact $P_{\ell-{1 \over 2}}^{1 \over 2} (Z) = P_{-\ell-{1 \over 2}}^{1 \over 2}(Z)$. 

Unlike the case of the two-point function for time-like separated points, the two-point functions for points that have space-like separation do not have any ambiguity. This is due to the fact that in this case, there is only one linear combination of radial wave function that is normalizable at the horizon.

Going back to the case of points with time-like separation, the two-point function of a $\{\b_{\n}\}$-vacuum can be expressed in terms of the Bunch-Davies two-point function (\ref{BDmode}). First, let us denote the radial wave function that oscillates as $R_{1+\n^2}\propto z^{i \n}$ near the horizon as
\EQ
{R^{\tilde{E}}_{1+\n^2}=R_{1+\n^2}^- + \b^E_{-\n} \, R_{1+\n^2}^+\, ,
}
with $\b^E_{-\n}$ can be read from Eq. (\ref{be}). Then, we can express the parameter $\b_{\n}$ as
\EQ
{\b_{\n} = {\b_{\n}^E + \a_{\n} \, \b_{-\n}^E \over 1 + \a_{\n}}\, ,
}
such that the symmetric two-point function is now given by
\EQ
{G_{\bar{\b}}(x,y) =  \int_{0}^{\infty}  d\n \,\left({1 \over 1 + \a_{\n}} \, G_{E}(x,y; \n) + {\a_{\n} \over 1 + \a_{\n}} \, {G_{E}(x,y; \n)}^*\right)\, , \label{galpha}
}
where $G_{E}(x,y; \n)$ is given in Eq. (\ref{BDmode}). We denote this two-point function as $G_{\a}(x,y)$, and instead of using $\{\b_{\n}\}$ to parametrize the infinite family of vacua, we will use $\{\a_{\n}\}$.

The two-point function of the $\{\a_{\n}\}$-vacuum has a singularity at $Z(x,y)=1$, {\it i.e.}, when the points are very close to each other or when they have a light-like separation. At short distance $d(x,y) \rightarrow 0$, the two-point function in an $\{\a_{\n}\}$-vacuum behaves like
\EQ
{G_{\bar{\a}}(x,y) = {N^2 \over {(2 \pi)}^3}  \, \int_{0}^{\infty}  d\n \,\left({\psi\left({3 -i\nu \over2}\right) + \a_{\n} \, \psi\left({3 +i\nu \over2}\right) \over 1 + \a_{\n}} - {1 -  \a_{\n} \over 1 +  \a_{\n}} \, {2i\nu  \over 1+\nu^2 }\right) \, \, \, {1 \over \big|d(x,y)\big|}\,.
}
Therefore, the two-point function of the $\{\a_{\n}\}$-vacuum differs from that of the Bunch-Davies vacuum by a factor
\EQ
{f_{\bar{\a}} = \frac{\int_{0}^{\infty}  d\n \,\left({\psi\left({3 -i\nu \over2}\right) + \a_{\n} \, \psi\left({3 +i\nu \over2}\right) \over 1 + \a_{\n}} - {1 -  \a_{\n} \over 1 +  \a_{\n}} \, {2i\nu  \over 1+\nu^2 }\right) }{\int_{0}^{\infty}  d\n \,\left({\psi\left({3 -i\nu \over2}\right) } - {2i\nu  \over 1+\nu^2 }\right) } \,.
}
It is tempting to identify an $\{\a_{\n}\}$-vacuum with an $\a$-vacuum of the weakly coupled theory by setting
\EQ
{f_{\bar{\a}} = {1+e^{\a+\a^*} \over 1-e^{\a+\a^*}} \, . \label{singular}
}
However, for a given $\a$, there is not a unique solution $\{\a_{\n}\}$ for this equation. Furthermore, for $\{\a'_{\n}\} \ne \{\a_{\n}\}$, but with $f_{\bar{\a}'} = f_{\bar{\a}}$, the two-point functions are not identical. Therefore, this identification is not valid.

\begin{savequote}[8pc]
\sffamily
D'oh!
\qauthor{Homer Simpson}
\end{savequote}
\chapter{Summary and Discussions} \label{chapter:disc1}

In this first part of the thesis, we have studied real time correlators in strongly
coupled ${\cal N}=4$ supersymmetric Yang-Mills theory on a time-dependent background,
namely $dS_3\times S^1$. 

In Chapter \ref{chapter:phases}, we have calculated the retarded scalar glueball correlators in the $\mathbb{Z}_N$-invariant phase. The retarded scalar glueball correlators have an infinite number of
poles in the lower half of the complex frequency plane, which
represent the topological black hole quasinormal frequencies. The
imaginary parts of these correlators are associated to the
Gibbons-Hawking temperature due to the cosmological horizon of
$dS_3$. These two facts suggest that the $\mathbb{Z}_N$ symmetric
phase of the boundary field theory corresponds to a deconfined plasma
in the exponentially expanding universe. 

Furthermore, in Chapter \ref{chapter:phases}, we have also calculated the retarded correlators of
scalar operators ${\cal O}_{\Delta}$, with conformal dimension $\Delta
\gg 1$, in both the $\mathbb{Z}_N$-invariant phase and $\mathbb{Z}_N$
broken bubble phase. Unlike the correlators of the $\mathbb{Z}_N$ symmetric
phase, the correlators in the $\mathbb{Z}_N$ broken phase feature an
infinite number of poles on the real frequency axis. These poles are
naturally associated
to bound glueball-like states, which suggests that this phase is a hadronized
phase, where the de Sitter temperature is too low to deconfine the
degrees of freedom (the Hubble parameter
is low compared to the dynamical scale of the effective, non-SUSY 3d theory
on $dS_3$). Since this geometry contains no horizon, Son-Starinets prescription \cite{Son:2002sd} is not applicable in this phase and we have restricted ourselves to the high frequency and large mass regime by using WKB approximation. Since the prescription proposed by Skenderis and van Rees \cite{Skenderis:2008dh,Skenderis:2008dg} does not rely on the existence of horizons in the geometry, it would be interesting to see whether one can obtain the retarded correlators beyond the high frequency limit using this prescription.

In Chapter \ref{chapter:plasma}, we have computed the retarded correlators for the spatial spherical
harmonics of the conserved R-currents using the Son-Starinets
approach. Here we encountered a subtle point wherein we had to include
in our mode expansions, the effects of 
real, normalizable, discrete solutions to the de Sitter mode
equations, in order to obtained a retarded Green's function. The
corresponding 
frequency space correlator, appropriately defined in de Sitter space,
also has an infinite number of
poles in the lower half of the complex frequency plane, but these do
not appear to correspond to diffusive poles. The lack of hydrodynamic behavior
of the system is presumably due to the fact that the expansion rate of
$dS_3$ is of the same order as the Gibbons-Hawking temperature. Here,
we did not calculate the correlators of the stress-energy
tensor. However, using the same argument as above, we expect not to
find any hydrodynamic poles there either.

It would be interesting to study further the properties of this strongly coupled plasma on $dS_3 \times S^1$ by introducing ``quark" degrees of freedom, which are in the fundamental representation of the $SU(N)$ gauge group. We can do so by first introducing a stack of $N_f$ D7-branes that shares the $3+1$ dimensions with the stack of $N$ D3-branes. For $N_f \ll N$, the gravity side of the correspondence is described by $N_f$ D7-branes probing the AdS topological black hole geometry. The static configurations for this system can be obtained by analytical continuation of the configuration for $N_f$ D7-branes probing the thermal AdS space found in \cite{Karch:2006bv}. The quark then corresponds to the endpoint of open string on the D7-branes. Therefore, the dynamics of the open string corresponds to the dynamics of the quark in the strongly coupled plasma \cite{Herzog:2006gh}. A work toward understanding the properties of the strongly coupled plasma on $dS_3 \times S^1$ using the quark as a test particle in in progress \cite{future}.


It is interesting to note that the relevant boundary condition on the horizon as prescribed by Son and Starinets \cite{Son:2002sd} that we have used in Chapters \ref{chapter:phases} and \ref{chapter:plasma} implies that the boundary theory is in the Euclidean or Bunch-Davies vacuum. In order to study the vacuum structure of the boundary field theory, in Chapter \ref{chapter:alpha}, using gauge/gravity correspondence, we have calculated the symmetric two-point functions of the scalar glueball operators of strongly coupled ${\cal N}=4$ Super Yang-Mills living on three-dimensional de Sitter space (times a circle). 

The two-point functions for points with space-like separation consist of a sum of functions parametrized by discrete ``frequency," while the two-point functions for points with time-like separation are described by an integral over a continuous spectrum of ``frequency." This is the explanation for the subtlety involving discrete normalizable mode functions in de Sitter space found in Chapter \ref{chapter:plasma}. The statement is that to obtain the retarded propagators of the inhomogeneous perturbations, one has to include the contributions from these discrete normalizable mode functions (see  Eqs. \ref{eq:subt1} and \ref{eq:subt2}), and the reason is because they correspond to contributions from points with space-like separation.

For points that are space-like separated, the two-point function is unique, but for those that have time-like separation, there is an infinite family of two-point functions $G_{\a_{\n}}(x,y)$, which is given by Eqs. (\ref{galpha}) and (\ref{BDmode}). This ambiguity arises as the consequence of the existence of infinite family of bulk radial wave functions that are normalizable at the horizon \cite{Balasubramanian:1998de}. This ambiguity implies that there is an infinite class of de Sitter invariant vacua for the strongly coupled theory on de Sitter space, which we have parametrized with a set of complex parameters $\{\a_{\n}\}$. The short-distance singularity of an $\{\a_{\n}\}$-vacuum differs from that of the Bunch-Davies vacuum by a factor $f_{\bar{\a}}$, which is defined in Eq. (\ref{singular}). As the short-distance behavior of the propagator of an $\a$-vacuum in the weakly coupled theory differs from the short-distance behavior of the propagator of the Bunch-Davies vacuum by a factor $\left(1+e^{\a+\a^*}\right)/\left(1-e^{\a+\a^*}\right)$, it is tempting to identify the $\{\a_{\n}\}$-vacuum at the strong coupling with the $\a$-vacuum at the weak coupling by setting these two factors to be equal. However, as for a given $\a$, there is not a unique set of $\{\a_{\n}\}$ that satisfies this condition, and as for $\{\a'_{\n}\} \ne \{\a_{\n}\}$, $G_{\a'_{\n}}(x,y) \ne G_{\a_{\n}}(x,y)$ even though $f_{\bar{\a}'} = f_{\bar{\a}}$, such identification is not valid. Since we do not have the tool to analyze the theory at intermediate coupling, we do not know how an $\a$-vacuum evolves into an $\{\a_{\n}\}$-vacuum as we go from the weak coupling regime to the strong coupling one. One possible explanation is that as one increases the coupling of the theory, there is a mixing between the $\a$-vacua. We note that this explanation is not contrary to the statement of the volume independence of the correlation functions of large $N$ gauge theories in the $\mathbb{Z}_N$ symmetric phase. This statement implies that for our present case, the correlation functions of perturbations which are homogeneous along the circle should be independent of $r_\x$, which means that the results should match with the result on $dS_3\times {\mathbb R}$. The latter is obtained
by a (double) Wick rotation of $S^3\times {\mathbb R}$.
Since $\Tr F^2$ is a chiral primary in the ${\cal N}=4$ 
theory and its correlator on $S^3\times {\mathbb R}$ is not renormalized
by interactions, one would expect this to be true also for Bunch-Davies correlators on $dS_3\times
{\mathbb R}$. However, one should not expect correlators for a general $\a$-vacuum to be non-renormalizable. Following the same logic, the transition amplitude between two different vacua should not be protected either, and we can expect it to deviate from its free or weak coupling value, which is zero.

In Section \ref{alpha}, we have mentioned that at weak coupling, there is a widely used interpretation, in which the extra contributions in the $\a$-vacuum propagator (\ref{alphaweak}) are thought of as arising from an image source at the antipodal point. At strong coupling, however, such interpretation does not arise. There is no natural way to express the symmetric two-point function in an $\{\a_{\n}\}$-vacuum (\ref{galpha}) in terms of the antipodal points. This suggests that the interpretation using an image source at the antipodal point is not the most general language that one can use to describe the vacuum structure of field theories in de Sitter space.

Let us also briefly comment on the connection between vacuum ambiguity in the bulk and vacuum ambiguity of the field theory on the boundary. Since the bulk spacetime, which is the topological AdS black hole, is a time-dependent geometry, one expects that there is also vacuum ambiguity in the bulk. However, all bulk vacua, with the exception of the bulk Euclidean vacuum, feature extra singularities that render them unphysical \cite{Ross:2004cb}. Ref. \cite{Ross:2004cb} then suggests that this might mean that there should be no vacuum ambiguity in the strongly coupled dual field theory in the de Sitter boundary either. Here, after explicitly constructing the symmetric two-point functions of the strongly coupled boundary field theory, we do not find any extra singularities that will deem a generic $\{\a_{\n}\}$-vacuum in the boundary unphysical. We would like to emphasize that this lack of extra singularities is only at the level of the two-point functions. As the higher $n$-point functions are more sensitive to the properties of the bulk spacetime (see for example \cite{Polchinski:1999ry,Gary:2009ae} and references therein), it is certainly worth studying the higher $n$-point functions of the strongly coupled ${\cal N}=4$ Super Yang-Mills on $dS_3 \times S^1$ and see whether the higher $n$-point functions of a generic $\{\a_{\n}\}$-vacuum feature any extra singularities that will render such a vacuum unphysical.

Lastly, as the de Sitter boundary is a three-dimensional de Sitter space, let us mention an odd property of odd dimensional de Sitter space, namely that at the level of free theory, there exists a vacuum with no particle production \cite{Bousso:2001mw}. It would be interesting to understand what the fate of this vacuum is as one goes to the strong coupling regime.

In this part of the thesis, we have used the gauge/gravity duality to understand the strongly coupled field theory on de Sitter space better. However, it would also be interesting to use the correspondence to understand the quantum gravity of anti de Sitter spacetimes with de Sitter boundary, such as the issues concerning:
\begin{itemize}
\item the nature and the interpretation of the horizon in the AdS topological black hole (\cite{Balasubramanian:2005bg} vs. \cite{Ross:2004cb});
\item the dynamics of the AdS topological black hole decaying into the AdS bubble of nothing.
\end{itemize}
In particular, the latter might provide us with some insights into understanding the quantum gravity of geometries with semi-classical instability in general. 

\part{Rolling Tachyon Backgrounds and Dyson Gases}

\begin{savequote}[15pc]
\sffamily
Time is nature's way of keeping everything from happening at once.
Space is what prevents everything from happening to me.
\qauthor{John Wheeler (1911 - 2008)}
\end{savequote}
\chapter{Introduction}
The behaviors of quantum fields in curved spacetime give hints that, as thermodynamic quantities, spacetime might be an emergent entity, emerging as an effective description of a system that does not contain it. String theory gives further motivations for this idea (for a review see \cite{Seiberg:2006wf}). Not only that, there have been a number of examples in which several dimensions of space emerge \cite{Klebanov:1991qa, Taylor:2001vb, Aharony:1999ti}.

What about the emergence of time? In a theory of quantum gravity, ``time'' should also be an emergent concept -- there should be an underlying formalism which does not explicitly include a timelike coordinate. This suggests that the fundamental system is a Eucledian or a statistical system. As a step toward understanding the emergence of time, we can consider a (heuristic) analogy with AdS/CFT correspondence. If we think of AdS/CFT as emergence of space from a configuration of branes that are localized in space, then a candidate setting for investigating emergence of time will be branes that are localized in \emph{time} \cite{Gutperle:2002ai}. This is provided by the unstable branes of bosonic and superstring theory \cite{Sen:2002nu, Sen:2002in, Larsen:2002wc} (for a review, see \cite{Sen:2004nf}). See also the initial proposal by Sen, where emergent time is viewed as the tachyon field that arises in the effective field theory description of the rolling tachyon \cite{Sen:2002qa}.

As we mentioned in Chapter \ref{overview}, K-theory classifications put the unstable D-branes of superstring theory on the same footing with the stable D-branes. The fact that D-brane charges on a manifold $Y$ take values in K-theory groups of $Y$ implies that the Ramond-Ramond form fields are also objects in K-theory and not merely differential forms \cite{Witten:1999vg,Moore:1999gb}. Let us recall that the low energy description of superstring theory on manifold $Y$ contains
\begin{equation}
{\cal L} \ni \int_Y dC_p \,\wedge\, \star dC_p, \label{oldaction}
\end{equation}
where $C_p$ is the Ramond-Ramond $p$-form field. The exterior derivative $d$, the wedge product $\wedge$ and the Hodge dual operator $\star$ are all well-defined objects in de Rham cohomology, however, once $C_p$ is reinterpreted as an object in K-theory, it is not clear how to even define the Lagrangian \ref{oldaction}. This crisis of Lagrangian formulation may already be suggested by the presence of a self-dual Ramond-Ramond field strength in Type IIB superstring theory and is further supported by the discovery of non-Lagrangian phases in the partition functions of various string theory and M-theory vacua (see for example \cite{Diaconescu:2000wz}).

A possible interpretation of the crisis of the Lagrangian formulation is that the subtle K-theory features of superstring theory and M-theory is an emergent phenomenon, with the objects charged under K-theory, such as D-branes and Ramond-Ramond field, emerging as composites of some more elementary degrees of freedom that admit a conventional Lagrangian description. Interestingly, the possibility of objects classified by K-theory emerging as derived objects is realized in condensed matter systems \cite{Horava:2005jt} and classification of novel phases of matters, the so-called topological insulators and superconductors, using K-theory has been proposed in \cite{Kitaev:2009mg}.

The analogy with AdS/CFT correspondence and the surprising connection between K-theory and statistical systems motivate us to look for dual descriptions of unstable D-branes in terms of statistical systems in thermal equilibrium. The hope is that we will learn about the nature of time through this dualities. The case of bosonic D-branes was investigated in \cite{Balasubramanian:2006sg}\footnote{For the so-called full S-brane -- a process of a creation and a subsequent decay of an unstable D-brane, see \cite{Jokela:2007wi}.}, where the worldsheet description of the brane decay is related to a sequence of points of thermodynamic equilibrium of a grand canonical ensemble of point charges on a circle, the Dyson gas. There, subsequent instants of time are related to neighboring values of the average particle number $\bar{N}$, and the free energy of the system decreases as $\bar{N}$ increases (corresponding to later times), thus defining a thermodynamic arrow of time. In this part of the thesis, we will study an extension of that work, \textit{i.e.}, the case of non-BPS branes in superstring theory and find analogous results. 

\begin{savequote}[15pc]
\sffamily
Heart knows neither duality nor the limitations of space and time.
\qauthor{Sathya Sai Baba (b. 1926)}
\end{savequote}
\chapter{The Decaying Brane as a Paired Dyson Gas} \label{chapter:gas}

In this chapter, we will establish the duality between the non-BPS branes of Type II superstring theory, \textit{i.e.}, the D-branes with the ``wrong" dimensionality, with a statistical system in thermal equilibrium.

\section{Partition Function of the Decaying Brane}

Non-BPS brane of Type II superstring theory can be described as the exactly marginal deformation
\begin{eqnarray}
\delta S_{\rm bdry} = - \sqrt{2} \pi \lambda \int \frac{dt}{2 \pi} \psi_0 \, e^{X^0/\sqrt{2}}\otimes \sigma_1 \label{eq:deform} \ ,
\end{eqnarray}
where $\psi^0$ is the time component of the worldsheet fermion and $\sigma_1$ is a Chan-Paton factor associated with the boundary tachyon  \cite{Sen:2004nf}. The worldsheet correlation functions in this background then take the form
\begin{eqnarray}
\bar{A}_n &=& \int DX^0 \, DX^1 \cdots DX^9  \,\, D\psi^0 \, D\psi^1 \cdots D\psi^9 \nonumber \\
& & \qquad D\tilde{\psi}^0 \, D\tilde{\psi}^1 \cdots D\tilde{\psi}^9 \,\, e^{- S} \prod_{a = 1}^{n} V_a (z_a, \bar{z}_a) \label{eq:correlation} \ ,
\end{eqnarray}
where the action $S$ includes the boundary deformation (\ref{eq:deform}) and the $V_a$ are vertex operators. 

We can adopt convenient gauge choices \cite{Lambert:2003zr, Hwang:1991an}, where the dependence on the time component $X^0$ of the bosonic field takes a simple form
\begin{eqnarray}
V_a = e^{i \omega_a X^0} \, V_a^{\perp}(X^i, \psi^i, \tilde{\psi}^i, \cdots)
\end{eqnarray}
in the NS-NS sector, and
\begin{eqnarray}
V_a = e^{i \omega_a X^0} \, \Theta_{s_0} \tilde{\Theta}_{\tilde{s}_0} \, V_a^{\perp}(X^i, \psi^i, \tilde{\psi}^i, \cdots) 
\end{eqnarray}
in the R-R sector. Here, the spin fields $\Theta_{s_0} = e^{i s_0 H^0}$ are in the bosonized form. The virtue of this gauge choice is that we can omit the trivial spatial part of the calculation of the correlations functions and just focus on the temporal part.

The worldsheet correlation function (\ref{eq:correlation}) can then be evaluated by isolating the zero mode $x^0$ from the fluctuations as $X^0 = x^0 + X'^0$, and by expanding the boundary perturbation $e^{- \delta S_{\rm bdry}}$ in power series. The results can be written in a form
\begin{eqnarray}
\bar{A}_n = \int dx^0 e^{i x^0 \sum_a \omega_a} A_n(x^0) \ ,
\end{eqnarray}
where $A_n(x^0)$ can be written as an infinite power series of expectation values in Circular Unitary Ensembles \cite{Jokela:2005ha} (see also \cite{Shelton:2004ij}).

Let us first focus on the disk partition function $Z_{\rm open} = A_0 (x^0)$. The result reads {}\footnote{Here we are following the conventions of \cite{Balasubramanian:2004fz, Polchinski:1998rr} with $\alpha' = 1$.}
\begin{eqnarray}
 Z_{\rm open} & = & \sum_N (-)^N \frac{(\pi \lambda e^{x^0/\sqrt{2}})^{2N}}{N! \, N!} \, \left(\int_{-\pi}^{\pi} \prod_{i = 1}^{N} \frac{dt_i}{2 \pi} \prod_{i<j} |e^{it_i} - e^{it_j}|^2 \right)^2 \ . \label{eq:diskcano}
\end{eqnarray}
Noticing that the integrand is the Haar measure for integration over $U(N)$ matrices, giving $N!$, we then have the following closed form
\begin{equation}\label{eq:opendisk}
Z_{\rm open} = \frac{1}{1 + \pi^2 \lambda^2 e^{\sqrt{2} \, x^0}} \ .
\end{equation}

\section{Partition Function of the Paired Dyson Gas}
Now, we are ready to describe the statistical mechanical system that is dual to the non-BPS brane. First, let us consider a gas of charged particles with infinitely heavy masses, confined to live on a unit circle on a two-dimensional plane, the Dyson gas. Pairs of these charges, having positions $e^{i t_j}$, interact through a two-dimensional Coulomb potential, given by
\begin{equation}
V^{\rm Dyson}(t_i, t_j) = - Q^2 \log|e^{i t_i} - e^{i t_j}| \ ,
\end{equation}
where $Q$ is the charge of the particles.

Let us consider now a system consisting of two separate subsystems: $N_+$ particles with charges $Q_+ = 1$ living on a boundary of a two-dimensional disk denoted by $v$, and $N_-$ particles with charges $Q_- = - 1$ living on a boundary of a different two-dimensional disk denoted by $w$, with $N_+ = N_- \equiv N$. Furthermore, particles living on different disks do not interact. One can imagine preparing the system with a perfect insulator wall separating the particles into two subsystems.

\begin{figure}[ht]
\begin{center}
\noindent
\includegraphics[width=0.75\textwidth]{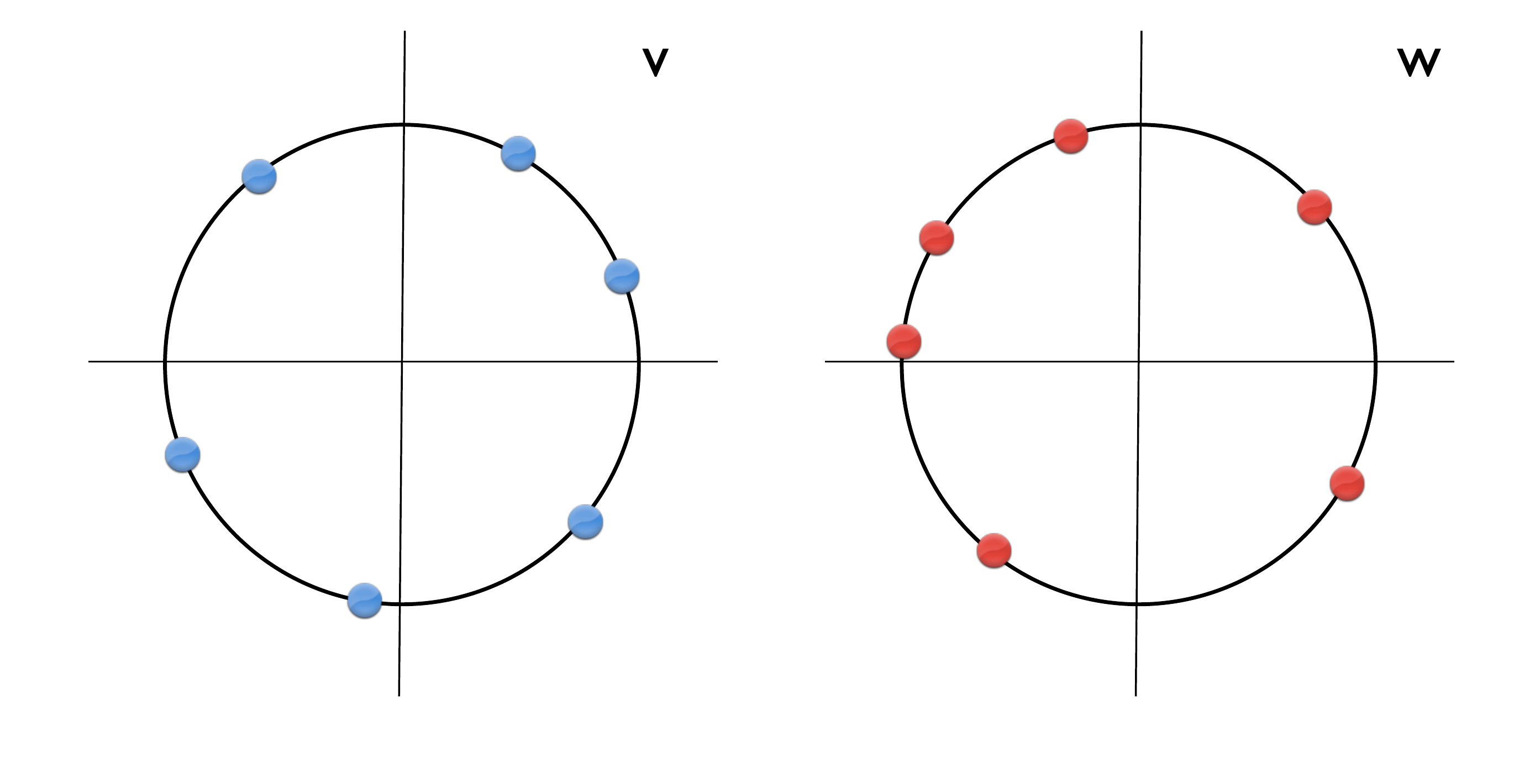}
\end{center}
\caption{\footnotesize Statistical dual of non-BPS brane: blue and red dots represent charges with $Q_+ = 1$ and $Q_- = - 1$, respectively.}
\label{fig:noinsertion}
\end{figure}

Let us suppose that this system is immersed into a heat bath at inverse temperature $\beta=1/T$. We can then aim to make contact with the disk partition function, by moving to consider the grand canonical ensemble with a particle reservoir with chemical potentials $\mu_+,\mu_-$. The grand canonical partition function of this system will then be
\begin{eqnarray}
Z_G &=& \sum_{N_+ = N_-} \left[\frac{{z_+}^{N_+}}{N_+!} \int_0^{2\pi} \prod_{i = 1}^{N_+} \frac{dt_i}{2\pi} \, e^{\left(- \beta \sum_{i<j} V^{\rm Dyson}(t_i,t_j) \right)} \right] \nonumber \\
& & \qquad \qquad \left[\frac{{z_-}^{N_-}}{N_-!} \int_0^{2\pi} \prod_{i = 1}^{N_-} \frac{d\tau_i}{2\pi} \, e^{\left(- \beta \sum_{k<l} V^{\rm Dyson}(\tau_k,\tau_l) \right)} \right] \label{eq:dysoncano} \ ,
\end{eqnarray}
where, as before, $V^{\rm Dyson}(x,y) = - \log |e^{ix} - e^{iy}|$. Here, $N_\pm!$ take into account that the system consists of identical particles of two types and $z_\pm = e^{\beta \mu_\pm}$ are the corresponding fugacities. At inverse temperature $\beta=2$ we then obtain
\begin{equation}\label{eq:dysonpart}
Z_G = \frac{1}{1 - z_+ z_-} \ .
\end{equation}

Comparing (\ref{eq:dysoncano}) with the open string disk partition function calculation (\ref{eq:diskcano}), by setting
\begin{equation}
z_+ = -z_- = \pi \lambda e^{x^0/\sqrt 2} \ ,
\end{equation}
one finds
\begin{equation}
 Z_G = \frac{1}{1+\pi^2\lambda^2 e^{\sqrt 2 x^0}} = Z_{\rm open} \ .
\end{equation} 
However, we would like to avoid having complex chemical potentials. If we want to end up with positive fugacity $z_-$, a simple remedy is to insert a $(-)^N$ into the sum (\ref{eq:diskcano}), and consider, instead of $Z_{\rm open}$, the quantity $Z'_{\rm open} \equiv \langle i^{\hat{N}_T} \rangle_{\rm brane}$. The virtue of this is that it can be related to the grand canonical partition function with positive fugacities. Another way of achieving this is to analytically continue $z_- \to -z_-$, while keeping $\lambda$ positive, so that we have one-to-one correspondence with time $x^0$ of the decaying brane and a positive fugacity,
\begin{equation}
 z \equiv |z_+| = |z_-| = \pi\lambda e^{x^0/\sqrt 2} \ .
\end{equation}
See \cite{Balasubramanian:2006sg} for more extensive discussion on this issue.
 
\section{Correlation Functions}

To further establish the relation between the non-BPS brane and the paired Dyson gas, we also need to prescribe how to calculate the worldsheet correlation functions in the Dyson gas picture. For clarity, in this subsection we will explicitly write $z_\pm$ in favor of $z$.

\subsection{NS-NS Vertex Operators}

The statistical dual will correspond to inserting a set of test charges with charges $q_+^a = i \omega_a /\sqrt{2}$ at $v = z_a$ and another set of test charges with charges $q_-^a = - i \omega_a /\sqrt{2}$ at $w = z_a$.

\begin{figure}[ht]
\begin{center}
\noindent
\includegraphics[width=0.75\textwidth]{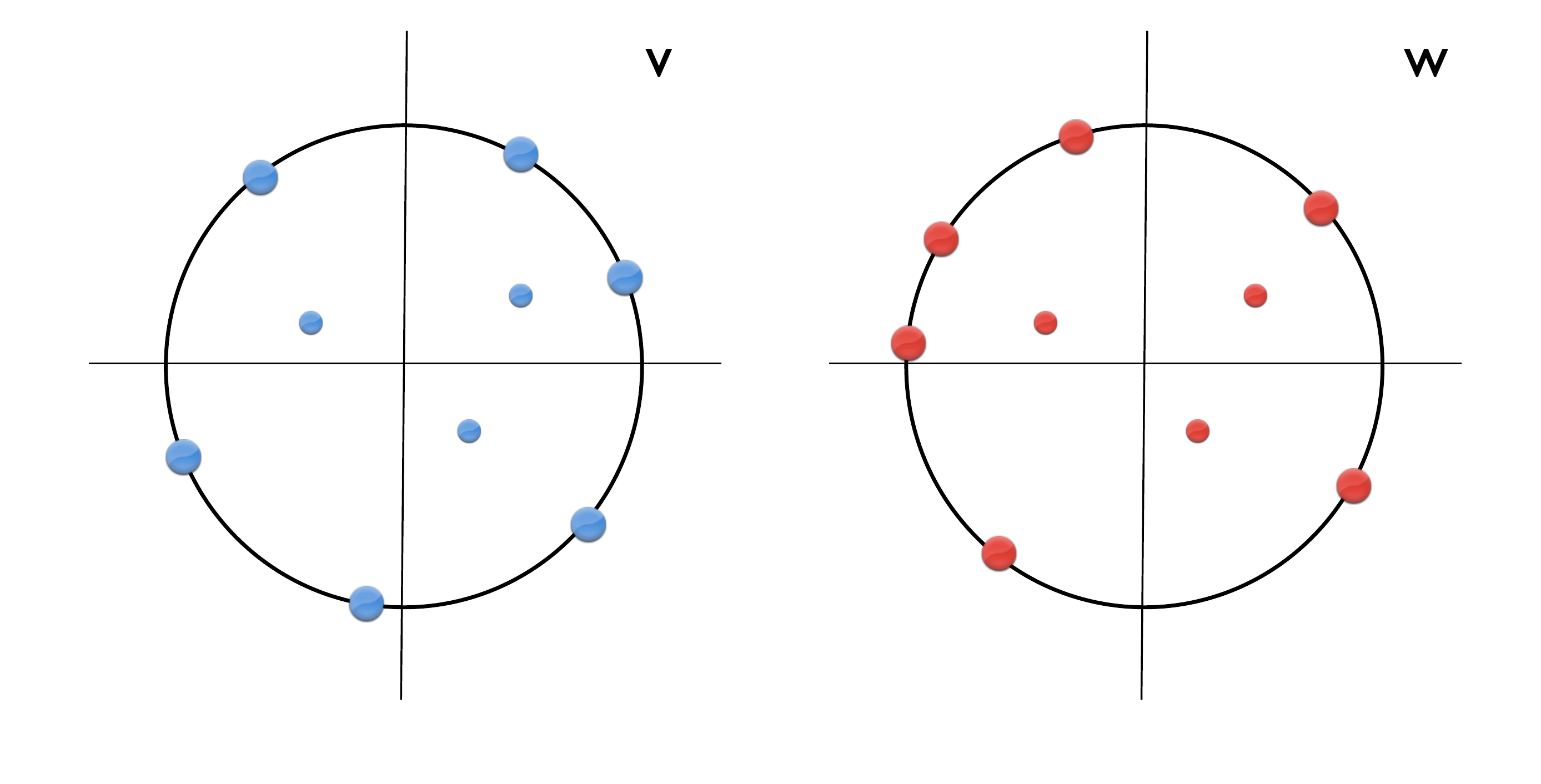}
\end{center}
\caption{\footnotesize Statistical dual of NS-NS vertex operators insertion: smaller blue and red dots represent test charges with $q_+^a = i \omega_a /\sqrt{2}$ at $v = z_a$, and $q_-^a = - i \omega_a /\sqrt{2}$ at $w = z_a$, respectively.}\label{fig:NS-NS}
\end{figure}

To be more precise, let us consider inserting vertex operators in the form of 
\begin{equation}
V_s = \prod_{a = 1}^n e^{i \omega_a X^0(z_a, \bar{z}_a)}. 
\end{equation}
To compute the correlation functions of this kind in the Dyson gas picture, we need to compute the grand canonical ``expectation value''  of the quantity $e^{- \beta \, V^{\rm bulk}}$, where $V^{\rm bulk}$ is the interaction among bulk charges. This is defined as
\begin{eqnarray}
\langle \prod_{a = 1}^n e^{i \omega_a X^0(z_a, \bar{z}_a)} \rangle &\equiv& \sum_N \left(\frac{z_+^N}{N!} \int \prod_{i = 1}^{N} \frac{dt_i}{2 \pi} \, e^{\left[ -\beta \left(V_+^{\rm Dyson} + V_+^{\rm bulk - Dyson} \right) \right]} \cdot e^{- \beta \, V_+^{\rm bulk}} \right) \nonumber \\
& &\qquad \left(\frac{z_-^N}{N!} \int \prod_{i = 1}^{N} \frac{d\tau_i}{2 \pi} \, e^{\left[ -\beta \left(V_-^{\rm Dyson} + V_-^{\rm bulk - Dyson} \right) \right]} \cdot e^{- \beta \, V_-^{\rm bulk}} \right), \nonumber \\
\end{eqnarray}
where $V^{\rm bulk - Dyson}$ is the interaction between the bulk charges and the Dyson gas particles.

\subsection{R-R Vertex Operators}

For correlation functions involving R-R sector, we need to do two things:
\begin{itemize}
\item add $|k|$ Dyson particles with charges $Q = {\rm sign(k)}$ to the respective two-dimensional plane (\emph{i.e.}, $Q = + 1$ to $v$-plane, $Q = - 1$ to $w$-plane). Here, $k$ is defined as $k \equiv - \sum_a (s_a + \tilde{s}_a) \in \mathbb{Z}$;
\item insert test charges with $q_+^a = (i \omega_a /\sqrt{2} + (s_a + \tilde{s}_a)/2)$ at $v = z_a$, and $q_-^a = (- i \omega_a /\sqrt{2} + (s_a + \tilde{s}_a)/2)$ at $w = z_a$, respectively.
\end{itemize}

\begin{figure}[ht]
\begin{center}
\noindent
\includegraphics[width=0.75\textwidth]{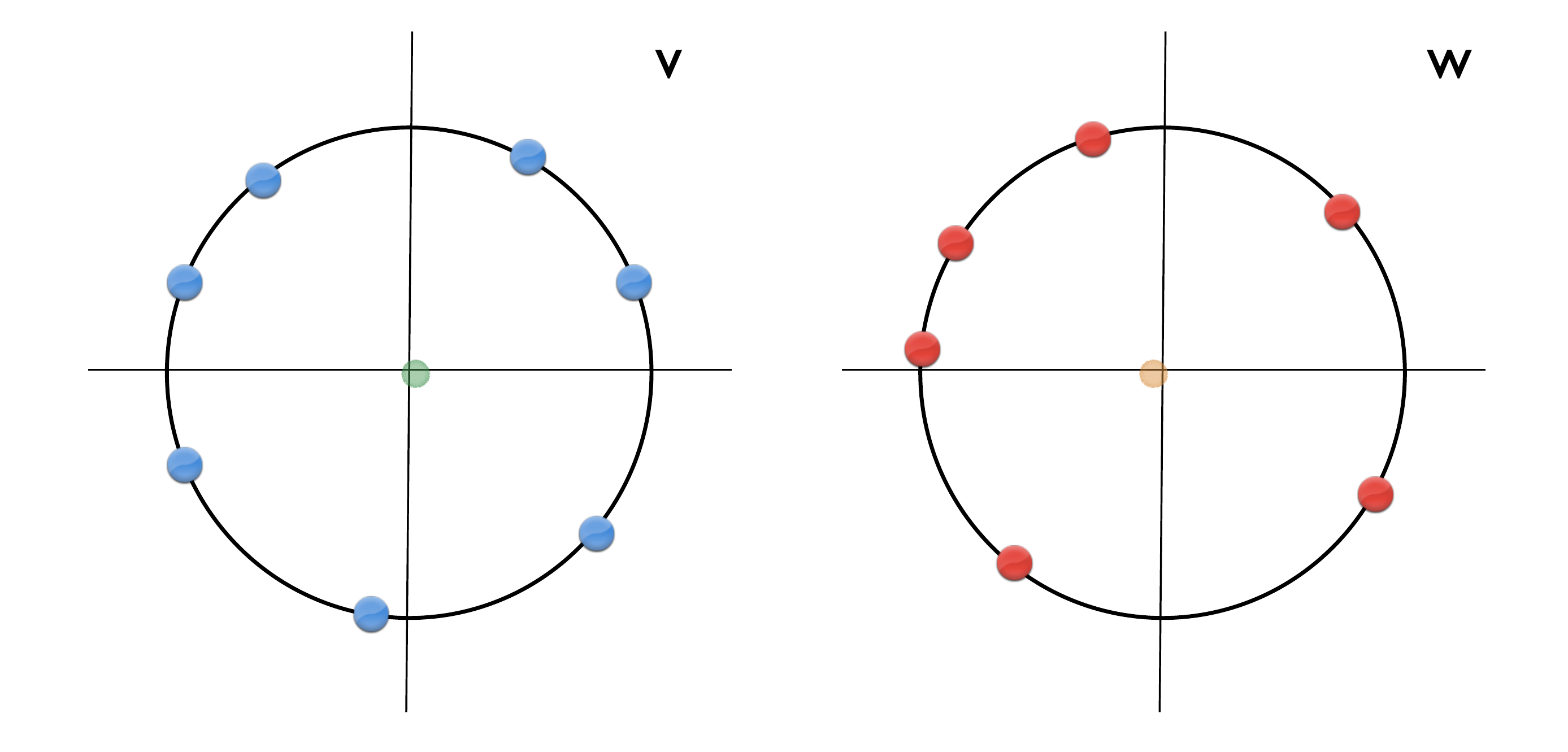}
\end{center}
\caption{\footnotesize Statistical dual of R-R vertex operators insertion: green and orange dots represent test charges with $q_+^a = (i \omega_a /\sqrt{2} + (s_a + \tilde{s}_a)/2)$ at $v = z_a$, and $q_-^a = (- i \omega_a /\sqrt{2} + (s_a + \tilde{s}_a)/2)$ at $w = z_a$, respectively. Here, as an example, we have taken $s_a = \tilde{s}_a = 1/2$ so that $k = 1$. Thus, there is an extra $q = 1$ Dyson particle on the boundary of $v$-disk.}\label{fig:R-R}
\end{figure}

Let us consider inserting vertex operators in the form of 
\begin{equation}
\prod_{a = 1}^n \Theta_{s_a}(z_a) \tilde{\Theta}_{\tilde{s}_a}(\bar{z}_a) \, e^{i \omega_a X^0(z_a, \bar{z}_a)}. 
\end{equation}
The prescription for calculating the expectation value in the Dyson gas picture then becomes
\begin{eqnarray}
\langle \prod_{a = 1}^n \Theta_{s_a}(z_a) \tilde{\Theta}_{\tilde{s}_a}(\bar{z}_a) e^{i \omega_a X^0(z_a, \bar{z}_a)} \rangle &\equiv& \sum_N \left(\frac{z_+^{N + |k|}}{(N + |k|)!} \int \prod_{i = 1}^{N + |k|} \frac{dt_i}{2 \pi} \, e^{\left[ -\beta \left(V_+^{\rm Dyson} + V_+^{\rm bulk - Dyson} \right) \right]}  \cdot \, e^{- \beta V_+^{\rm bulk}}\right) \nonumber \\
& & \qquad \left(\frac{z_-^N}{N!} \int \prod_{i = 1}^{N} \frac{d\tau_i}{2 \pi} \, e^{\left[ -\beta \left(V_-^{\rm Dyson} + V_-^{\rm bulk - Dyson} \right) \right]} \cdot e^{- \beta V_-^{\rm bulk}} \right) \nonumber \\
\end{eqnarray}
for $k \geq 0$, or
\begin{eqnarray}
\langle \prod_{a = 1}^n \Theta_{s_a}(z_a) \tilde{\Theta}_{\tilde{s}_a}(\bar{z}_a) e^{i \omega_a X^0(z_a, \bar{z}_a)} \rangle &\equiv& \sum_N  \left(\frac{z_-^{N + |k|}}{(N + |k|)!} \int \prod_{i = 1}^{N + |k|} \frac{d\tau_i}{2 \pi} \, e^{\left[ -\beta \left(V_-^{\rm Dyson} + V_-^{\rm bulk - Dyson} \right) \right]} \cdot \, e^{- \beta V_-^{\rm bulk}}\right) \nonumber \\
& & \qquad \left(\frac{z_+^{N}}{N!} \int \prod_{i = 1}^{N} \frac{dt_i}{2 \pi} \, e^{\left[ -\beta \left(V_+^{\rm Dyson} + V_+^{\rm bulk - Dyson} \right) \right]} \cdot e^{- \beta V_+^{\rm bulk}} \right) \nonumber \\
\end{eqnarray}
for $k \leq 0$.

\begin{savequote}[19pc]
\sffamily
What then is time?\\
If no one asks me, I know what it is.\\
If I wish to explain it to him who asks, I do not know.
\qauthor{Augustine of Hippo (354 - 430)}
\end{savequote}
\chapter{The Emergence of Time} \label{chapter:time}

The dualities between rolling tachyon backgrounds and Dyson gas systems, such as prescribed in the previous chapter, have been proven useful as tools for calculating observables in the rolling tachyon backgrounds \cite{Jokela:2008zh,Jokela:2009zz,Jokela:2009fd}. However, these dualities are not only tools for calculation, they can also be thought of as reformulation of the unstable D-branes, which are K-theory objects and whose original descriptions include time, in terms of Dyson gas systems, which are statistical systems in thermal equilibrium that consist of objects without K-theory charges and whose descriptions do not include time.

In our present case of the paired Dyson gas picture, the time of the target spacetime $x^0$ has been replaced by the chemical potential $\mu$ of the statistical system. Different points of thermodynamic equilibrium characterized by different values of the chemical potential correspond to the different instants in time for the decaying non-BPS brane.

\section{Time as the Average Particle Number}
We can now give an interpretation to time as the average particle number in the statistical system. The average particle numbers for both sectors of the Dyson gases are
\begin{equation}
 \bar{N} = \frac{ z^2}{1 - z^2} = \frac{\pi^2 \lambda^2 e^{\sqrt{2} x^0}}{1 - \pi^2 \lambda^2 e^{\sqrt{2} x^0}}.
\end{equation}
Solving for $x^0$ we get
\begin{equation}
x^0 = \frac{1}{\sqrt{2}} \log \frac{\bar{N}}{1 + \bar{N}} - \sqrt{2} \log \pi \lambda \ .
\end{equation}

At the past infinity, $x^0 \rightarrow - \infty$, we have vanishing fugacity and average particle number. Then $\bar{N}$ increases monotonically as a function of the fugacity corresponding to later time values of $x^0$.

$\bar{N}$, like time, is a continuous quantity. However, the underlying physical quantity is $N$, which is discrete. $N$ fluctuates around $\bar{N}$, and the fluctuation is given by
\begin{equation}
\frac{\delta N}{\bar{N}} = \frac {\sqrt{\langle N^2 \rangle - \langle N \rangle^2}}{\bar{N}} = \frac{1}{z}  = \frac{e^{- x^0/\sqrt{2}}}{\pi \lambda} \ .
\end{equation}
At early times, $\delta N$ is large, but later $\bar{N}$ becomes more sharply defined. We can interpret this as a continuous time emerging from an underlying discrete variable $N$ in the large $N$ limit.

\section{Thermodynamic Arrow of Time}
Let us consider the grand potential of the statistical system
\begin{equation}
\Omega (z, T, V) = - T \log Z_G.
\end{equation}
Using Legendre transformation, we can get the Helmholtz free energy
\begin{eqnarray}
A(\bar{N}, T, V) = - \frac{1}{2} \left[\log (\bar{N} + 1)^{(\bar{N} + 1)} - \log \bar{N}^{\bar{N}} \right].
\end{eqnarray}
We can then rewrite $A = - T I_S$ where $I_S$ is the Shannon entropy. Shannon entropy in the particle number distribution is given by
\begin{eqnarray}
I_S = - \sum_n p(n) \log p(n); \,\,\,\,\, p(n) = \frac{(z^2)^n}{Z_G},
\end{eqnarray}
where $p(n)$ is the probability that there are $n$ positively charged particles on $v$-disk and $n$ negatively charged particles on $w$-disk.  Since $I_s$ increases as $\bar{N}$ increases, the free energy decrease as $\bar{N}$ increases. Recalling that the increase of $\bar{N}$ marks the passage of time, we can interpret this as a thermodynamic arrow of time.

\begin{savequote}[12pc]
\sffamily
When will I learn? The answers to life's problems aren't at the bottom of a bottle, they're on TV!
\qauthor{Homer Simpson}
\end{savequote}
\chapter{Discussions}
In this part of the thesis, we have reformulated the worldsheet description of non-BPS brane decay in terms of the statistical mechanics of the paired Dyson gas. As in the case of bosonic D-brane decay, the progress of time was marked by the increase of the average number of particles in the gas. The decrease of free energy marked the passage of time and thus gave rise to a thermodynamic arrow of time.

We would like to emphasize that even though one might have expected a non-equilibrium flow to realize the flow of time, this expectation is somewhat misleading. This is because there is already a physical time in non-equilibrium statistical mechanics. Thus, one can not expect to have time emerging from a non-equilibrium statistical mechanical system. Instead, in this set-up, we found that each instant of time was described by a different point of equilibrium in a statistical system.

In the known examples of emergence spatial dimensions, we learned that the classical or geometrical picture of spacetime is only valid in a certain limit. Here, we can also see that it is true, in the sense that the classical geometry of time, \emph{i.e.}, $\mathbb{R}$ emerges in the limit of large number of particles. Furthermore, we would also like to emphasize that this classical geometry of time does not correspond to a thermal state of the paired Dyson gas, but corresponds to the \emph{whole} ensemble of the gas (\emph{c.f.}, black hole does not correspond to the thermal state but the whole ensemble of its microstates \cite{Balasubramanian:2007qv}).

If the notion of spacetime being emergent is true, then it is a possibility that spacetime symmetries are emergent too. In our case, the spacetime picture of the homogeneously decaying unstable D-brane is believed to be supersymmetric at late times, while on the other hand the Dyson gas is not. Then, one could ask whether the spacetime supersymmetry must be emergent too and how this is realized in Dyson gas language. This certainly deserves more study.







\bibliographystyle{utphys}
\renewcommand{\bibname}{References} 

\bibliography{References}









\end{document}